\pdfoutput=1
\documentclass[pdflatex,sn-nature]{sn-jnl}% Style for submissions to Nature Portfolio journals

\usepackage{graphicx}%
\usepackage{multirow}%
\usepackage{amsmath,amssymb,amsfonts}%
\usepackage{amsthm}%
\usepackage{mathrsfs}%
\usepackage[title]{appendix}%
\usepackage{xcolor}%
\usepackage{textcomp}%
\usepackage{manyfoot}%
\usepackage{booktabs}%
\usepackage{algorithm}%
\usepackage{algorithmicx}%
\usepackage{algpseudocode}%
\usepackage{listings}%
\usepackage{subcaption}% 2x3 sub-figure layout for Morris mu* heatmaps (Figure 2)

\usepackage{tikz}% Study-design flowchart (Figure 1)
\usetikzlibrary{arrows.meta,positioning}%
\usepackage{float}% [H] placement to pin the study-design figure near its reference

\newcommand{\mainref}[2]{#2}% number as written in the supplement
\usepackage{flafter}% loaded by supplement.tex; applies to the whole document

\begin{document}
\addtocontents{toc}{\protect\setcounter{tocdepth}{-10}}%

\title[Reliability Atlas: Pulse Physiology Engine]{An Open Reliability Atlas of Sensitivity and Failure Modes in the Pulse Physiology Engine}

%%%%%%%%%%%%%%%%%%%%%%%%%%%%%%%%%%%%%%%%%%%%%%%%%%%%%%%%%%%%%%%%%%%%%%
%% AUTHOR + AFFILIATIONS                                            %%
%% Replicated byte-for-relevant-byte from ../preprint/manuscript.md %%
%% lines 3-19 (after the YAML frontmatter and the Affiliations      %%
%% block). Single corresponding author across four affiliations.    %%
%% Underscore-S in 'UMR_S' escaped as 'UMR\_S'. French accents      %%
%% rendered with LaTeX escapes (\`e, \'e). Street and postcode      %%
%% omitted in \orgaddress because the monograph does not record     %%
%% them; \city / \state / \country are sufficient per sn-jnl.       %%
%%                                                                  %%
%% Corresponding-author primary affiliation: marker on \affil*[4]   %%
%% (CHU Sainte-Justine / Universite de Montreal PICU) per           %%
%% maintainer directive in S028 Step 2 (2026-05-23). Affiliations   %%
%% 1-4 retain their monograph ordering; only the * marker moved.    %%
%%%%%%%%%%%%%%%%%%%%%%%%%%%%%%%%%%%%%%%%%%%%%%%%%%%%%%%%%%%%%%%%%%%%%%

\author*[1,2,3,4,5]{\fnm{Santino} \sur{Nanini}}\email{santino.nanini@umontreal.ca}

\author[3,4,6]{\fnm{Arnaud} \sur{Wiedemann}}

\author[3]{\fnm{Yassir} \sur{Mamouni}}

\author[4,5]{\fnm{Cyril} \sur{Schweitzer}}

\author[1,2,3]{\fnm{Philippe} \sur{Jouvet}}

\author[1,2,3]{\fnm{Michael} \sur{Sauthier}}

\affil*[1]{\orgdiv{Pediatric Intensive Care Unit, Department of Pediatrics}, \orgname{CHU Sainte-Justine, Universit\'e de Montr\'eal}, \orgaddress{\city{Montreal}, \state{QC}, \country{Canada}}}

\affil[2]{\orgdiv{Clinical Decision Support System Artificial Intelligence Health Cluster in Acute Child Care (PEDIATRICS)}, \orgname{CHU Sainte-Justine Centre Hospitalier Universitaire M\`ere-Enfant}, \orgaddress{\city{Montr\'eal}, \country{Canada}}}

\affil[3]{\orgdiv{Research Center}, \orgname{Sainte-Justine University Hospital}, \orgaddress{\city{Montreal}, \state{Quebec}, \country{Canada}}}

\affil[4]{\orgdiv{Pediatric Intensive Care Unit, Department of Pediatrics}, \orgname{Children Hospital, CHRU Nancy}, \orgaddress{\city{Nancy}, \country{France}}}

\affil[5]{\orgdiv{UR 3450 DevAH}, \orgname{Universit\'e de Lorraine}, \orgaddress{\city{Nancy}, \country{France}}}

\affil[6]{\orgdiv{INSERM UMR\_S 1256, Nutrition-G\'en\'etique et Exposition aux Risques Environnementaux}, \orgname{Universit\'e de Lorraine, Facult\'e de M\'edecine de Nancy}, \orgaddress{\city{Vand\oe{}uvre-l\`es-Nancy}, \country{France}}}

%%%%%%%%%%%%%%%%%%%%%%%%%%%%%%%%%%%%%%%%%%%%%%%%%%%%%%%%%%%%%%%%%%%%%%
%% ABSTRACT                                                         %%
%% 150 words (npj cap 150). 8 sentences (S028 #3 + 2026-06 retouch, %%
%% + 6403787 sensitivity sentence, + 2026-08-18 healthy-lung scope). %%
%% refinement #3. Three headline findings, one number each. Every   %%
%% numerical claim is registered in paper_to_monograph_map.md       %%
%% with its monograph claim_id and expected_occurrences count.      %%
%%%%%%%%%%%%%%%%%%%%%%%%%%%%%%%%%%%%%%%%%%%%%%%%%%%%%%%%%%%%%%%%%%%%%%

\abstract{Pulse is an actively maintained open-source adult-physiology simulator, yet its maintainers report no formal sensitivity analysis of its operational inputs. We present the first open reliability atlas across its critical-care envelope: Morris elementary-effects sampling of seven inputs, four severity scenarios, male and female bodies (1{,}280 simulations). Sensitivity is skewed: three inputs carry a median 93~\% of each output's total. A four-flag classifier reveals failure modes: two operator-actionable, one sex-stratified. Tidal-volume oscillation (20.6~\% of runs) is consistent with patient--ventilator dyssynchrony; locking the VC-AC trigger to $-2$~cmH$_2$O collapses that variability. At FiO$_2$~=~1.0, transient hyperoxia makes 120-second runs underestimate plateau PaO$_2$; 300-second healthy-lung reruns give a median 643~mmHg. Solver divergence was all-male in an input-extreme regime (20/1{,}280); reducing the male blood-volume baseline to the 4{,}082~mL floor eliminated it. Runs and flags are released openly, showing users which regions need checking before training on Pulse output.}


\maketitle

%%%%%%%%%%%%%%%%%%%%%%%%%%%%%%%%%%%%%%%%%%%%%%%%%%%%%%%%%%%%%%%%%%%%%%
%% BODY -- Results, Discussion, Methods.                            %%
%%%%%%%%%%%%%%%%%%%%%%%%%%%%%%%%%%%%%%%%%%%%%%%%%%%%%%%%%%%%%%%%%%%%%%

\section{Introduction}\label{sec:intro}

Longitudinal health data that capture a patient's evolving state are often scarce and incomplete~\cite{katsoulakis2024digitaltwins}. Digital twins for health (virtual, often mechanistic models of human physiology used for in-silico trials, medical-device design, and synthetic-patient generation) are proliferating~\cite{katsoulakis2024digitaltwins}, and synthetic data can help address data scarcity, account for under-represented groups and simulate unseen scenarios~\cite{vanbreugel2024synthetic}. For example, simulator data that include infrequent adverse conditions have been used to pre-train a glucose-forecasting and hypoglycemia-detection model later evaluated on real-world datasets~\cite{hwang2025sim2real}. Outside medicine, a recent technical report on a large multimodal model attributes most of its post-training gains, under a fixed optimisation procedure, to the scale, diversity and verifiability of its synthesised training data and environments~\cite{deepseek2026v41flash}. In biomedicine, a perspective on synthetic data argues that ``the value of synthetic data will likely depend less on generation alone and more on careful integration with real-world data, layered validation and digital twin systems''~\cite{victoriano2026sim2real}.

Mechanistic simulators generate physiological time series from explicitly specified inputs, with model-internal states accessible by construction, but a mechanistic origin does not make their outputs automatically reliable. The Pulse Physiology Engine (Kitware, Apache-2.0) is an actively maintained open-source whole-body simulator~\cite{bray2019}. It represents each physiological system as a small network of coupled compartments (a lumped-parameter approach) and simulates cardiovascular, respiratory, gas-exchange and pharmacological dynamics in real time (Supplement~\S\,\ref{sec:s1}). Pulse has been used for ventilator-safety what-if analysis~\cite{webb2020}, validation of its own ventilator model against a physical reference simulator~\cite{webb2022vent}, medical education and simulation training~\cite{pulseinaction}, general-purpose computational physiology~\cite{bray2019}, and as a source of simulated data for learning and benchmarking ventilator-control policies in Acute Respiratory Distress Syndrome (ARDS)~\cite{ward2024ventilator}. A review of synthetic data in biomedicine cites Pulse as an example of a mechanistic generator~\cite{vanbreugel2024synthetic}.

Yet, to our knowledge, no public map of how the engine responds across its user-facing \emph{input} envelope has been published. The maintainers state on the project's public FAQ that ``we have not performed a systematic forward propagation or inverse quantification of model uncertainty, nor have we conducted a formal sensitivity analysis''~\cite{pulsefaq}. The only sensitivity study in the upstream repository (2020, tagged \texttt{STUDY\_SENSITIVITY\_ANALYSIS\_1\_0\_0}) varies internal model parameters that cannot be measured clinically (cardiovascular and respiratory resistances and compliances) for calibration purposes, not operational inputs such as heart rate, blood pressure or ventilator settings. Other assessments probe the engine's \emph{outputs}, either by validating them against reference data~\cite{webb2022vent} or, in the study by Cornish et al.~\cite{cornish2023}, by causal falsification on MIMIC-III sepsis trajectories, which located intervention-response errors mainly in electrolyte, metabolic, acid--base and dermal outputs. Webb et al.~\cite{webb2020} did sweep inputs (12{,}642 patient and ventilator combinations spanning respiratory compliance, diffusion impairment and positive end-expiratory pressure) and rank parameters by outcome correlation, but to score the \emph{mismatch} between two equal-sized 70~kg bodies sharing one ventilator against a single composite safety outcome, under pressure-mode ventilation with respiratory rate, I:E ratio and respiratory resistance held fixed. That design resolves neither each input's influence on each physiological output, nor input clamping, nor solver divergence. Without a shared public map, downstream users may need to establish a use-case-specific working envelope independently.

We focus on the ARDS severity axis for two reasons. First, Pulse models mechanical ventilation of an impaired lung, a regime whose downstream reliability remains incompletely characterised. Second, mapping the adult engine's sensitivity and failure modes is a prerequisite for adapting it to pediatric ARDS, where we ultimately aim to use simulated cohorts to train machine-learning models that predict ventilation-dependent blood gases in critically ill children; an adaptation cannot be assessed against an unmapped adult substrate.

Our objective was therefore to build the first open reliability atlas of Pulse's adult critical-care operational envelope, around three questions: (1)~how strongly does each input drive each output, and where does that influence become non-linear or interactive; (2)~does the engine run a submitted setting as given, or silently clamp it; and (3)~where does the solver diverge or fail to reach steady state?

We varied seven user-facing operational inputs across their accepted ranges: heart rate (HR), systolic and diastolic blood pressure (SBP, DBP), respiratory rate (RR), tidal volume (Vt), inspired oxygen fraction (FiO$_2$) and positive end-expiratory pressure (PEEP), using a Morris elementary-effects design~\cite{morris1991,campolongo2007} (a global screening method that varies one input at a time along randomised trajectories). Six steady-state outputs were tracked: HR, mean arterial pressure (MAP), delivered tidal volume, peripheral oxygen saturation (SpO$_2$), and arterial oxygen and carbon dioxide partial pressures (PaO$_2$, PaCO$_2$). Four clinical scenarios (a healthy baseline and three ARDS severities) crossed with two adult reference bodies, male and female, gave 1{,}280 simulations, collectively called the \emph{sweep}. A four-flag classifier labelled every run as refused at setup, diverged, physiologically implausible or not at steady state; flagged runs are categorised rather than discarded, so the atlas stays complete. We release this run-level classifier and a per-cell map that sets review priority. Neither is an external physiological validation, and the runs that pass screening are not a representative subset. The core sampling design, classifier thresholds and primary analysis plan were preregistered, deviations are reported explicitly, and the pipeline is reproducible from a single Docker image with a fixed seed (open-source, Apache-2.0).

\section{Results}\label{sec:results}

\begin{table}[!h]
\centering
\caption{\textbf{Key terms used in this paper.} Morris terminology (trajectory, elementary effect, $\mu^{*}$, clean pair) is defined in Supplement~\S\,\ref{sec:s3-terms}.}
\label{tab:terms}
\begin{tabular}{lp{0.70\textwidth}}
\toprule
Term & Definition \\
\midrule
\textbf{Scenario} & One of four impairment-fraction settings: \textbf{Healthy} baseline (0.00), \textbf{Mild} (0.26), \textbf{Moderate} (0.60) and \textbf{Severe} ARDS (0.95). Mild, Moderate and Severe are Berlin Definition bands, set by the arterial PaO$_2$/FiO$_2$ ratio at PEEP $\geq 5$~cmH$_2$O~\cite{ards2012}. The impairment fraction is Pulse's own severity setting, calibrated on the male reference body so that PaO$_2$/FiO$_2$ at a fixed reference ventilation falls in the matching band (Supplement~\S\,\ref{sec:s5-berlin}; Methods~\S\ref{sec:methods-sweep}). Data files and run identifiers code these four scenarios S1--S4, in this order. \\
\addlinespace
\textbf{Cell} & One (scenario, sex) slice of the atlas, labelled by scenario and body (\textbf{M} male, \textbf{F} female), e.g.\ Healthy-F (S1\_F in the data files) for the Healthy scenario on the female body. The $4 \times 2 = 8$ cells each contain 160 runs and form the rows of the curation map (Methods~\S\ref{sec:methods-traffic}). \\
\addlinespace
\textbf{Simulation (run)} & One Pulse run at a single point of the seven-input space (HR, SBP, DBP, RR, Vt, FiO$_2$, PEEP), recorded for 120~s after a 120~s ventilator-settling phase; the six outputs (HR, MAP, Vt delivered, SpO$_2$, PaO$_2$, PaCO$_2$) are sampled every 5~s (Methods \S\ref{sec:methods-sweep}). \\
\bottomrule
\end{tabular}
\end{table}

\subsection{Overview of the reliability atlas}\label{sec:results-overview}

\begin{figure}[!t]
\centering
\includegraphics[width=0.748\textwidth]{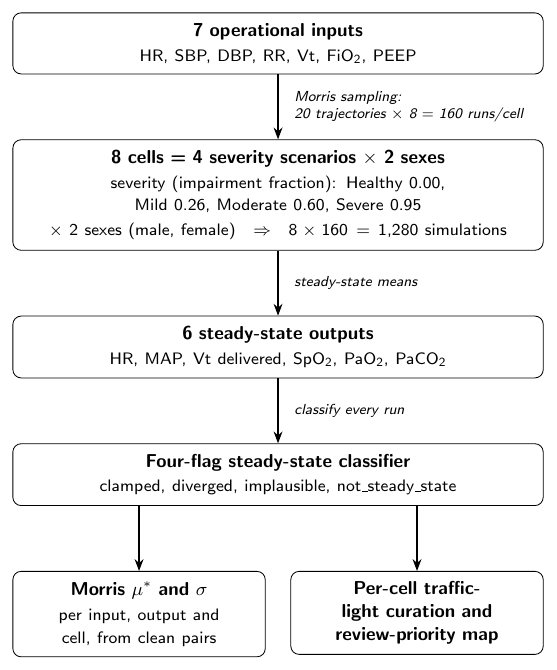}
\caption{\textbf{Study design of the 1{,}280-run sweep.} Seven user-facing operational inputs are screened with a Morris elementary-effects design of 20 trajectories per cell, each visiting $7{+}1 = 8$ input points (160 runs per cell). The same Morris sample is replayed in all eight cells (four severity scenarios with impairment fractions 0.00 / 0.26 / 0.60 / 0.95, crossed with the male and female reference bodies; $8 \times 160 = 1{,}280$ simulations), so differences between cells reflect the scenario and body, not where the sampler happened to look. Each step changes exactly one input, except for the pulse-pressure clip (Supplement~\S\,\ref{sec:s3-clip-cost}). A four-flag classifier labels every run before analysis, and Morris indices use only clean pairs (the two runs on either side of one Morris step, neither flagged). Cell-label key: Table~\ref{tab:terms}.}
\label{fig:design}
\end{figure}

The atlas comprises 1{,}280 simulations in eight cells of 160 runs, one cell per clinical scenario (Healthy, Mild, Moderate and Severe ARDS; Table~\ref{tab:terms}) and sex (Figure~\ref{fig:design}). Every run carries the four preregistered flags (\texttt{clamped}, \texttt{diverged}, \texttt{implausible} and \texttt{not\_steady\_state}; Table~\ref{tab:flag_defs}), and Methods~\S\ref{sec:methods-detector} lists the hard physiological bounds behind \texttt{implausible}.

Three of the four preregistered hypotheses (H1--H4, Methods~\S\ref{sec:methods-stats}) are resolved by the flag counts (Table~\ref{tab:flag_summary}). H3 and H4 are supported: solver divergence occurs outside the previously characterised ARDS instability band (20 / 1{,}280 runs) and only on the male body (\S\ref{sec:results-sex}), an asymmetry that alone satisfies H4's criterion of a sex difference in clamping, divergence or $\mu^{*}$ ranking. H2 (silent input clamping) is \emph{not supported}: the \texttt{clamped} flag never fired (0 / 1{,}280), and because this log-based flag cannot detect silent substitution, a post hoc input-instantiation audit provided the complementary check. H1, which concerns input importance rather than any flag, is addressed in \S\ref{sec:results-morris}.

The audit compared every submitted value with the value Pulse serialises after ingestion and found exact agreement for all 10{,}160 input contracts (one value submitted in one configuration block) that reached ingestion. The remaining 80 of the 10{,}240 contracts are the ventilator contracts of the 20 runs that failed during patient stabilisation, before the ventilator action was submitted, and were excluded. Both results hold only within the sampled envelope. Pulse makes one explicit downstream adjustment: where FiO$_2$~=~1.0 was requested and the ventilator action was reached (195 runs), the ventilator model caps the inspired oxygen fraction at 0.9996 and logs an error, because the ambient mixture already carries CO$_2$ (Supplement~\S\,\ref{sec:s3-clamp}).

\begin{figure}[!htbp]
\centering
\includegraphics[width=0.95\textwidth]{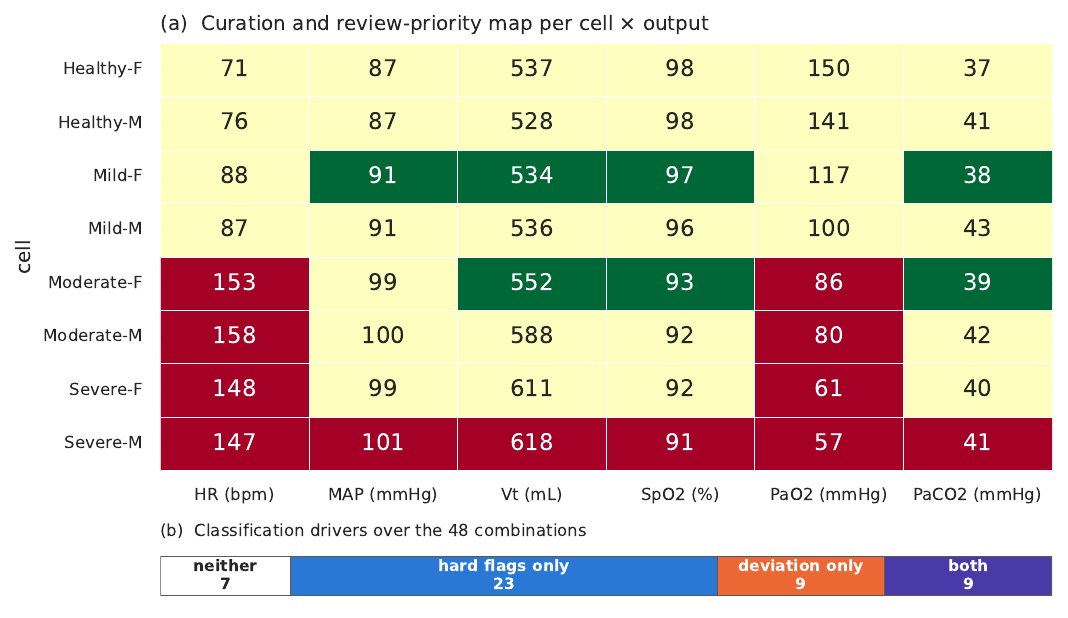}
\caption{\textbf{Per-cell curation and review-priority map.} The colour indicates review priority, not reliability, and does not encode the value printed in the tile. \textbf{(a)} Clean-run mean of each output (columns; SpO$_2$ as a percentage) in each cell (rows). The \emph{hard flags} are \texttt{diverged} and \texttt{implausible}; the \emph{deviation} is the relative gap between the cell's clean-run mean and the Healthy-M clean-run baseline. \textit{Green}: no hard flag \emph{and} a deviation within $10~\%$. \textit{Yellow}: any hard flag \emph{or} a deviation of $10$--$30~\%$. \textit{Red}: a deviation above $30~\%$ \emph{or} a hard-flag rate above $30~\%$. Totals: 7 green, 29 yellow, 12 red. \textbf{(b)} Criterion behind each colour: 7 \emph{neither} (green), 23 \emph{hard flags only}, 9 \emph{deviation only}, 9 \emph{both}. Of the 12 red tiles, 6 are red on deviation only, 4 on hard flags only and 2 on both. The six deviation-only red tiles (HR and PaO$_2$ in Moderate-M, Moderate-F and Severe-F) show tachycardia and lower PaO$_2$, the direction expected in ARDS; an expected direction does not, however, validate the magnitude. Per-cell drivers, hard-failure rates, \texttt{not\_steady\_state} attribution and signed deviations: Supplement~\S\,\ref{sec:s3-decomposition}. Cell-label key: Table~\ref{tab:terms}.}
\label{fig:traffic_light}
\end{figure}

Retention, the number of runs kept after screening, depends strongly on both the cell and the screen (Table~\ref{tab:flag_summary}). Under the four-flag screen, 663 of 1{,}280 runs (51.8~\%) are clean, with a 3.43-fold range across cells, from 144 of 160 in Moderate-M to 42 of 160 in Healthy-F. The hard-failure-free screen, which also admits runs flagged only \texttt{not\_steady\_state}, keeps 1{,}134 of 1{,}280 (88.6~\%), a 1.51-fold range. The two screens deplete opposite ends of the severity axis; Supplement~\S\,\ref{sec:s4} decomposes the \texttt{not\_steady\_state} runs that separate them. In the Healthy cells, the low four-flag retention mostly reflects the tidal-volume oscillation of \S\ref{sec:results-vcac} and SpO$_2$ pinned at saturation (Supplement~\S\,\ref{sec:s4}). Figure~\ref{fig:traffic_light} gives the resulting curation map: one colour per (scenario, sex, output) combination, set by the cell's \texttt{diverged}-plus-\texttt{implausible} rate and by the deviation of the combination's clean-run mean from Healthy-M (Methods~\S\ref{sec:methods-traffic}).

\begin{table}[h]
\centering
\caption{\textbf{Definitions and trigger conditions of the four steady-state flags.} Flags are independent, so a run can fire several. Trigger conditions were locked at preregistration; full bounds and drift thresholds are given in Supplement~\S\ref{sec:s3-hard-bounds} and \S\ref{sec:s3-drift}.}
\label{tab:flag_defs}
\begin{tabular}{lp{0.38\textwidth}p{0.30\textwidth}}
\toprule
Flag & Fires when & What it means \\
\midrule
\texttt{clamped} & Pulse's setup log matches one of the locked rejection patterns (e.g.\ ``baseline of \ldots\ is too high'' or ``must be modeled by adding/using a condition''), i.e.\ a setup input was not accepted as given & the submitted setup value was explicitly rejected, so the run does not proceed under the requested setting \\
\addlinespace
\texttt{diverged} & the run does not finish with finite numbers: NaN/Inf in the output, a missing or truncated CSV, or a non-zero exit code & Pulse's numerical solver failed: there is no usable output for that run \\
\addlinespace
\texttt{implausible} & a steady-state output mean lands outside the locked hard physiological bounds (e.g.\ PaO$_2$ above 600~mmHg) & the run completed, but a value is non-physiological under the preregistered bounds, so it cannot be trusted as-is \\
\addlinespace
\texttt{not\_steady\_state} & in the final 60~s window, at least one tracked output flatlines at a hard bound (frozen with near-zero variation), varies too much (coefficient of variation $>$ 0.05; most often tidal volume), or drifts past its locked per-channel limit & the run completed but did not satisfy the steady-state criterion, so its 60~s mean is not a reliable summary \\
\bottomrule
\end{tabular}
\end{table}
\begin{table}[h]
\centering
\caption{\textbf{Per-cell flag counts and retention across the eight scenario $\times$ sex cells.} The \texttt{clamped} flag never fired (0 of 1{,}280). The number of clean runs ranges from 144 of 160 in Moderate-M to 42 of 160 in Healthy-F. A run can fire several flags, so a row can sum to more than 160; \texttt{clean} means no flag. Retention is the number of a cell's 160 runs that a screen keeps: \emph{four-flag clean} (the \texttt{clean} column) keeps only runs firing no flag; \emph{hard-failure-free} keeps runs firing neither \texttt{diverged} nor \texttt{implausible}, treating \texttt{not\_steady\_state} as a label for review rather than an automatic exclusion. Counts are given rather than percentages because, with $n = 160$, every percentage is a multiple of $0.625$ and nine of the sixteen retention values would fall on an exact rounding tie at one or two decimals. Neither subset is distribution-neutral. Analyses that need one category per run use the priority order of Methods~\S\ref{sec:methods-detector}. Flag definitions: Table~\ref{tab:flag_defs}. Cell-label key: Table~\ref{tab:terms}.}
\label{tab:flag_summary}
\setlength{\tabcolsep}{4pt}
\begin{tabular}{lrrrrrrr}
\toprule
Cell & Runs & clamped & diverged & implausible & \shortstack[r]{not\_steady\_\\state} & clean & \shortstack[r]{hard-failure-\\free} \\
\midrule
Healthy-M & 160 & 0 & 4 & 11 & 107 & 49 & 145 \\
Healthy-F & 160 & 0 & 0 & 23 & 118 & 42 & 137 \\
\midrule
Mild-M & 160 & 0 & 4 & 0  & 69  & 87 & 156 \\
Mild-F & 160 & 0 & 0 & 0  & 68  & 92 & 160 \\
\midrule
Moderate-M & 160 & 0 & 4 & 0  & 12  & 144 & 156 \\
Moderate-F & 160 & 0 & 0 & 0  & 38  & 122 & 160 \\
\midrule
Severe-M & 160 & 0 & 8 & 46 & 44  & 67 & 106 \\
Severe-F & 160 & 0 & 0 & 46 & 74  & 60 & 114 \\
\midrule
All cells & 1{,}280 & 0 & 20 & 126 & 530 & \shortstack[r]{663\\(51.8~\%)} & \shortstack[r]{1{,}134\\(88.6~\%)} \\
\bottomrule
\end{tabular}
\end{table}

\subsection{Primary input-sensitivity rankings}\label{sec:results-morris}

\begin{figure}[!htbp]
\centering
\begin{subfigure}[b]{0.467\textwidth}\includegraphics[width=\textwidth]{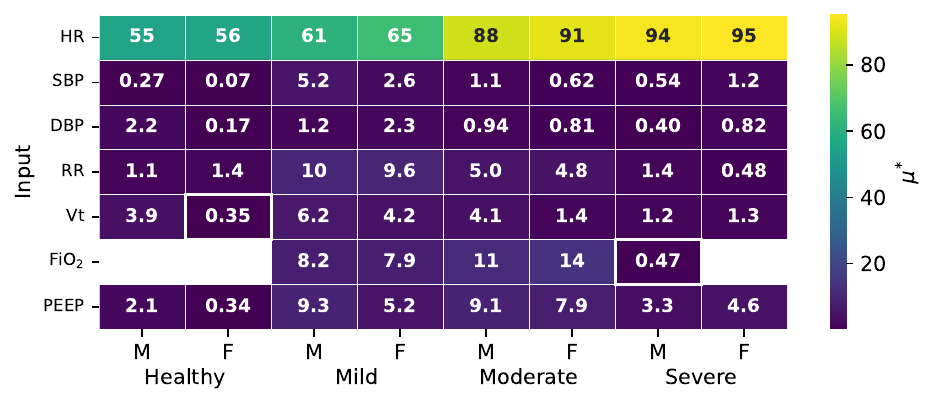}\caption{HR}\label{fig:morris-hr}\end{subfigure}\hfill
\begin{subfigure}[b]{0.467\textwidth}\includegraphics[width=\textwidth]{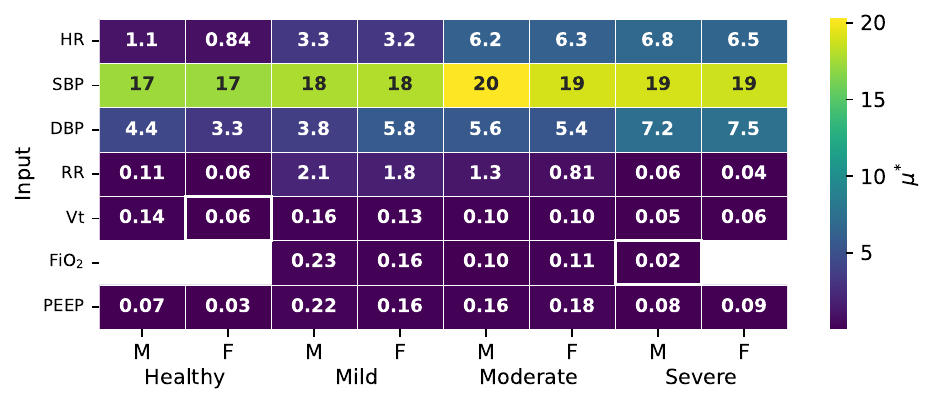}\caption{MAP}\label{fig:morris-map}\end{subfigure}
\\[-0.6em]
\begin{subfigure}[b]{0.467\textwidth}\includegraphics[width=\textwidth]{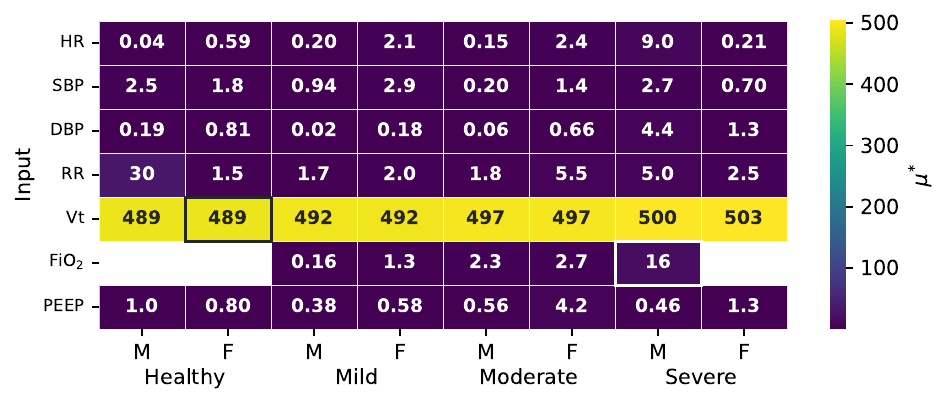}\caption{Vt delivered}\label{fig:morris-vt}\end{subfigure}\hfill
\begin{subfigure}[b]{0.467\textwidth}\includegraphics[width=\textwidth]{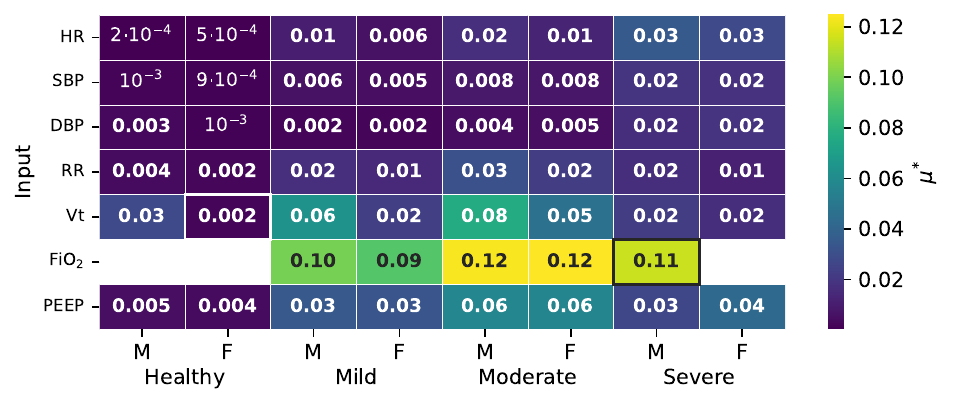}\caption{\texorpdfstring{SpO$_2$}{SpO2}}\label{fig:morris-spo2}\end{subfigure}
\\[-0.6em]
\begin{subfigure}[b]{0.467\textwidth}\includegraphics[width=\textwidth]{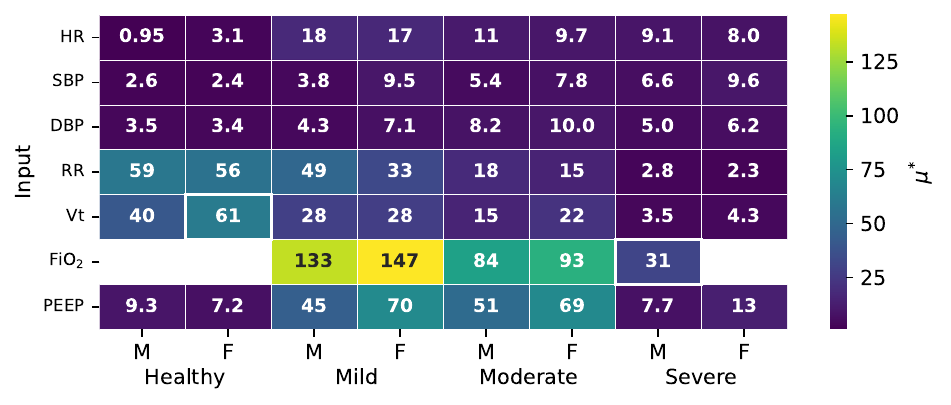}\caption{\texorpdfstring{PaO$_2$}{PaO2}}\label{fig:morris-pao2}\end{subfigure}\hfill
\begin{subfigure}[b]{0.467\textwidth}\includegraphics[width=\textwidth]{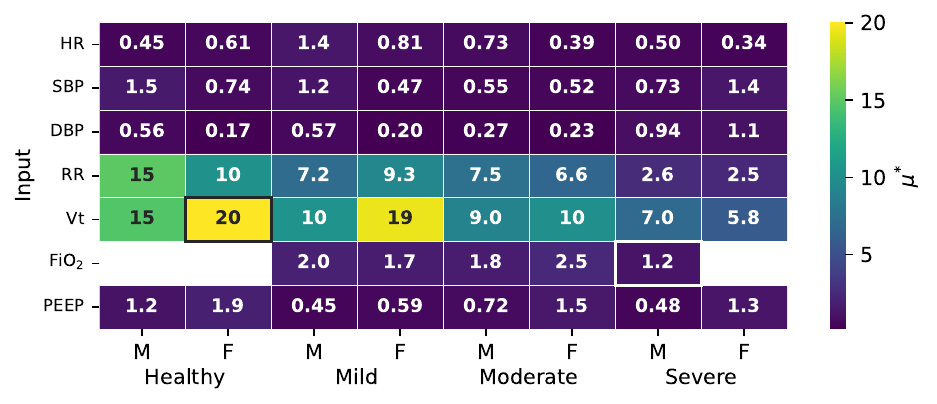}\caption{\texorpdfstring{PaCO$_2$}{PaCO2}}\label{fig:morris-paco2}\end{subfigure}
\caption{\textbf{Morris $\mu^{*}$ heatmaps for the six steady-state outputs.} \textbf{(a)}~HR, \textbf{(b)}~MAP, \textbf{(c)}~Vt delivered, \textbf{(d)}~SpO$_2$, \textbf{(e)}~PaO$_2$, \textbf{(f)}~PaCO$_2$. Rows are the seven swept inputs and columns the eight (scenario $\times$ sex) cells. Colour intensity is $\mu^{*}$, the average absolute change in the output per full-range move of the input, on a panel-specific scale in the output's units (for SpO$_{2}$, Pulse's fractional 0--1 scale, so 0.12 corresponds to 12 percentage points). In every panel, $\mu^{*}$ concentrates in a few rows (H1; median top-three share 0.928). The dominant inputs differ between outputs, and for gas exchange also with severity, so a ranking read from one panel does not transfer to another. White tiles mark combinations with no estimable $\mu^{*}$, not $\mu^{*} = 0$; this occurs only for FiO$_2$ in Healthy-M, Healthy-F and Severe-F, where every FiO$_2$ elementary-effect pair has at least one flagged endpoint. Under the hard-failure-free screen, FiO$_2$ becomes estimable in all three and is then the rank-1 input for PaO$_2$ and SpO$_2$, so the white tiles hide the input that dominates oxygenation (Supplement~\S\,\ref{sec:s13-screen}). Outlined tiles (twelve) rest on a single elementary-effect pair, so their $\sigma$ is zero by construction. Cell-label key: Table~\ref{tab:terms}.}
\label{fig:morris}
\end{figure}

Figure~\ref{fig:morris} shows the Morris $\mu^{*}$ (input importance) of the seven swept inputs for each output and cell. H1, which predicts that the three most influential inputs carry at least half of each output's summed $\mu^{*}$, holds in all 48 cell $\times$ output combinations (range 0.658--0.998; median 0.928). The lowest fractions (0.658--0.727) all occur for gas-exchange outputs in Severe, where sensitivity spreads across more inputs. Cardiovascular outputs are driven by cardiovascular inputs (output HR by input HR, and MAP by SBP, in every cell; the SBP dominance is not an artefact of the pulse-pressure clip, Supplement~\S\,\ref{sec:s3-clip-cost}), whereas gas-exchange outputs are ventilator-driven: FiO$_2$ dominates PaO$_2$ in Mild and Moderate. In Severe, its $\mu^{*}$ on PaO$_2$ falls sharply; it is still rank 1 in Severe-M, but on a single elementary-effect pair, and is not estimable in Severe-F under the four-flag screen. Ventilator settings also reach the cardiovascular outputs, but unevenly: their $\mu^{*}$ reaches 13.9~bpm on output HR (FiO$_2$ in Moderate-F) but stays at or below 2.1~mmHg on MAP wherever it is estimable (Figure~\ref{fig:morris}).

Influence is not proportionality. Of the 144 top-three inputs, 23 (spread over 17 of the 48 combinations) exceed the preregistered $\sigma / \mu^{*} \geq 0.6$ threshold, where $\sigma$ is the standard deviation of an input's elementary effects, with a 95~\% bootstrap interval at or above it under each of five independent seeds. All seven inputs and all four scenarios are represented, so departures from a linear, additive response are common rather than localised. This ratio is the standard Morris proxy for non-linearity \emph{or} input--input interaction and does not separate the two. Delivered tidal volume is the most affected output (8 of the 23), but this map is bounded by run survival, because the delivered-Vt oscillation that dominates the \texttt{not\_steady\_state} flags (\S\ref{sec:results-vcac}) removes the affected runs from the analysis. In the Healthy scenario, the delivered-Vt coefficient of variation exceeds the 0.05 threshold on 139 of 320 runs and none of the 8 falls there, whereas in the Moderate scenario it does so on 36 of 320 runs and 3 of the 8 fall there. The regime in which delivered volume oscillates most is thus the one in which that oscillation is least measurable.

Run survival also limits the data behind each index. Across the 144 top-three entries, the median index is computed from 7 of the 20 possible elementary-effect pairs; 25 rest on three or fewer pairs and 9 on a single pair, five of them their combination's rank-1 input. For these nine, all from tidal volume in Healthy-F and FiO$_2$ in Severe-M, $\sigma$ is zero by construction and no bootstrap interval is estimable. Per-entry pair counts are released with the indices.

The rankings were tested in two ways. The top-three input sets were cross-checked against a SHAP surrogate fitted on the same runs (Supplement~\S\,\ref{sec:s13}) and re-derived under the hard-failure-free screen (Supplement~\S\,\ref{sec:s13-screen}); H1 holds in all 48 combinations under both screens. In the 30 combinations whose candidate set (the inputs with at least one retained pair) is the same under both screens, the rank-1 input is preserved in all 30, whereas ranks two and three change in 11. Under this screen comparison, rank 1 was therefore more stable than the ordering below it, and the full ordering is not resolved by this design.

\subsection{\texorpdfstring{VC-AC tidal-volume oscillation and the $-2$~cmH$_2$O trigger lever}{VC-AC tidal-volume oscillation and the -2 cmH2O trigger lever}}\label{sec:results-vcac}

The largest single source of \texttt{not\_steady\_state} flags is tidal-volume (Vt) oscillation under volume-controlled assist-control (VC-AC) ventilation, measured as the coefficient of variation (CV) of delivered Vt over the final 60~s of the 120~s recording. It is the \emph{primary} flag on 20.6~\% of runs: delivered volume keeps changing from breath to breath, a genuine oscillation rather than a solver failure. The effect concentrates in healthy lungs, is strongest at high set respiratory rate and normal compliance, and falls markedly with increasing ARDS severity. This pattern is consistent with \emph{patient--ventilator dyssynchrony}, well documented clinically in volume-control ventilation~\cite{figueroacasas2016}; the conditions that produce it here are an in-silico finding of this study.

A follow-up experiment isolated the mechanism. In 64 affected configurations, setting Pulse's \texttt{InspirationPatientTriggerPressure} (the airway-pressure drop below PEEP at which a patient effort triggers a breath) to $-2$~cmH$_2$O reduced the median Vt CV from 0.351 to 0.016, and 78.1~\% of reruns fell below the 0.05 steady-state threshold (Figure~\ref{fig:corrections}a). This lock is a data-generation lever, not a clinical recommendation. By default, the assisted breaths of VC-AC are triggered by the inspiratory efforts of the engine's respiratory model. The lock replaces this respiratory-model trigger with a fixed pressure threshold, loosening the coupling between breath delivery and the modelled respiratory drive that produces effort-dependent breath-to-breath variability in the default configuration. In a real patient, the threshold is a clinical choice: with any fixed threshold, this one included, efforts too weak to reach it fail to trigger a breath (ineffective triggering), a form of patient--ventilator dyssynchrony~\cite{thille2006} that is clinically undesirable~\cite{blanch2015}. Further detail, including the two most affected configurations: Supplement~\S\,\ref{sec:s8}.

\subsection{\texorpdfstring{Transient hyperoxia regime at FiO$_2$~=~1.0}{Transient hyperoxia regime at FiO2 = 1.0}}\label{sec:results-hyperoxia}

\begin{figure}[!htbp]
\centering
\includegraphics[width=0.95\textwidth]{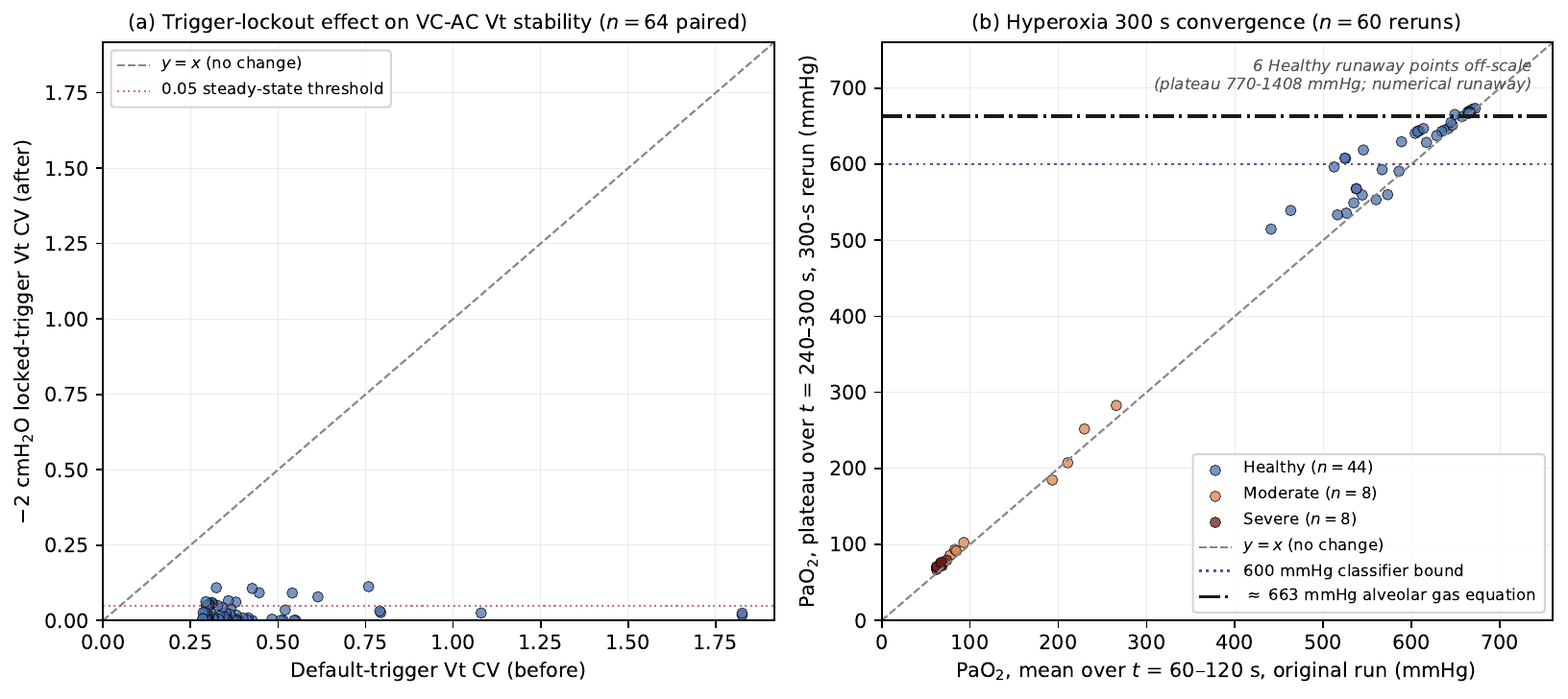}
\caption{\textbf{Two operator-actionable corrections surfaced by the atlas.} Paired before/after scatters, one dot per configuration; the dashed diagonal marks no change. \textbf{(a)} Trigger lock and VC-AC tidal-volume stability ($n = 64$ paired reruns): delivered-Vt CV under Pulse's default trigger ($x$) and after locking \texttt{InspirationPatientTriggerPressure} at $-2$~cmH$_2$O ($y$). The dotted line at $0.05$ is the steady-state threshold. The lock reduces the median CV from 0.351 to 0.016 and brings 78.1~\% of reruns below the threshold; it is a data-generation lever, not a clinical recommendation. \textbf{(b)} Hyperoxia convergence ($n = 60$ reruns in the Healthy, Moderate and Severe scenarios at FiO$_2 = 1.0$): mean PaO$_2$ over $t = 60$--$120$~s of the original run ($x$) and plateau over $t = 240$--$300$~s of a 300-second rerun ($y$). The dotted line marks the 600~mmHg classifier bound and the dash-dotted line the alveolar-gas-equation prediction at sea level ($\approx$~663~mmHg)~\cite{west2020respiratory}. Axes are limited to $[0, 760]$~mmHg; six Healthy reruns that ran into a numerical runaway (plateau up to $\approx$1{,}408~mmHg; Supplement~\S\,\ref{sec:s9}) are omitted. Most Healthy dots lie above the diagonal, because PaO$_2$ was still rising when the 120-second window closed, but below the prediction. Scenario key: Table~\ref{tab:terms}.}
\label{fig:corrections}
\end{figure}

A second cluster of flags concentrates at the highest inspired-oxygen settings. Within the 120-second window, steady-state PaO$_2$ exceeds the 600~mmHg upper bound on 34 of the 126 \texttt{implausible} runs, 91~\% of them at FiO$_2 \geq 0.95$. To separate an engine defect from a window-length artefact, 60 configurations were re-run from time zero for 300~s, with the plateau taken over the final 60~s ($t = 240$--$300$~s). The result supports the window-length artefact: in the Healthy scenario, 40 of 44 reruns (90.9~\%) converged to a stable plateau under the preregistered drift criterion (Figure~\ref{fig:corrections}b), with a median of 643~mmHg over all 40, approximately 20~mmHg \emph{below} the alveolar-gas-equation prediction~\cite{west2020respiratory}. This rules out a cohort-wide oxygenation runaway, although validating the plateau level itself would require measured arterial data. Three of the 40 converged runs nevertheless plateau above 1{,}400~mmHg, which is physically impossible at sea level (760~mmHg); they belong to a six-run solver-instability regime (below), so 37 of 44 reruns (84.1~\%) converge below the wider 700~mmHg bound used in Supplement~\S\,\ref{sec:s9}.

Hyperoxia studies on Pulse should therefore record at least 300 seconds before extracting a steady-state plateau. In the audited hyperoxia and Vt-oscillation runs, window-length sensitivity was concentrated in PaO$_2$ and delivered Vt, and threshold exceedances were uncommon for HR, MAP and PaCO$_2$ (Supplement~\S\,\ref{sec:s16-window}). Because extended runs on healthy lungs at FiO$_2$ = 1.0 can also enter a separate solver-instability regime, far more often on the female than on the male body, each PaO$_2$ trajectory should be screened individually (Supplement~\S\,\ref{sec:s9}).

\subsection{Sex-stratified solver divergence and blood-volume dependence}\label{sec:results-sex}

\begin{figure}[!htbp]
\centering
\includegraphics[width=0.95\textwidth]{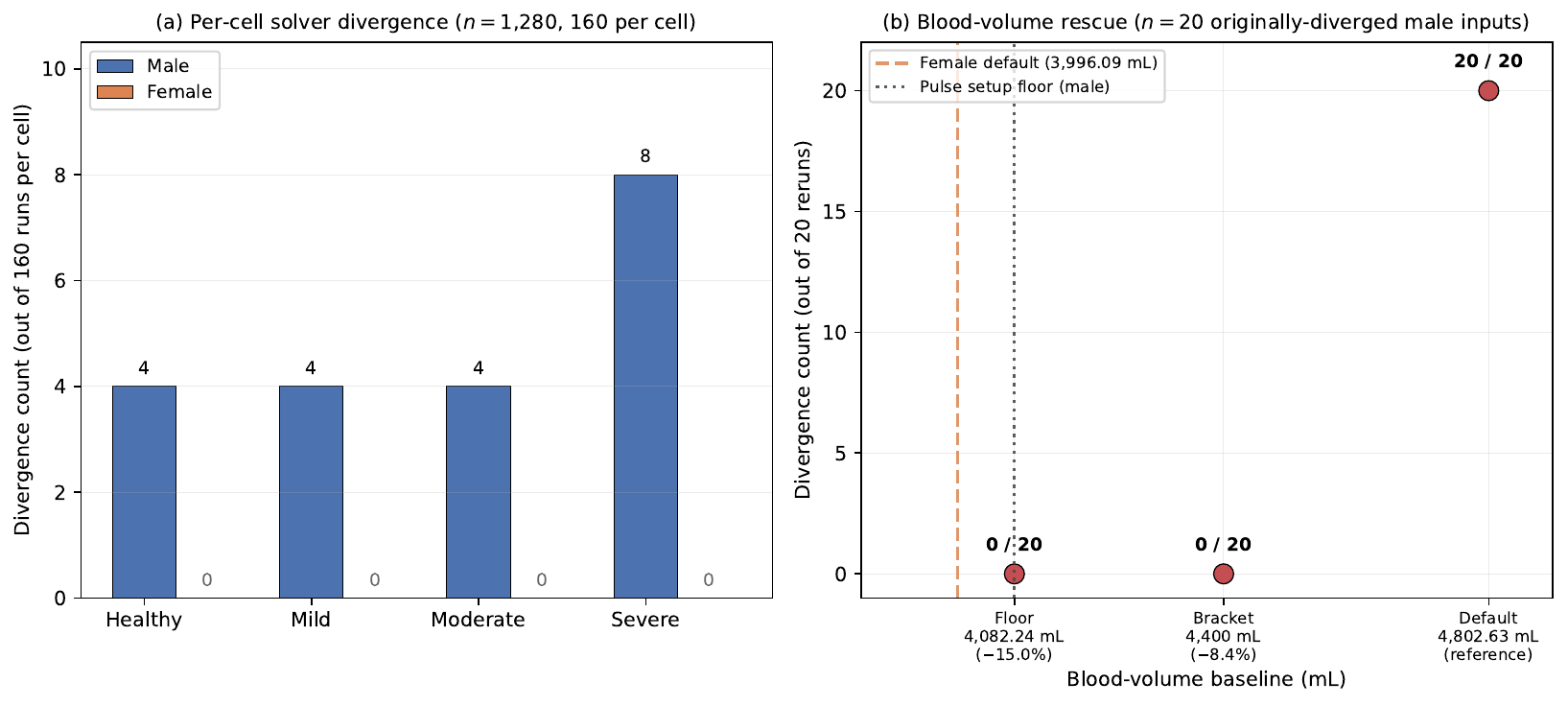}
\caption{\textbf{Sex-stratified solver divergence and blood-volume rescue.} \textbf{(a)} Divergences per (scenario $\times$ sex) cell: all 20 occur on the male body, whereas the female body, run on the same Morris grid, has none. The asymmetry is specific to this input-extreme regime; at midrange inputs across the instability band, both bodies diverge (Supplement~\S\,\ref{sec:s5-regimes}). \textbf{(b)} Divergence count (out of 20) in the 20 originally diverged male configurations against baseline blood volume ($x$-axis): Pulse's default (4{,}803~mL) or an imposed \texttt{BloodVolumeBaseline} of 4{,}400~mL or 4{,}082~mL (the setup floor for this body), i.e.\ reductions of $8.4$ and $15.0~\%$. Both reductions eliminate divergence (20 $\rightarrow$ 0), showing that a lower baseline is sufficient, without identifying the numerical pathway of the failure. The dashed line marks the female body's computed volume (3{,}996~mL), 86~mL below the male floor, so a one-way blood-volume transplant cannot make the two reference bodies equal in blood volume. Cell-label key: Table~\ref{tab:terms}.}
\label{fig:sex_bv}
\end{figure}

Solver divergence is rare but strikingly sex-stratified: \emph{all 20 diverged runs (1.6~\%) were on the male reference body}, and none of the 640 runs on the female body diverged (Figure~\ref{fig:sex_bv}a). The 20 events span only 8 distinct input vectors. FiO$_2$ was requested at its 1.0 boundary in 13 of them, although every divergence preceded the ventilator action that would have applied it (\S\ref{sec:results-overview}). The asymmetry is specific to these input extremes: at midrange inputs across the instability band (\S\ref{sec:results-secondary}), both bodies diverge, the female body mainly in the lower part of the band and the male body in its upper part (Supplement~\S\,\ref{sec:s5-regimes}).

A follow-up experiment identified a blood-volume dependence. Pulse's patient-setup code (\texttt{SetupPatient.cpp}) computes baseline blood volume from body weight alone, with the same allometric formula for both sexes: 4{,}803~mL for the 67.3~kg male body and 3{,}996~mL for the 56.2~kg female body. Both bodies thus have about the same blood volume per kilogram ($\approx$71~mL\,kg$^{-1}$, within standard clinical estimates~\cite{nadler1962}) and differ only in absolute volume. We re-ran the 20 diverged male configurations with the baseline imposed through Pulse's \texttt{BloodVolumeBaseline} patient setting. Divergences fell from 20 / 20 at the default to 0 / 20 both at an intermediate 4{,}400~mL and at the engine's setup floor for this body, 4{,}082~mL (Figure~\ref{fig:sex_bv}b). Removing divergence does not by itself make a configuration usable: 12 of the 20 are clean at both levels, whereas 8 keep a flag already defined by the classifier (SpO$_2$ pinned at saturation, the tidal-volume oscillation of \S\ref{sec:results-vcac}, or the high-FiO$_2$ PaO$_2$ modes of \S\ref{sec:results-hyperoxia}; Supplement~\S\,\ref{sec:s10}).

Steady-state values separate the two bodies far less than divergence does. Across 24 (scenario $\times$ output) comparisons, the Wasserstein distance between male and female clean-run distributions exceeds its permutation null in six, five of which survive Benjamini--Hochberg correction (Methods~\S\ref{sec:methods-stats}), all for gas exchange (PaCO$_2$ in Healthy, Mild and Moderate; SpO$_2$ and PaO$_2$ in Severe). Heart rate, MAP and delivered tidal volume separate no further than under random relabelling of the pooled runs. Input-importance rankings move even less: in Mild and Moderate, the only scenarios where the comparison is computable, male and female Morris rankings are concordant (Spearman $\rho$ point estimates 0.75--1.00 across 12 combinations), and only delivered tidal volume in Mild has a confidence interval that includes zero (Supplement~\S\,\ref{sec:s6-rho}). Between the two bodies, where the solver fails thus differs far more than which inputs dominate.

These cross-body contrasts call for caution. They are computed on clean runs, whose number itself differs by sex (e.g.\ 144 vs 122 of 160 in Moderate), so a surviving distance cannot separate an effect on the physiology from an effect on which configurations pass the flag filter. Gas-exchange contrasts also include a systematic ventilation-dose difference: at the same absolute tidal volume (the sweep sets Vt in millilitres, not per kilogram, and identically for both bodies), the female body receives 19.8~\% more per kilogram of actual body weight and 36.1~\% more per kilogram of ideal body weight (sex-specific Devine relation applied by Pulse at setup; Supplement~\S\,\ref{sec:s6-dose}). Moreover, Pulse's sex flag simultaneously changes body composition, basal metabolic rate, haemoglobin, cardiovascular sizing and, through ideal body weight, anatomic dead space (Supplement~\S\,\ref{sec:s10-breadth}), so no cross-body distance can be attributed to a single physiological pathway. None of this affects the divergence finding, which rests on a direct intervention on absolute blood volume rather than on a cross-body contrast.

\subsection{Secondary observations}\label{sec:results-secondary}

Four further observations bear on interpretation (Supplementary Information, indexed in \S\,\ref{sec:s17-summary}): Pulse's ARDS impairment-fraction solver-instability band; a compensatory tachycardia that lifts mean output heart rate above the sampled input range in moderate and severe ARDS, although input heart rate remains its dominant within-scenario determinant; a preregistered Sobol analysis (which apportions output variance among inputs and their interactions) that was triggered but not executed because the triggered design exceeded its locked compute budget; and a limitation of the flatline rule itself.

\section{Discussion}\label{sec:discussion}

This work delivers the first open reliability atlas of the Pulse Physiology Engine across several adult critical-care scenarios, with a preregistered core design and primary analysis plan. The 1{,}280-run sweep yields two operator-actionable corrections and one sex-stratified failure mode. First, a single configuration setting markedly stabilises VC-AC tidal-volume delivery for synthetic-cohort generation, at the cost of suppressing much of the effort-dependent variability that Pulse models; it is a data-generation tool, not a clinical recommendation (\S\ref{sec:results-vcac}). Second, hyperoxia simulations at FiO$_2 = 1.0$ should be recorded for at least 300~s, rather than the 120~s used here, before a steady-state plateau is taken (\S\ref{sec:results-hyperoxia}). Third, a previously uncharacterised solver-divergence cluster (runs that end without usable output) is confined to the male body in an input-extreme regime and depends on blood volume (\S\ref{sec:results-sex}).

The per-cell traffic-light map (Figure~\ref{fig:traffic_light}) is the principal deliverable for downstream users. Its yellow--red boundary deviates materially from the preregistered wording; this post hoc change is declared with a sensitivity analysis in \S\ref{sec:methods-traffic}. The map is a curation and review-priority instrument, not a physiological validation of the engine. Credibility frameworks in biomedical modelling treat sensitivity analysis and a bounded domain of applicability as components of assessment, but validation requires comparison against measured data~\cite{viceconti2021insilico}, which this work does not perform. Of a proposed five-stage validation scheme for synthetic biomedical data, running from internal plausibility through downstream utility, out-of-distribution generalisation and simulation-to-reality transfer to biological or clinical coupling~\cite{victoriano2026sim2real}, the atlas addresses only part of the first stage: it tests whether runs are internally plausible, not whether they match real-world distributions.

The four-flag classifier can also screen Pulse-generated synthetic cohorts before they are used to train machine-learning models, a role comparable to the verification and filtering stages of the synthetic-data pipeline cited in the Introduction, where each synthesised agent task (one the model performs by acting in a software environment) carries its own verification system~\cite{deepseek2026v41flash}. Unlike those stages, the classifier checks how the simulator behaved, not whether its outputs agree with patient data. Because its \texttt{clamped} flag sees only the rejections that Pulse logs, this screening relies on the null result of \S\ref{sec:results-overview}: within the sampled envelope, no submitted input was silently altered at ingestion. This concerns setup-time input handling only; it says nothing about how physiological states evolve during a run or about Pulse's behaviour outside the sampled envelope. The map sets cell-level review priority, but run-level screening remains necessary whatever the cell colour, because a tile's colour does not count \texttt{not\_steady\_state} runs, and every use remains subject to use-case-specific validation. The green set is the same under both non-degenerate readings of the rule, and green does not establish physiological validity. The map also serves educators: Pulse is embedded in training products that present healthcare learners with real-time physiological and pharmacological data~\cite{pulseinaction}, and instructors can use it to avoid regimes in which the engine diverges or never settles.

Four findings call for care when Pulse-generated data are reused. First, pooling male and female runs from the input-extreme regime without flag-based filtering introduces a spurious sex covariate through the divergence cluster (\S\ref{sec:results-sex}). Second, input and output heart rates are not interchangeable in ARDS scenarios because of compensatory tachycardia (\S\ref{sec:results-secondary}), even though input heart rate still drives output heart rate within each scenario; a model trained on such cohorts would otherwise learn an active response to hypoxemia as a steady-state feature. Third, sex-conditioned distributional differences persist even among clean runs (\S\ref{sec:results-sex}); they may be physiological, but they have no single mechanism and no scenario-independent direction of effect, so they cannot be read as validated sex effects. Fourth, the four-flag screen excludes every run showing the tidal-volume oscillation of \S\ref{sec:results-vcac}, most often in healthy lungs, so a cohort filtered that way under-represents it (the hard-failure-free screen admits runs flagged only \texttt{not\_steady\_state}; \S\ref{sec:results-overview}).

For the computational-physiology community, the atlas directly addresses the maintainers' statement that no formal sensitivity or uncertainty quantification has been performed on user-facing operational inputs~\cite{pulsefaq}, and shows that sensitivity is concentrated in a few inputs in every cell $\times$ output combination tested (\S\ref{sec:results-morris}). The method also carries over to other engines: a Morris screen of the operational inputs paired with a classifier that sorts every run into a category instead of discarding failures. To our knowledge, no comparable public reliability atlas combining Morris screening with explicit clamping analysis has been published for BioGears~\cite{baird2020biogears} or HumMod~\cite{hester2011hummod}; the same recipe could be applied to them by wrapping each engine's own inputs and outputs.

Prior falsification work on MIMIC-III is complementary rather than overlapping. Of the 85 Hoeffding-inequality rejections reported by Cornish et al.~\cite{cornish2023}, 80 fall on channels the atlas does not assess, 2 on the partially overlapping pressure channel SysBP, and only 3 on the four channels assessed by both studies (HR, MAP, PaO$_2$, PaCO$_2$; Supplement~\S\,\ref{sec:s7}). Cornish et al.\ test Pulse's chemistry and dermal outputs against a real cohort undergoing clinical interventions, whereas the atlas tests its cardio-respiratory response to operational inputs. Neither maps how operational inputs drive the electrolyte channels most often flagged by Cornish et al., so a metabolic-and-renal reliability atlas is the natural next step.

For the engine's maintainers, the atlas provides, to our knowledge, the first systematic characterisation of an ARDS impairment-fraction solver-instability band ($[0.879, 0.946]$), together with engine-level observations that target tooling and validation infrastructure rather than Pulse's underlying physiology (Supplement~\S\,\ref{sec:s11}).

Six scope limits bound the conclusions (details in Supplement~\S\,\ref{sec:s2-scope}). (i)~The atlas is pinned to Pulse v4.3.1; results apply identically to the build-only v4.3.2 and may not carry over unchanged to v4.4.0, released after the sweep with revised physiology models. (ii)~Only two young, lean adult reference bodies were used (age 18, the youngest age Pulse accepts; body-mass index about 21), so variation in age, and in mass and height within each sex, is not characterised. (iii)~Only adult VC-AC ventilation was covered; pressure-targeted modes and within-run transients are out of scope by design. (iv)~The Morris estimator discards any pair containing a flagged run, which removes every FiO$_2$ pair in Healthy-M, Healthy-F and Severe-F; these FiO$_2$ entries are blank in Figure~\ref{fig:morris}, and the cross-sex Spearman comparison is confined to Mild and Moderate. (v)~The hard physiological bounds judge whether a \emph{sustained} steady-state mean is plausible, not whether a value is instantaneously attainable; they are a screening standard, not a claim that exceptional clinical trajectories are impossible. (vi)~Pulse enforces DBP/SBP $\leq$ 0.75, and the resulting clip shortened 9 of 20 DBP steps, so some SBP and DBP effects are not strict fixed-step Morris estimates. In addition, no trajectory-count convergence analysis was performed, and shunt fraction was not recorded, so shunt-mediated interpretations remain physiologically motivated rather than directly tested.

The atlas is intended as a baseline for further work. The classifier can be applied to new runs recorded with the same protocol, and the traffic-light map to configurations within its eight cells, without re-running the sweep. Validating the cell-level classifications against clinical databases such as MIMIC-IV would test them against real patient care; it would require replaying recorded patient states and ventilator settings in the engine and comparing the result with the next observed state. Simulated and real-world cohorts are complementary rather than interchangeable: mechanistic simulation provides controlled inputs and directly accessible model states, whereas real-world data retain the messiness of clinical practice. Whether a model trained on simulated cohorts performs on real patients is a further question. One sim-to-real study quantified the distributional mismatch between its simulated training data and real-world datasets and evaluated its model on the latter~\cite{hwang2025sim2real}; the atlas does neither. Models trained only on simulated data and tested on real patients, for ventilator-asynchrony detection and aortic pressure estimation, performed reasonably, though below models trained on real data~\cite{bakkes2023asynchrony,bikia2020aortic}.

Pulse-generated cohorts are already used as machine-learning training data~\cite{ward2024ventilator}, and ventilation automation is reaching the bedside: in a recent multicentre randomised trial, closed-loop ventilation improved overall ventilation quality yet showed no difference in patient-centred outcomes~\cite{sinnige2026active}. Characterising a simulator's operational limits is therefore a precondition for using it to train models that may inform such care. The longer-term aim is models that adapt to each patient's own responses, as clinicians do at the bedside, rather than applying cohort averages.

\section{Methods}\label{sec:methods}

\subsection{Pulse engine and configuration}\label{sec:methods-pulse}

All simulations used Pulse v4.3.1 (Kitware, Apache-2.0)~\cite{bray2019}, a real-time-capable lumped-parameter whole-body physiology engine whose respiratory sub-model is a 5-compartment lung~\cite{pulserespmethod}. The engine ran inside a single \texttt{pulse-atlas-engine} Docker image, pinned for the 1{,}280-run sweep so that every simulation runs against bit-identical engine binaries. A thin Python layer around Pulse's command-line interface wrote one JSON configuration per Morris input vector (reference body, ARDS scenario, ventilator actions and the seven sampled inputs) and launched one container per run.

Re-running the pipeline reproduces the canonical outputs bit-for-bit at three successive stages, each checked against a committed artefact: the input sample, the per-cell aggregate tables and the derived statistics. Mechanical ventilation was restricted to volume-controlled assist-control (VC-AC) mode so that all runs remain directly comparable; the sweep runs used Pulse's default patient trigger. Hardware and environment details are given in Supplement~\S\,\ref{sec:s12-env}.

\subsection{Patient reference bodies}\label{sec:methods-patients}

Two adult reference bodies were used: male (age 18, 67.3~kg, 180~cm) and female (age 18, 56.2~kg, 163~cm). Both weights are the U.S.\ Centers for Disease Control and Prevention (CDC) growth-chart 50th-percentile weight-for-age values for each sex~\cite{kuczmarski2002cdc} (\texttt{Agemos}~=~216.5). The female height matches the CDC 50th percentile and Pulse's standard female height; the male height was fixed by the researchers (full provenance: Supplement~\S\ref{sec:s3-bodies}).

\subsection{Sweep design}\label{sec:methods-sweep}

Seven user-facing operational inputs were swept across the following ranges: HR 50--110~bpm, SBP 90--120~mmHg, DBP 60--80~mmHg (subject to the pulse-pressure clip below), RR 8--20~breaths\,min$^{-1}$, Vt 300--800~mL, FiO$_2$ 0.21--1.0 and PEEP 0--20~cmH$_2$O. The HR, blood-pressure and RR ranges are Pulse's own setup acceptance bands, whereas the Vt, FiO$_2$ and PEEP ranges are clinical ranges fixed in the preregistration. MAP was derived as $\mathrm{DBP} + (\mathrm{SBP} - \mathrm{DBP})/3$ rather than swept independently. The swept heart-rate and blood-pressure values are setup values honoured at ingestion, not setpoints maintained against the engine's own reflexes: Pulse re-baselines them to the stabilised state at the end of stabilisation, and no action used here rewrites them.

Pulse refuses some setup values outright, and the sampler ranges were set inside the two constraints that act on swept inputs: a gate on systolic and diastolic blood pressure, and acceptance bands on the initial heart and respiratory rates (both verified against the Pulse source and by boundary-probe runs before the sweep; Supplement~\S\,\ref{sec:s2}). A third constraint, Pulse's limit on the pulse-pressure ratio, was deliberately not imposed on the Morris sample, because doing so would distort the sample geometry on which the elementary-effect estimator depends. It was applied instead when each configuration was written (a sampled DBP above 0.75~$\times$~SBP was lowered to 0.75~$\times$~SBP), and every run records the requested value, the applied value and a clip flag (Supplement~\S\,\ref{sec:s3-clip-cost}). Every configuration submitted to Pulse therefore lies inside the accepted envelope. A fourth constraint, a per-patient floor on baseline blood volume, is not exercised by the sweep and is characterised for the male body in \S\ref{sec:results-sex}.

Four scenarios spanned the adult critical-care severity range: a healthy baseline (Healthy, ARDS impairment fraction = 0) and three ARDS severities (Mild, 0.26; Moderate, 0.60; Severe, 0.95). At a low-tidal-volume reference ventilation, these impairment fractions place each scenario in a Berlin P/F band for the male reference body, anchoring them to a clinically recognisable state rather than to arbitrary engine settings (calibration and its sweep-wide FiO$_2$ dependence: Supplement~\S\,\ref{sec:s5-berlin}). The mild, moderate and severe labels name the band each scenario was calibrated to, not a diagnosis of each run, since individual runs vary FiO$_2$ and PEEP. The three severities also bracket the 0.91--0.945 solver-instability zone documented at design time (Supplement~\S\,\ref{sec:s5}), which our continuous-severity sweep later widened to $[0.879, 0.946]$ (\S\ref{sec:results-secondary}). Pulse represents severity as a single per-lung scalar driving four hard-coded, adult-calibrated interpolation curves (Supplement~\S\,\ref{sec:s5-mapping}).

Sensitivity was screened with the Morris elementary-effects design~\cite{morris1991, campolongo2007}, whose cost is linear in the number of inputs, so that the same runs both map where the engine fails and rank which inputs drive each output. The sample was generated with SALib~\cite{iwanaga2022,herman2017salib} using 20 trajectories, 4 grid levels per input and random seed 42. Each trajectory contributes $7 + 1 = 8$ model evaluations, giving 160 runs per (scenario $\times$ sex) cell and 1{,}280 runs in total. The trajectory count was fixed at 20 in the preregistration for computational tractability, not through a convergence analysis. Each run was recorded for 120 seconds of simulated time after patient stabilisation and a 120-second ventilator-settling phase, with outputs sampled every 5 seconds (13 samples in the final 60-second steady-state window).

Follow-up runs used fixed rules. The trigger-lock reruns (\S\ref{sec:results-vcac}) took the 64 non-diverged sweep runs with the highest delivered-Vt CV. The 300-second reruns (\S\ref{sec:results-hyperoxia}) took the 22 Morris points requested at FiO$_2$ = 1.0 that diverged in no cell, in both Healthy cells, and the first four, in sample order, in the Moderate and Severe cells. The continuous-severity sweep cited in \S\ref{sec:results-sex} ran 100 impairment fractions, drawn at random and more densely above 0.85, on both bodies, with the seven inputs at the midpoints of their swept ranges (FiO$_2$ rounded to 0.6). Details: Supplement~\S\,\ref{sec:s8-protocol}, \S\,\ref{sec:s9-trajectories} and \S\,\ref{sec:s5-cont}.

\subsection{Steady-state classifier}\label{sec:methods-detector}

Every simulation was classified by a four-flag steady-state classifier before any statistical analysis.

\texttt{clamped} fires when Pulse's setup log contains one of the messages Pulse writes when it rejects a setup input (the name comes from preregistered hypothesis H2, which predicted silent clamping). This log-based detector identifies \emph{explicit} setup rejection only; silent substitution was checked by the post hoc input-instantiation audit reported in \S\ref{sec:results-overview}. The sweep was classified with 4 message patterns; after the sweep, one pattern was widened and a fifth added to catch Pulse's heart-rate lower-bound rejection. Replaying both pattern sets over all 3{,}349 recorded setup logs reclassifies no run of this sweep (Supplement~\S\,\ref{sec:s3-clamp}).

\texttt{diverged} fires when a run does not finish with finite output: a non-zero exit code, a missing or truncated CSV, or NaN or infinite values. Although a setup rejection would also exit non-zero, in this sweep each of the 20 diverged runs carries Pulse's own \texttt{Could not converge to provided Convergence} message in its setup log, and none of the 1{,}260 completed runs does (\texttt{results/divergence\_timeline.csv}).

\texttt{implausible} fires when any steady-state output mean falls outside the locked hard physiological bounds: HR [20, 220]~bpm, MAP [30, 180]~mmHg, Vt [50, 1{,}500]~mL, SpO$_2$ [50, 100]~\%, PaO$_2$ [20, 600]~mmHg, PaCO$_2$ [10, 120]~mmHg. Pulse imposes no acceptance envelope on these outputs; it does stop a run on some states far outside physiology (e.g.\ an aortic pressure below $-2$ or above 700~mmHg), but no run of this sweep reached one. The bounds are therefore our own post-run plausibility standard, not a constraint the engine enforces (rationale and literature sources: Supplement~\S\,\ref{sec:s3-hard-bounds}).

\texttt{not\_steady\_state} fires when, over the final 60-second window, any tracked output flatlines, drifts beyond its per-channel threshold (Supplement~\S\,\ref{sec:s3-drift}) or has a coefficient of variation above 0.05. A flatline is a standard deviation below 1~\% of the mean \emph{and} a mean within 0.1~\% of a hard bound ($\geq$~99.9~\% for the SpO$_2$ ceiling). The rule was written to catch an output pinned numerically at a bound, but a physiologically saturated SpO$_2$ meets it too (Supplement~\S\,\ref{sec:s4}).

The four flags are independent, so one run can fire several. Analyses that need one category per run keep the highest-ranked flag, in an order set after the sweep: \texttt{clamped} $>$ \texttt{diverged} $>$ \texttt{implausible} $>$ \texttt{not\_steady\_state} $>$ \texttt{clean}, which follows where a run fails (at setup, during the run, then in its output; Supplement~\S\,\ref{sec:s3-detector-order}). The [60, 120]~s window was preregistered before any sweep data were collected, and a 22-run pilot on the Healthy-M cell verified five preregistered gating criteria, including detector self-consistency and Pulse determinism.

\subsection{Traffic-light curation map}\label{sec:methods-traffic}

The curation map (Figure~\ref{fig:traffic_light}) gives each (scenario, sex, output) combination one colour (green, yellow or red) that combines the cell's \texttt{diverged}-plus-\texttt{implausible} rate with the deviation of the combination's clean-run mean from the corresponding clean-run mean of the Healthy-M cell. The $10~\%$ and $30~\%$ deviation bands were preregistered before any data were collected. The legend to Figure~\ref{fig:traffic_light} gives the full colour criteria; scoring the female cells against Healthy-F instead changes no colour (Supplement~\S\,\ref{sec:s3-traffic}).

\textbf{Post hoc deviation from the preregistered traffic-light rule.} The preregistered rule gave red when the deviation exceeded $30~\%$ or when ``any flag'' was raised. The implemented rule counts only the hard flags, \texttt{diverged} and \texttt{implausible}: they make a combination at least yellow, and red only when their combined rate exceeds $30~\%$. Because the repository cannot establish whether this rule was defined before or after the sweep outputs were inspected, we record the altered yellow--red boundary as a material deviation. Read literally, with ``flag'' taken as the two hard flags, the preregistered rule keeps the same seven green and twelve red combinations and turns 26 of the 29 yellow ones red. We retain the implemented rule as the primary display because it preserves a graded distinction without adding a single green combination. The classifier, the map and the reported counts were not changed in response; Supplement~\S\,\ref{sec:s3-tl-sensitivity} reports the preregistered readings, including the four-flag one, as a sensitivity analysis.

\subsection{Statistical analyses}\label{sec:methods-stats}

Morris $\mu^{*}$ and $\sigma$ were computed per (scenario, sex, input, output) using only elementary-effect pairs whose two runs are both clean. Each elementary effect divides the output change by the input step, with each input's swept range mapped to $[0,1]$, so $\mu^{*}$ is an output change per full-range move of the input. Indices resting on at least two trajectories carry non-parametric 95~\% bootstrap confidence intervals from 1{,}000 trajectory-level resamples. Four hypotheses were preregistered before data collection. H1: the top-three inputs account for at least half of the summed $\mu^{*}$ for each output. H2: at least 5~\% of the Morris sample hits at least one silent clamp. H3: at least one solver-divergence zone exists outside the previously characterised 0.91--0.945 band within the seven-input adult envelope. H4: the male and female reference bodies differ in clamping, divergence or $\mu^{*}$ ranking on at least one output, with no direction prespecified. H1 was evaluated per cell $\times$ output on point estimates of $\mu^{*}$; its summed $\mu^{*}$ covers only inputs with at least one clean pair (six rather than seven in the 18 combinations where FiO$_2$ has none; Figure~\ref{fig:morris}), so the denominators differ across cells. H2, H3 and H4 are resolved in \S\ref{sec:results-overview}.

Sex differences in steady-state distributions were measured per (scenario $\times$ output) by the Wasserstein distance (the minimum ``work'' needed to reshape one probability distribution into the other) between the male and female clean-run distributions. Each distance was tested against a permutation null of 2{,}000 random relabellings at the original group sizes, and the 24 tests were corrected by Benjamini--Hochberg control of the false-discovery rate at $q = 0.05$. Cross-sex agreement of input-importance rankings was quantified per (scenario, output) by the Spearman rank correlation between the male and female $\mu^{*}$ vectors.

A Sobol analysis was preregistered as a conditional escalation for any top-three input whose $\sigma / \mu^{*}$ ratio reaches 0.6, within a compute envelope of 1{,}500 additional runs. Applied as preregistered, the trigger fires on 72 of the 144 top-three inputs (43 of the 48 combinations), and the design it selects needs 6{,}144 runs (3{,}584 for the first-order and total indices only), 2.4 to 4.1 times that envelope; it was therefore reported rather than executed (Supplement~\S\,\ref{sec:s3-sobol}). Because the preregistered trigger rests on the point ratio alone, we added a post hoc sensitivity analysis: for each top-three input, trajectories were resampled 1{,}000 times and $\sigma / \mu^{*}$ was recomputed on each draw. Inputs whose 95~\% interval has a lower bound at or above 0.6 under each of five independent seeds are reported in \S\ref{sec:results-morris} (full decomposition: Supplement~\S\,\ref{sec:s3-sobol}).

The sampler, sweep driver, classifier, statistical-analysis pipeline and audit ledger are released open-source. The audit ledger, developed after data collection, recomputes the paper's computed claims from the saved output files and checks them against the text.

\subsection{Use of AI tools}\label{sec:methods-ai}

\noindent Large language models (LLMs) were used in this work, under direct supervision by S.N., for drafting, editing and shortening the manuscript, for writing, refactoring and debugging the code that runs and analyses the sweep, and for in-repository replications and primary-source audits. During iterative review, two LLMs worked in separate sessions: always a Claude model (Anthropic; Opus 4.7, 4.8, 5 and 5.5 in succession), and Google Gemini Pro 3.1, later OpenAI ChatGPT (GPT-5.6 Sol). The Claude model in particular was often asked to re-check its own output, claim by claim and number by number, and to correct it; the two models then critiqued each other's latest output in turn. These checks often surfaced errors or missed points, which were then corrected. S.N.\ used the critiques to decide whether to run another round or accept the result. M.S.\ used a Claude model (Anthropic) to propose editorial revisions during his review; each proposal was checked against the data and sources before S.N.\ accepted or rejected it. Supplement~\S\,\ref{sec:s18} gives each model's exact version, reasoning level, phases and tasks.

Responsibility rests with all the authors. S.N.\ personally reviewed and, as far as practicable, verified AI-assisted outputs, including editorial revisions, consistency checks and methodological interpretations, and reviewed and committed AI-assisted code changes. Verification was an iterative process of checks. S.N.\ authored and locked the preregistered thresholds, hypotheses, statistical-test choices and experimental decisions before the corresponding analyses were run. Every cited reference was checked against its source. Although the atlas code was written with LLM assistance, the simulations, the classifier and the audit ledger run without any LLM. Quantitative claims are traceable, as applicable, to primary sources, the preregistration, or entries in the open audit ledger released alongside the code, whose commit history records the audit findings, their evidence and the resulting corrections.

%%%%%%%%%%%%%%%%%%%%%%%%%%%%%%%%%%%%%%%%%%%%%%%%%%%%%%%%%%%%%%%%%%%%%%
%% MANDATORY CLOSING SECTIONS                                       %%
%% Per npj Digital Medicine policy: Data availability, Code         %%
%% availability, Acknowledgements, Author contributions, Competing  %%
%% interests.                                                       %%
%%%%%%%%%%%%%%%%%%%%%%%%%%%%%%%%%%%%%%%%%%%%%%%%%%%%%%%%%%%%%%%%%%%%%%

\hypertarget{tgt:data-avail}{}\bmhead{Data availability}

\noindent The raw simulation outputs, derived parquet artefacts and curation-map figures will be deposited at Zenodo, together with the code and the companion technical monograph (Supplement~\S\,\ref{sec:s14}); pre-deposit access is available from the corresponding author. This paper and its Supplementary Information are also available as a preprint on arXiv; the Zenodo DOI will be added upon deposit. The per-cell curation map is also released as CSV and JSON (\texttt{results/reliability\_traffic\_light.csv} and \texttt{.json}), so that users can apply it directly as a review-priority layer on synthetic cohorts drawn from its eight cells; neither the map nor the run-level labels is a distribution-neutral automatic filter (Table~\ref{tab:flag_summary}). The deposited data and the authors' own monograph content are licensed under Creative Commons Attribution 4.0 (CC-BY-4.0); verbatim third-party excerpts, notably from the Pulse documentation and source (Apache-2.0), retain their original licensing terms.

\hypertarget{tgt:code-avail}{}\bmhead{Code availability}

\noindent The complete atlas codebase, including the audit ledger and the per-experiment lab-notebook entries, is released under the Apache-2.0 license at \url{https://github.com/sadc-lab/pulse_reliability_atlas}. Y.M.\ reproduced the full 1{,}280-run sweep and the Healthy-M pilot on a second, independent x86\_64 computer (AMD CPU, Linux under WSL2; reference machine: Intel CPU, Windows~11): every run-level flag and steady-state value was bit-identical, and derived statistics agreed to floating-point round-off (reproducibility recipe: Supplement~\S\,\ref{sec:s12}).

\bmhead{Acknowledgements}

\noindent This work benefited from government funding managed by the National Research Agency under France 2030 via the ENACT AI Cluster (ANR-23-IACL-0004). This work also relies on the financial support provided by the R\'egion Grand Est, France, for which we express our gratitude. We gratefully acknowledge the support of the CHU Sainte-Justine Foundation in Montreal and the Artificial Intelligence Applied to Pediatric Acute Care Research Group (R-SADC) in Montreal, Canada. These four funders played no role in study design, data collection, analysis and interpretation of data, or the writing of this manuscript. We thank the BIOSE Doctoral School of the Universit\'e de Lorraine for its support. We thank the Kitware Pulse engineering team for the open-source release of the Pulse Physiology Engine and its accompanying documentation, without which this work would not have been possible. We also thank Jama Hussein for productive and inspiring discussions at the Mila --- Quebec AI Institute. S.N.\ dedicates this work to the memory of U.N.\ --- for the knock at the door, at one, at two, at three in the morning.

\bmhead{Author contributions}

\noindent S.N.\ conceived and led the study, whose overarching research direction was shaped through clinical discussions with M.S.\ and P.J.; designed the Morris sweep and the four-flag steady-state classifier; performed all computational runs and statistical analyses; supervised the two-model LLM cross-checking loop (Methods~\S\ref{sec:methods-ai}); and wrote the manuscript. A.W.\ contributed to the conceptualization and, through sustained methodological discussion, to the iterative scientific refinement of the analysis and its framing (notably the ARDS severity-to-Berlin calibration and the reliability-classification transparency), and provided clinical supervision and manuscript review. Y.M.\ contributed to validating the simulation environment through independent cross-platform and cross-vendor reproductions, and to manuscript review and editing. C.S., M.S.\ and P.J.\ secured the funding, provided clinical supervision, contributed to interpreting the findings in their clinical context and performed the global clinical review of the manuscript; M.S.\ also provided a detailed clinical review and contributed the clinical interpretation of the VC-AC inspiration-trigger caveat. All authors approved the final version. AI tools were used as described in Methods~\S\ref{sec:methods-ai}.

\bmhead{Competing interests}

\noindent The authors declare no competing interests.

\bmhead{Ethics approval and consent to participate}

\noindent Not applicable. This work is a computational simulation study using a publicly distributed physiology engine; no human subjects, identifiable human data, or animal subjects were involved.

\bmhead{Consent for publication}

\noindent Not applicable. No individual person's data, images, or videos are included in this work.

%%%%%%%%%% SUPPLEMENTARY INFORMATION (merged for the arXiv single-PDF deposit) %%%%%%%%%%
\clearpage
\setcounter{section}{0}\setcounter{figure}{0}\setcounter{table}{0}
\renewcommand{\thesection}{S\arabic{section}}
\renewcommand{\thefigure}{S\arabic{figure}}
\renewcommand{\thetable}{S\arabic{table}}
\renewcommand{\theHsection}{S\arabic{section}}
\renewcommand{\theHfigure}{S\arabic{figure}}
\renewcommand{\theHtable}{S\arabic{table}}
\renewcommand{\floatpagefraction}{0.7}
\makeatletter
\renewcommand\numberline[1]{\hb@xt@ 3em{#1\hfil}}
\makeatother
\addtocontents{toc}{\protect\setcounter{tocdepth}{\arabic{tocdepth}}}%
\section*{Supplementary Information}

\noindent This Supplementary Information provides the full data tables, parameter provenance, reproducibility recipe, supplementary analyses, extended results and further detail on the follow-up experiments cross-referenced in the main manuscript. Table~\ref{tab:s16-index} (\S\ref{sec:s17-summary}) indexes the findings stated in full only here. Numbering follows the Nature Portfolio convention (sections \S\,S1, figures S1, tables S1, etc.). Quantitative claims are traceable, as applicable, to primary sources, the preregistration or the audit ledger (\texttt{scripts/audit\_claims.py}) that also anchors the main manuscript. References back to the main manuscript give both the section number and the phrase ``of the main manuscript'', to avoid confusion with S-numbered items. Three-digit identifiers such as S011, S014 or S021 name entries of the experiment log indexed in the companion technical monograph (\S\,\ref{sec:s14}), not sections of this supplement.

\tableofcontents

%%%%%%%%%%%%%%%%%%%%%%%%%%%%%%%%%%%%%%%%%%%%%%%%%%%%%%%%%%%%%%%%%%%%%%
%% S1 -- Pulse engine context
%%%%%%%%%%%%%%%%%%%%%%%%%%%%%%%%%%%%%%%%%%%%%%%%%%%%%%%%%%%%%%%%%%%%%%

\section{Pulse engine in context}\label{sec:s1}

The Pulse Physiology Engine~\cite{bray2019} is an actively maintained open-source whole-body human physiology simulator. It represents the body as a network of coupled electrical-circuit analogues (vascular resistances as resistors, vessel and chamber compliances as capacitors, the heart as a time-varying capacitor~\cite{pulsecardiomethod} and the respiratory muscles as a time-varying pressure source), solved as a system of ordinary differential equations at a fixed 20~ms time step. The cardiovascular, respiratory (a 5-compartment lung model: carina, left and right anatomic dead space, left and right alveoli)~\cite{pulserespmethod}, gas-exchange, endocrine, renal and drug sub-models share state at each step, producing a self-consistent whole-body trajectory at about 5--10$\times$ real time, depending on the machine~\cite{pulsefaq}. The C++ codebase has bindings for Java, C\# and Unity~\cite{pulseabout} and an interface for Python~\cite{webb2022vent}, and runs on Windows, Mac, Linux and ARM~\cite{pulseabout}.

Four implementation choices bear directly on how the atlas is read. First, the circuits are built from paths carrying idealised elements (resistances, capacitances, diodes), and the resulting non-linear networks are solved by one generic circuit solver shared across all systems~\cite{pulsecircuit}. Second, the engine advances on a fixed transient-analysis time step of 0.02~s (50~Hz), recalculating every system state at each step~\cite{pulsesystem}; this is the same 20~ms step as above, confirmed by an independent source. Third, execution is deterministic, so one configuration replays to one trajectory, which allows the per-cell curation map to be reproduced rather than merely reported (\S\ref{sec:s12}). Fourth, patients and diseases enter at setup rather than continuously: body dimensions drive allometric baselines (quantities that scale with body size) bounded by hard acceptance limits~\cite{pulsepatientmethod}, and a single ARDS severity scalar on $[0,\,1]$ drives the respiratory modifiers through piecewise-linear functions calibrated at five nodes (0.0, 0.3, 0.6, 0.9 and 1.0)~\cite{pulserespmethod}. This five-node calibration is why the main design fixes severity at scenario points rather than sweeping it (\S\ref{sec:s5-mapping}); the continuous-severity micro-sweep of \S\ref{sec:s5-cont} is a separate, targeted experiment.

Pulse was forked in 2017 from the U.S.\ Department of Defense--funded BioGears engine by members of the original BioGears team and is now maintained by Kitware, Inc.\ under the Apache-2.0 license~\cite{pulseabout,pulsefaq}. The two projects have diverged substantially since then: Pulse v4.3.1 models mechanoreceptor feedback and lung recruitment, two features we did not identify in the examined BioGears respiratory or nervous-system implementations or in its respiratory methodology documentation.

The Pulse documentation characterises the engine as using ``a low fidelity clinical approach'' and states it is ``not necessarily for high fidelity predictive purposes''~\cite{pulsesystem}. This self-description frames the atlas's contribution (Section~\mainref{sec:intro}{1} of the main manuscript): a quantitative characterisation of Pulse's operational envelope that helps users distinguish regions meeting the stated screening criteria from those requiring scenario-specific review. \texttt{REL\_4\_3\_2} (released 2025-08-12) differs from \texttt{REL\_4\_3\_1} only in its build system (7 files / +46 / -17 lines: web-WASM dockcross support and a .NET binding tweak); no physiology, validation, patient-setup or solver code changed, so results on v4.3.1 apply identically to v4.3.2. The next release, \texttt{REL\_4\_4\_0} (2026-09-30, after the sweep), changes the physiology. Its release notes~\cite{pulse440} list, among other changes, a distinction between anatomic and physiologic pulmonary shunting, revised respiratory resistance and compliance calculations, refactored respiratory compartments and volume distribution, and the removal of an inflection point in the CO$_2$ saturation calculation at high dissolved CO$_2$; the Eigen library used by the circuit solver moves from version 3.4.0 to 5.0.1. In its source, the patient-setup acceptance limits used here are unchanged, as is the rule that enables the respiratory-model patient trigger by default in VC-AC; the ARDS-severity-to-shunt mapping of \S\ref{sec:s5-mapping} is rewritten, and the cardiovascular and nervous-system model sources also change. The results of this atlas may therefore not carry over unchanged to v4.4.0.

%%%%%%%%%%%%%%%%%%%%%%%%%%%%%%%%%%%%%%%%%%%%%%%%%%%%%%%%%%%%%%%%%%%%%%
%% S2 -- Verified Pulse 4.3.1 limitations
%%%%%%%%%%%%%%%%%%%%%%%%%%%%%%%%%%%%%%%%%%%%%%%%%%%%%%%%%%%%%%%%%%%%%%

\section{Verified Pulse 4.3.1 limitations}\label{sec:s2}

The atlas is calibrated against twelve Pulse limitations (Table~\ref{tab:limits}), each verified against a primary source (Pulse FAQ, the upstream code repository or an experiment-log entry in the companion technical monograph) before any data were collected.

\begin{table}[htbp]
\centering
\caption{\textbf{Verified Pulse v4.3.1 limitations against which the atlas is calibrated.} Each row was verified against a primary source before the sweep launched. Two earlier assertions from internal probing in the companion technical work (a solver crash at FiO$_2$ = 1.0 and the absence of any sex-specific physiology) could not be confirmed before preregistration was locked (see text). The rightmost column points to the companion-monograph experiment-log entry that established each limitation.}
\label{tab:limits}
\footnotesize
\begin{tabular}{rp{8.5cm}p{3.5cm}}
\toprule
\# & Limitation & Primary source \\
\midrule
1 & No formal SA / UQ on user-facing operational inputs, addressed by this paper. The 2020 \texttt{STUDY\_SENSITIVITY\_ANALYSIS\_1\_0\_0} repository tag performs Sobol analysis on internal lumped-parameter circuit values (cardiovascular and respiratory resistances and compliances) for calibration, a different parameter class. & Pulse FAQ~\cite{pulsefaq} + upstream repo audit (monograph S016) \\
\midrule
2 & Blood-pressure gate: SBP outside 90--120 or DBP outside 60--80 mmHg rejected at setup with explicit \texttt{[ERROR]} log and exit; pulse-pressure ratio DBP/SBP $\leq 0.75$ also enforced. & Pulse source audit + monograph S005 deliberate-clamp probes \\
\midrule
3 & Adult-only restriction: age $\geq 18$ years hard-gated in \texttt{SetupPatient.cpp}. & Source audit (monograph S003) \\
\midrule
4 & ARDS solver instability in the impairment-fraction band 0.91--0.945. The locked Severe scenario at 0.95 is intentionally above this band; a continuous-severity micro-sweep confirms it and widens the band to 0.879--0.946 (Supplement~\S\,\ref{sec:s5}). & Three independent sources (monograph Appendix B.3) \\
\midrule
5--6 & Heart-rate initialisation accepted only in the range 50--110~bpm; respiratory-rate initialisation accepted only in the range 8--20~bpm. Sampler ranges match. & Boundary probes in the companion translation work (monograph S003) \\
\midrule
7 & PC-AC delivered tidal volume deviates from the physical reference by 11.3--25.3~\% across healthy, restrictive, and obstructive lung profiles, against 0.07--2.25~\% for VC-AC, per Pulse's own validation tables~\cite{pulsevent} (the ventilator model is described in~\cite{webb2022vent}; that paper reports qualitative ASL-5000 waveform agreement, not these quantitative percentages). The atlas restricts to VC-AC accordingly. & Pulse documentation~\cite{pulsevent} \\
\midrule
8 & Upper-airway resistance is a single fixed trachea resistance, between the pharynx and carina compartments; the nose is not modelled. & Pulse circuit source + source audit (monograph S003) \\
\midrule
9 & No regional ventilation--perfusion mismatch is modelled (the lung is 5-compartment lumped-parameter; intra-compartment $\dot V$/$\dot Q$ is uniform). & Pulse documentation~\cite{pulserespmethod} + structural design (monograph S003) \\
\midrule
10 & Pulse's respiratory methodology reports a standard-patient vital capacity of 4.93~L against a literature value of about 4.6~L (about 7~\% higher); its current patient validation table lists 4{,}819~mL for the standard male (about 5~\% higher). & Pulse documentation~\cite{pulserespmethod,pulsepatientmethod} \\
\midrule
11 & Lumped-parameter limits: no inertance, linear resistance only, idealised valves. & Pulse documentation~\cite{pulserespmethod,pulsecardiomethod} \\
\midrule
12 & No dicrotic notch, no retrograde flow; ECG P-wave and T-wave merge at high heart rate. & Pulse documentation~\cite{pulsecardiomethod} \\
\bottomrule
\end{tabular}
\end{table}

Rows 5--6 were later re-derived within this repository. On 2026-07-26, eight single-purpose probes on the male reference body, run under the atlas's own engine image, tested both sides of each documented bound; both bounds are enforced at patient setup, before any ventilation begins. Heart rates of 49 and 111~bpm and respiratory rates of 7 and 21~bpm were each rejected with an explicit \texttt{[ERROR]} and exit code 1, with no output written, whereas 50, 110, 8 and 20 were accepted and ran to completion. Pulse names the bound in its rejection text (e.g.\ ``Maximum heart rate baseline allowed is 110 bpm''), and the corresponding branches are visible in \texttt{SetupPatient.cpp}. This does not alter the design-time provenance recorded in Table~\ref{tab:limits}, which justified the sampler ranges when they were locked, but it removes any reliance on an external repository for re-verifying the two bounds.

Two earlier assertions from our internal probing in the companion technical work were checked before preregistration was locked and could not be confirmed. (i)~Pulse does not crash at FiO$_2$ = 1.0: Docker probing showed that it accepts FiO$_2$ = 1.0 and produces a stable trajectory (the transient hyperoxia regime is characterised in Section~\mainref{sec:results-hyperoxia}{2.4} of the main manuscript and in Supplement~\S\,\ref{sec:s9}). (ii)~The claim that Pulse has no sex-specific physiology beyond weight and height was withdrawn, but not because of blood volume: \texttt{SetupPatient.cpp} computes the blood-volume baseline from weight alone, so the blood-volume dependence of Section~\mainref{sec:results-sex}{2.5} of the main manuscript and Supplement~\S\,\ref{sec:s10} is silent on sex.

A subsequent source-code audit identified other sex-dependent patient-setup pathways. In particular, \texttt{SetupPatient.cpp} sets a sex-dependent basal metabolic rate using the Roza--Shizgal revision of the Harris--Benedict equations~\cite{roza1984metabolic} (not the 1919 originals, whose constants differ). Under the resting, normothermic conditions of the atlas, the engine's energy model uses this value both to initialise total metabolic rate and as the set point toward which it is regulated; total metabolic rate in turn drives tissue oxygen consumption and carbon-dioxide production. Because the 1{,}280-run sweep varies sex jointly with body mass and height, it cannot by itself quantify the independent effect of the sex flag.

\subsection{Scope limits of the present atlas}\label{sec:s2-scope}

The limitations above are Pulse's, verified against its source and documentation. Those below belong to the present study: they bound what the atlas establishes and are summarised in the Discussion of the main manuscript.

\textbf{(i) Engine version.} The atlas pins to Pulse v4.3.1. Results apply identically to the build-only v4.3.2 release; v4.4.0, released after the sweep with revised physiology models (\S\ref{sec:s1}), and newer physiology branches will require re-mapping.

\textbf{(ii) Patient anatomy.} Only two young, lean adult reference bodies, male and female, were used (age 18, the youngest age Pulse accepts; body-mass index about 21); variation in age, and in body mass and height within each sex, is not characterised.

\textbf{(iii) Operational envelope.} Only adult critical-care ventilation in VC-AC mode is covered. Two areas are out of scope by design: pressure-targeted ventilator modes such as pressure-controlled and pressure-support; and transient within-run dynamics, particularly in runs that never reach steady state. The mode restriction also limits transfer to the ventilator-control application of Ward et al.~\cite{ward2024ventilator}, which drives the engine through a pressure-targeted action space: only three of its six ventilator settings (FiO$_2$, respiratory rate and PEEP) are inputs swept here.

\textbf{(iv) Discarded input pairs.} The Morris elementary-effect estimator discards any input pair in which either run is flagged (Section~\mainref{sec:methods-detector}{4.4} of the main manuscript). This removes all clean FiO$_2$ pairs in the two Healthy cells and the Severe-F cell, because every FiO$_2$ step has at least one flagged endpoint: in healthy lungs, high-FiO$_2$ runs flagged \texttt{not\_steady\_state} or \texttt{implausible} by SpO$_2$-saturation pinning and, at FiO$_2$ = 1.0, transient hyperoxia; in Severe-F, low-FiO$_2$ runs flagged \texttt{implausible} or \texttt{not\_steady\_state} by severe hypoxemia or a still-drifting SpO$_2$. The FiO$_2$ rows of Figure~3 of the main manuscript are therefore blank in these cells, and the cross-sex Spearman comparison is limited to Mild and Moderate (\S\ref{sec:s6-rho}). These are deterministic consequences of the preregistered implausibility bound and steady-state window at the edges of the envelope, not a Pulse defect. More trajectories cannot recover the discarded pairs, and re-running the grid with the 300-second window that resolves the hyperoxia plateau would break preregistration integrity. An independent same-size Morris resample of one cell (Mild-M, seed 44) preserved the highest-$\mu^{*}$ input for all six outputs but showed sample dependence in the finer ranking, and changed that cell's divergence count from 4 / 160 to 7 / 160, so the number of divergent configurations a sweep encounters is itself sample-dependent (\S\ref{sec:s3-resample}). No trajectory-count convergence analysis was performed, so the stability of $\mu^{*}$, $\sigma$ and the full top-three ordering at larger trajectory counts is not established.

\textbf{(v) What the hard bounds judge.} The hard physiological bounds behind the \texttt{implausible} flag (\S\ref{sec:s3-hard-bounds}) judge whether a \emph{sustained steady-state mean} is biologically plausible over the evaluation window, not whether a value is instantaneously attainable; rare clinical extremes, often transient or peri-mortem, can legitimately exceed them. The classifier keeps such runs and records the flag they fired rather than discarding them. The bounds are also adjustable choices rather than fixed biological constants: the deliberately conservative PaO$_2$ ceiling was re-examined under a wider, physiologically anchored limit (\S\ref{sec:s3-pao2-sensitivity}). They provide a run-level screening standard for synthetic training cohorts, not a claim that exceptional clinical trajectories are impossible.

\textbf{(vi) A non-rectangular input domain.} Pulse enforces DBP/SBP $\leq$ 0.75, which makes the sampled SBP--DBP domain non-rectangular. Configuration-time clipping shortened 9 of 20 DBP steps and made applied DBP co-vary in 8 of 20 SBP steps, so these effects are not strict fixed-step, one-at-a-time Morris estimates. SBP nevertheless remained rank 1 for MAP in all eight cells under six sensitivity estimators (\S\ref{sec:s3-clip-cost}), which establishes ranking robustness but not causal isolation of the clipping contribution. Shunt fraction was not recorded in the atlas, so shunt-mediated interpretations throughout this work remain physiologically motivated rather than directly tested.

%%%%%%%%%%%%%%%%%%%%%%%%%%%%%%%%%%%%%%%%%%%%%%%%%%%%%%%%%%%%%%%%%%%%%%
%% S3 -- Full Methods detail
%%%%%%%%%%%%%%%%%%%%%%%%%%%%%%%%%%%%%%%%%%%%%%%%%%%%%%%%%%%%%%%%%%%%%%

\section{Parameter provenance and supplementary analyses}\label{sec:s3}

The Methods section of the main manuscript states the constants on which the atlas runs: the hard physiological bounds and drift thresholds of the four-flag classifier, the clamp patterns, the sampler settings and the two reference bodies. This section does not restate them. It documents where each comes from (the clinical guideline, physiological reference or code artefact behind it) and then reports analyses that characterise rather than define the design: the cost of the pulse-pressure clip on diastolic elementary effects, the fate of the preregistered Sobol bolt-on, and robustness checks on the traffic-light baseline and on the most contestable hard bound.

\subsection{Hard physiological bounds: per-bound provenance}\label{sec:s3-hard-bounds}

The six \texttt{(lower, upper)} pairs that gate the \texttt{implausible} flag are stated in Section~\mainref{sec:methods-detector}{4.4} of the main manuscript and held in the \texttt{HARD\_BOUNDS} dictionary of \texttt{src/detector.py}. All were locked in the preregistration before any data were collected, and none has been revised post hoc. Each is regex-extracted from the source file at audit time by its \texttt{T1.06p.*} ledger entry, so any drift between code and manuscript is caught immediately. Pulse reports SpO$_2$ as a fraction in [0, 1], so the bound written as [50, 100]~\% in the main text is held in the code as 0.50 / 1.00.

Each bound belongs to one of two categories: (a)~a physical or definitional ceiling, whose only instance is the SpO$_2$ upper bound at hemoglobin saturation; or (b)~a clinical-plausibility threshold for a 60-second steady-state mean, i.e.\ a screening standard, not a claim that values beyond it are impossible in a patient (\S\ref{sec:s2-scope}). No bound rests on an equipment specification. Table~\ref{tab:hard-bounds-rationale} gives the rationale for each bound, with its primary-literature anchor where one exists; the citation set mirrors the per-bound rationale audited in the companion technical monograph (§2.6).

\begin{table}[htbp]
\centering
\small
\caption{\textbf{Per-bound rationale for the \texttt{HARD\_BOUNDS} table.} Each row gives the rationale for the lower and upper extremes of the corresponding output and, where one exists, the clinical guideline or physiological reference behind it.}
\label{tab:hard-bounds-rationale}
\begin{tabular}{lp{0.66\textwidth}}
\toprule
Output & Primary-literature rationale (lower / upper) \\
\midrule
HR [20, 220]~bpm &
Deliberately permissive at both extremes. Upper: sinus rates exceeding $(220 - \mathrm{age})$ per minute reduce cardiac output and myocardial perfusion~\cite[Ch.~1]{marini2018critical}, which gives 202~bpm on the 18-year-old reference bodies; 220~bpm is the intercept of that rule and therefore sits above it. Lower: pathological bradycardia depresses cardiac output and oxygen delivery~\cite[Ch.~1]{marini2018critical}, but no numeric floor is asserted there, so 20~bpm is a conservative outer bound. \\
MAP [30, 180]~mmHg &
Lower: well below both the 70~mmHg at which the Sepsis-3 SOFA cardiovascular score first registers organ dysfunction~\cite{singer2016} and the Surviving Sepsis Campaign initial MAP target of 65~mmHg~\cite{evans2021sepsis}. Upper: ESH 2023 hypertensive emergency, i.e.\ grade~3 hypertension (SBP $\geq$ 180 and/or DBP $\geq$ 110~mmHg) with acute symptomatic organ damage~\cite{mancia2023esh}. \\
Vt [50, 1{,}500]~mL &
Lower: below adult anatomical dead space ($\sim$~150~mL), implying zero alveolar ventilation~\cite[Ch.~2]{west2020respiratory}. Upper: $\sim 3$--$4.5\times$ the ARDS Network lung-protective tidal volume (6~mL/kg PBW $\approx$ 450 and 330~mL on the male and female reference bodies)~\cite{brower2000}, and above the 10--15~mL/kg tidal volumes that were standard practice before lung-protective ventilation~\cite[Ch.~25]{marino2013icu}. \\
SpO$_2$ [50, 100]~\% &
Lower: oxyhemoglobin dissociation curve P50 places SpO$_2$ = 50~\% at PaO$_2 \approx 27$~mmHg~\cite{severinghaus1979}. Upper: hemoglobin saturation ceiling (definitional). \\
PaO$_2$ [20, 600]~mmHg &
Lower: PaO$_2$ = 20~mmHg corresponds to SpO$_2 \approx 0.32$ on the standard Severinghaus dissociation curve~\cite{severinghaus1979}, well inside the Berlin severe-ARDS oxygenation territory~\cite{ards2012}. Upper: deliberately conservative below the alveolar-gas-equation sea-level ceiling of $\approx$~663~mmHg on FiO$_2$ = 1.0~\cite[Ch.~5]{west2020respiratory}; the supplementary \texttt{HARD\_BOUNDS\_PHYSIOLOGICAL} dictionary in \texttt{src/detector.py} raises the PaO$_2$ upper bound to 700~mmHg and powers the sensitivity-analysis reclassification reported in \S\ref{sec:s3-pao2-sensitivity}. \\
PaCO$_2$ [10, 120]~mmHg &
Lower: profound respiratory alkalosis from sustained maximal hyperventilation. Upper: deliberately permissive rather than clinical. One multinational ARDS cohort (1{,}899 patients, 927 ICUs, 40 countries) categorised the maximum PaCO$_2$ recorded in the first 48~h of ventilation; its cutoffs place approximately the 95th percentile at 70~mmHg~\cite{nin2017}. A 120~mmHg ceiling therefore leaves severe but clinically observed hypercapnia unflagged, and marks only values well beyond the reported range. \\
\bottomrule
\end{tabular}
\end{table}

\subsection{Drift thresholds for the not\_steady\_state flag}\label{sec:s3-drift}

The \texttt{DRIFT\_MAX\_DELTA\_60S} dictionary (Table~\ref{tab:drift}) sets, for each channel, the maximum tolerated $|\Delta| = |\text{slope} \times 60~\text{s}|$ over the final 60 seconds of the steady-state window before a run is flagged \texttt{not\_steady\_state} for drift. Drift is estimated by ordinary least-squares regression on the 13 samples of the [60, 120]~s window (5-second interval); the $R^2$ of the fit is logged as diagnostic metadata but does not gate the flag. Beyond these thresholds, a bedside clinician would read the trajectory as actively changing rather than at homeostasis. Unlike the hard bounds, they are atlas-internal calibration choices anchored in standard critical-care monitoring practice, not biological survival limits; the citations in Table~\ref{tab:drift-rationale} support this monitoring framing, not the specific drift magnitudes. Like the hard bounds, each value is regex-extracted from \texttt{src/detector.py} at audit time.

\begin{table}[htbp]
\centering
\caption{\textbf{Maximum tolerated drift per 60-second window for the \texttt{not\_steady\_state} flag.} SpO$_2$ is a fraction in [0, 1], hence 0.02 corresponds to 2 percentage points. Each row is regex-extracted from \texttt{src/detector.py DRIFT\_MAX\_DELTA\_60S} via the corresponding \texttt{T1.06q.*} ledger entry.}
\label{tab:drift}
\begin{tabular}{lrl}
\toprule
Output & Drift threshold per 60 s & Ledger anchor \\
\midrule
HR (bpm) & 5 & \texttt{T1.06q.hr\_drift} \\
MAP (mmHg) & 5 & \texttt{T1.06q.map\_drift} \\
Vt (mL) & 50 & \texttt{T1.06q.vt\_drift} \\
SpO$_2$ (fraction) & 0.02 & \texttt{T1.06q.spo2\_drift} \\
PaO$_2$ (mmHg) & 15 & \texttt{T1.06q.pao2\_drift} \\
PaCO$_2$ (mmHg) & 5 & \texttt{T1.06q.paco2\_drift} \\
\bottomrule
\end{tabular}
\end{table}

\begin{table}[htbp]
\centering
\small
\caption{\textbf{Per-channel clinical-meaningfulness rationale for the \texttt{DRIFT\_MAX\_DELTA\_60S} table.} Each row identifies the standard critical-care monitoring context that anchors the drift magnitude for the corresponding channel; the citation set mirrors the per-channel rationale audited in the companion technical monograph (drift-threshold subsection). The citations frame that monitoring context rather than endorse the specific magnitudes, which are atlas-internal calibration choices (see text).}
\label{tab:drift-rationale}
\begin{tabular}{lp{0.55\textwidth}}
\toprule
Channel \& drift threshold & Clinical-meaningfulness anchor \\
\midrule
HR (5 bpm / 60~s) &
A preregistered convention, not a sourced threshold: the smallest heart-rate excursion treated as an active trajectory rather than monitor noise. No numeric drift criterion for heart rate was found in the cited critical-care sources. \\
MAP (5 mmHg / 60~s) &
The clinically decisive MAP band is narrow: the Sepsis-3 SOFA cardiovascular score registers organ dysfunction below 70~mmHg~\cite{singer2016}, the Surviving Sepsis Campaign targets 65~mmHg on resuscitation, and organ perfusion declines below approximately 60~mmHg, tracking linearly with MAP~\cite{evans2021sepsis}. These landmarks lie 5~mmHg apart, so a drift of that size is clinically consequential rather than nominal. \\
Vt (50 mL / 60~s) &
Approximately 0.7--0.9~mL/kg ideal body weight across the two reference bodies, within the noise band of breath-to-breath Vt measurement at ARDS Network protective ventilation (6~mL/kg predicted body weight)~\cite{brower2000}. \\
SpO$_2$ (2 percentage points / 60~s) &
A preregistered convention, not a sourced threshold: two percentage points is the smallest saturation change treated as a genuine desaturation rather than monitor noise. No numeric desaturation-drift criterion was found in the cited critical-care sources. \\
PaO$_2$ (15 mmHg / 60~s) &
A 15--71~\% fraction of one Berlin P/F band depending on FiO$_2$: sub-band at the high-FiO$_2$ end of the sweep but approaching one band at room air; clinically meaningful oxygenation change over a 60-second window~\cite{ards2012}. \\
PaCO$_2$ (5 mmHg / 60~s) &
Approximately one alveolar-ventilation step under ARDS Network protective ventilation per minute; a clinically meaningful CO$_2$ change~\cite{west2020respiratory}. \\
\bottomrule
\end{tabular}
\end{table}

Two further \texttt{not\_steady\_state} triggers complement the drift check. The coefficient-of-variation threshold \texttt{CV\_THRESHOLD = 0.05} fires when the standard deviation of any output across the steady-state window exceeds 5~\% of its mean. The flatline-at-bound criterion fires when \texttt{std / |mean| < 0.01} AND the mean is within 0.1~\% of a hard bound; it is intended to flag numerical pinning at a bound, although physiological saturation can satisfy it too. Both constants are locked under the same preregistration discipline as the bounds and drift thresholds, but as detector conventions: unlike the 0.1~\% bound tolerance, which was calibrated against existing Pulse runs, neither has a recorded derivation, source or empirical calibration. The coefficient-of-variation threshold applies identically to all six outputs. For delivered tidal volume, where it fired most often, the median coefficient of variation across the 64 paired reruns of \S\ref{sec:s8} fell from 0.351 before trigger locking to 0.016 afterwards, so the 0.05 convention lies between the two medians; prevalence estimates conditioned on that threshold, including the 20.6~\% primary-flag rate, nevertheless remain threshold-dependent.

\subsection{Clamp patterns: provenance and the post-sweep patch}\label{sec:s3-clamp}

The \texttt{clamped} flag fires when Pulse's setup-rejection log matches any pattern in a locked set of regular expressions, each written to match a Pulse rejection message; one of them, \texttt{is outside the valid range}, matches no message in the Pulse 4.3.1 source and fired on no log on disk. The set printed below is the one currently in \texttt{src/detector.py}. The 1{,}280-run sweep was classified with the earlier \emph{four}-pattern version, calibrated against deliberate-clamp probes in the pilot before the full sweep; it replaced an overly broad regex set that produced 100~\% false-positive clamping by matching Pulse's normal solver-convergence messages. The note after the listing records what changed afterwards and why the published classification is unaffected:

\noindent\begin{minipage}{\linewidth}
\begin{lstlisting}[basicstyle=\ttfamily\small, breaklines=true]
r"baseline of [\d\.]+ (?:\S+ )?is too (?:high|low)"
r"maximum [\w\s]+ baseline allowed is"
r"minimum [\w\s]+ baseline allowed is"
r"must be modeled by adding/using a condition"
r"is outside the valid range"
\end{lstlisting}
\end{minipage}

\noindent On the calibration runs, the four-pattern set fired on both deliberate-clamp probes of the pilot (SBP = 140 and DBP = 85~mmHg, each rejected at setup with exit code 1) and on none of the eight smoke-test runs; its behaviour on every Pulse log on disk is replayed in the note below.

\noindent \textbf{Patch of 2026-07-26, applied after the sweep.} Two changes produce the five-pattern set above: the unit token between the value and ``is'' became optional, and a \texttt{minimum \ldots\ baseline allowed is} pattern was added to mirror the existing \texttt{maximum} one. Of the ten baseline-rejection strings built by Pulse's \texttt{SetupPatient.cpp}, nine include the unit; the exception, the heart-rate lower bound, meant that a heart rate below 50~bpm was rejected at setup without raising \texttt{clamped}. Replaying both pattern sets over all 3{,}349 Pulse logs on disk changes exactly one run (the probe written to expose the gap) and no run of the published sweep, whose requested heart rates never leave [50, 110]~bpm. The patch therefore has no effect on the reported classification; it is printed so that users of this detector do not inherit the gap. No run was lost to the gap in any case: a setup rejection exits non-zero with no CSV, so \texttt{diverged} fires instead.

\noindent \textbf{One explicit downstream adjustment, outside this flag by construction.} Pulse makes a single deterministic change to a submitted ventilator setting in this sweep, and it is neither silent nor a setup-time event. When the requested inspired-oxygen fraction plus the non-nitrogen ambient fractions exceeds one, the ventilator model writes \texttt{[ERROR] FiO2 setting + ambient fractions other than N2 is greater than 1.0. Setting FiO2 to max value of 0.9996.} to the log and sets the ventilator compartment's oxygen volume fraction to 0.9996. The cause is compositional rather than a bound on the input: Pulse's ambient gas mixture carries carbon dioxide at $4 \times 10^{-4}$, so an oxygen fraction of exactly 1.0 leaves no room for it. Of the 208 runs requesting FiO$_2$ = 1.0, 195 reach the ventilator action and \emph{all 195} are adjusted once, when the action is applied; the other 13 diverge during patient stabilisation, before the action is submitted (Section~\mainref{sec:results-sex}{2.5} of the main manuscript). No requested FiO$_2$ below 1.0 is affected, and the substituted value is identical in every affected run. The adjustment (0.04~\% in relative terms) moves the sea-level alveolar-gas ceiling from 663.0 to 662.7~mmHg, so the $\approx$663~mmHg used elsewhere remains correct. It reclassifies no run as \texttt{clamped}, a category preregistered on setup-time rejections matched by the locked patterns, and the serialised ventilator action still carries FiO$_2$ = 1.0 (ingestion is exact for all 10{,}160 contracts that reached it). Nor does it bear on hypothesis H2, which concerns \emph{silent} clamping: this adjustment is announced in the log.

\subsection{Body-dimension provenance for the two adult reference patient bodies}\label{sec:s3-bodies}

The two adult reference bodies are male (age 18, 67.3~kg, 180~cm) and female (age 18, 56.2~kg, 163~cm). Both weights are U.S.\ Centers for Disease Control and Prevention (CDC) growth-chart 50th-percentile weight-for-age values for each sex~\cite{kuczmarski2002cdc}, taken from the row \texttt{Agemos}~=~216.5 of the primary CDC data file, whose 50th percentiles (67.28992603~kg for males, 56.22969564~kg for females) round to the preregistered 67.3 and 56.2~kg. CDC lists age at the half-month point standing for the whole month, so \texttt{Agemos}~=~216.5 denotes 216.0--216.99 months, the interval beginning at exactly 18.0 years. The preregistration's provenance erratum, which describes these weights as the CDC values \emph{at 18 years (216.5 months)}, is therefore correct.

The female height matches both the CDC 50th-percentile value~\cite{kuczmarski2002cdc} and Pulse's \texttt{heightStandardFemale\_cm = 163.0} reference constant in \texttt{SetupPatient.cpp}. The male height is a clinically standard adult value chosen to fix the male body mass index at 20.8; it is not Pulse's \texttt{heightStandardMale\_cm = 177.0} constant, which would have yielded a BMI close to the CDC 50th percentile. Pulse computes the blood-volume baseline via the allometric relation $\text{BV} = 65.6 \cdot \text{weight}_{\text{kg}}^{1.02}$, giving 4{,}802.63~mL for the male body and 3{,}996.09~mL for the female body, the 17~\% gap whose effect on the input-extreme divergence cluster is characterised in Section~\mainref{sec:results-sex}{2.5} of the main manuscript.

\subsection{Sampler constants in the code base}\label{sec:s3-morris}

The sampler settings of Section~\mainref{sec:methods-sweep}{4.3} of the main manuscript (20 trajectories, 4 grid levels per input, random seed 42) are fixed as \texttt{SWEEP\_TRAJECTORIES}, \texttt{SWEEP\_LEVELS} and \texttt{SWEEP\_SEED} in \texttt{scripts/run\_full\_sweep.py} and regex-extracted at audit time by ledger entries \texttt{T1.06e}, \texttt{T1.06f} and \texttt{T1.06g}, so any drift between the manuscript and the code that produced the sweep is caught mechanically.

\subsection{Four-flag classifier: priority order and rationale}\label{sec:s3-detector-order}

The detector's priority order \texttt{clamped > diverged > implausible > not\_steady\_state > clean} follows the chronology of failure modes in a Pulse simulation, from input handling through solver execution to output evaluation. \texttt{clamped} is an input-contract breach: Pulse did not accept at least one setup input as submitted, so the run no longer corresponds to the requested Morris grid point and no downstream flag would refer to the input being studied. \texttt{diverged} is a mathematical failure (non-zero exit code, NaN or infinite values in the CSV, missing CSV or truncated simulation) that precludes any biological interpretation. \texttt{implausible} is a plausibility failure: the computation ran cleanly, but a steady-state mean falls outside the hard bounds of Section~\mainref{sec:methods-detector}{4.4} of the main manuscript (a screening outcome; \S\ref{sec:s3-hard-bounds}). \texttt{not\_steady\_state} is a dynamic failure: the run completed with finite output but did not satisfy the steady-state criterion in the 60-second window (flatline at a bound, coefficient of variation above 0.05, or drift beyond the thresholds of Table~\ref{tab:drift}), so its 60-second mean is not treated as a steady-state summary.

\subsection{Cost of the pulse-pressure clip on diastolic elementary effects}\label{sec:s3-clip-cost}

Clipping DBP when each configuration is written, rather than when the sample is drawn, keeps the Morris sample geometrically intact but introduces a quantifiable distortion, reported here: for the affected steps, the engine receives a smaller diastolic change than the sample records, whereas the elementary-effect estimator divides by the full nominal step.

Of the 160 sampled points, 35 request a DBP above the 0.75 ratio and are clipped. This has two consequences, both fixed by the sample and therefore identical in all eight cells. First, 9 of the 20 diastolic steps are shortened: the nominal DBP step is two-thirds of the 20~mmHg range (13.33~mmHg); five of the nine fall to 6.25~\% of it (0.83~mmHg), one to 56.25~\% and three to 62.5~\%, and none to zero. Second, in 8 of the 140 steps the changed input is SBP, which moves the $0.75 \times \mathrm{SBP}$ threshold and hence the applied DBP, so two inputs change at once, violating the one-at-a-time assumption. The remaining 123 of 140 steps (87.9~\%) are unaffected. After flag filtering, 41 of 78 clean diastolic pairs are shortened and 19 of 68 clean systolic pairs are affected, leaving 49 systolic pairs free of concomitant diastolic movement (per cell: 4, 4, 7, 6, 10, 9, 4 and 5 for Healthy-M, Healthy-F, Mild-M, Mild-F, Moderate-M, Moderate-F, Severe-M and Severe-F). Because each trajectory carries exactly one systolic step, dropping the contaminated pairs or the trajectories that carry them removes the same systolic pairs, so these counts apply to both the uncontaminated-step and trajectory-matched estimators: the systolic robustness checks below rest on as few as four elementary-effect pairs in Healthy-M, Healthy-F and Severe-M.

The distortion does not overturn the reported findings. H1 holds in 48 of 48 cell--output combinations under either estimator (median top-3 fraction 0.928 nominal versus 0.910 applied-step; minimum 0.658 versus 0.696), and output MAP remains SBP-dominated in all eight cells. What changes is the standing of DBP itself: under the applied-step estimator, it enters the top three in 9 of the 48 combinations from which the reported ranking excludes it, and its $\mu^{*}$ rises by a median factor of 2.4 (maximum 15.4). The reported DBP $\mu^{*}$ should therefore be read as a lower bound relative to the applied-step estimator on the same pairs (dividing by a shorter step can only raise $|\mathrm{EE}|$, so this direction is arithmetic rather than empirical), not as a lower bound on DBP's influence in general, which the flag-filtered pair set does not establish.

We do not substitute the applied-step estimator for the preregistered one. Dividing by the applied step corrects the denominator on the shortened diastolic transitions, but it does not restore the one-at-a-time property on the eight systolic transitions where the ratio constraint also moves DBP, and the effective diastolic step then ranges from 0.83 to 13.33~mmHg, which is no longer the fixed-step design on which the Morris index is defined. Step length is also structurally confounded with position in the sampled space: the shortest steps occur only at SBP = 90~mmHg, whereas all 11 full-length diastolic steps occur at SBP $\geq$ 100~mmHg, so no difference between the two estimators can be attributed to step length alone. The applied-step $\mu^{*}$ is therefore reported as evidence on the direction of the clip's effect, not as a term-by-term corrected index. Imposing the ratio at sampling time instead would forfeit the sample geometry the design was built to protect.

Two further estimators test whether the reported systolic dominance depends on the affected steps. The \emph{uncontaminated-step} estimator drops every clean pair in which the applied DBP co-varies with SBP; the \emph{trajectory-matched} variant also restricts every input to the 12 of 20 trajectories whose systolic step is uncontaminated, so that all inputs are estimated on the same amount of data. Two \emph{combined} estimators apply the applied-step denominator and the contaminated-pair exclusion together. SBP remained rank 1 for MAP in all eight cells under all six estimators (nominal-step, applied-step, uncontaminated-step, trajectory-matched, hybrid-combined and matched-combined). Under the combined estimators, DBP is the runner-up in all eight cells, with a smallest margin of 1.22-fold ($\mu^{*}$ ratio) in Mild-M. Under the hybrid-combined estimator, DBP would overtake SBP in a cell only if its applied-to-nominal $\mu^{*}$ ratio (range 1.131--3.289) exceeded that cell's uncontaminated-SBP-to-nominal-DBP ratio (range 2.266--4.669), which occurred in no cell. The matched-combined analysis independently kept SBP at rank 1 in all eight cells.

A descriptive transition-stratified comparison narrows, without removing, the confounding between exclusion and position. Three systolic transitions carry both contaminated and uncontaminated steps, so the two groups can be compared at identical systolic start and end values (Table~\ref{tab:transition-stratified}). In each of the three, contaminated steps had the higher median absolute MAP elementary effect, and the stratified contrast remained positive in every leave-one-transition-out analysis. Requested DBP is perfectly collinear with contamination: structurally for the 100-to-120 and 120-to-100 transitions, whose 75~mmHg ceiling no lower grid level can exceed, and by chance in the realised Morris draw for 110-to-90. The ordering does not support a simple monotonic explanation by mean applied-DBP level alone: the largest contrast occurred for 110-to-90 despite its lower mean endpoint DBP, which is in any case only a coarse proxy for the level at which the elementary effect is evaluated. These results are consistent with an influence of applied-DBP co-variation but do not establish a causal contribution.

\begin{table}[htbp]
\centering
\caption{\textbf{Transition-stratified comparison of systolic elementary effects on MAP.} Counts are clean elementary-effect pairs pooled across the eight atlas cells, not trajectories of the shared 20-trajectory design. The comparison is descriptive: matching the systolic transition does not control the requested DBP, the other five inputs, or position within the trajectory. The 90-to-110 transition carries no contaminated step and is therefore not evaluable. \emph{Midpoint} is the mean of the two endpoint values of the applied-DBP path, a coarse proxy for the diastolic level at which the elementary effect is evaluated.}
\label{tab:transition-stratified}
\footnotesize
\begin{tabular}{lrrrrrrr}
\toprule
SBP & \multicolumn{2}{c}{Clean pairs} & \multicolumn{2}{c}{Median $|\mathrm{EE}|$} & Diff. & \multicolumn{2}{c}{Applied DBP (mmHg)} \\
\cmidrule(lr){2-3}\cmidrule(lr){4-5}\cmidrule(lr){7-8}
transition & contam. & uncontam. & contam. & uncontam. & & path & midpoint \\
\midrule
100-to-120 & 3 & 9 & 20.44 & 15.74 & $+4.70$ & $75 \to 80$ & 77.5 \\
110-to-90 & 15 & 16 & 24.94 & 17.26 & $+7.68$ & $80 \to 67.5$ & 73.75 \\
120-to-100 & 1 & 19 & 20.32 & 16.74 & $+3.59$ & $80 \to 75$ & 77.5 \\
90-to-110 & 0 & 5 & --- & --- & --- & --- & --- \\
\bottomrule
\end{tabular}
\end{table}

\subsection{Sobol bolt-on: preregistered decision}\label{sec:s3-sobol}

A second-tier variance-based (Sobol) analysis was preregistered as a conditional bolt-on to the Morris screen. The rule locked in \texttt{preregistration.md}~\S~Sampling method evaluates the top-three $\mu^{*}$-ranked inputs of each cell $\times$ output combination by their $\sigma / \mu^{*}$ ratio, the standard Morris proxy for non-linearity and input--input interaction: at or below 0.4 the response is treated as linear and no Sobol analysis is run; at or above 0.6 the input is promoted; values in between form a declared gray zone.

Applied as written, the trigger fires on 72 of the 144 top-three inputs, spanning 43 of the 48 cell $\times$ output combinations. The preregistration capped the bolt-on at 1{,}500 additional runs, with a cost model of $64 \times (2k + 2)$, where $k$ is the number of promoted inputs. A Sobol sample cannot be shared across cells, because each scenario--sex cell is a different model, but it is shared across the six outputs of a cell, which come from the same Pulse runs. The 43 triggered combinations fall in eight cells, each with three to six distinct promoted inputs, for 6{,}144 runs under the formula as written. Computing only the first-order and total indices required by the analysis plan, at $64 \times (k + 2)$, lowers this to 3{,}584; a full seven-input design in every firing cell would cost 8{,}192 and 4{,}608, respectively. All these figures exceed the preregistered envelope, by a factor of 2.4 to 5.5, so the bolt-on is reported rather than executed.

The released code base keeps a restricted exploratory Sobol demonstrator (\texttt{scripts/s019\_sobol\_bolton.py}), written during an earlier, dominant-input reading of the trigger (six cells selected on the dominant input, all seven swept inputs, base sample 256, first-order indices only; 13{,}824 runs). It shares the locked input space of the Morris sweep but does not implement the preregistered design: the 6{,}144-run figure above is the cost of that design, not of the demonstrator.

The preregistered rule conditions on the point ratio without any uncertainty qualifier, so we add a post hoc sensitivity analysis, labelled as such: for each top-three input, trajectories are resampled 1{,}000 times and $\sigma / \mu^{*}$ is recomputed jointly on each draw, so that the interval covers the ratio itself. Bootstrapping $\sigma$ alone and dividing by a fixed $\mu^{*}$ would ignore their covariance and distort the interval in an uncontrolled direction. Under each of five independent bootstrap seeds, 23 of the 144 top-three inputs keep the lower bound of this joint interval at or above 0.6, and repeating the procedure with 500, 2{,}000 and 4{,}000 resamples returns the same 23 inputs. This stable set spans 17 of the 48 cell $\times$ output combinations, all seven swept inputs, all four scenarios and both reference bodies (12 female, 11 male); PEEP is the most frequent input (8 of 23), followed by respiratory rate (5).

One structural limit governs how this map should be read. Because the elementary-effect estimator uses only unflagged runs, a regime unstable enough to flag its own runs cannot supply the pairs needed to characterise it. Delivered tidal volume is the clearest case: it is the most affected output in the stable set (8 of 23), yet none of the 8 falls in Healthy, the scenario where its oscillation is most prevalent (delivered-Vt coefficient of variation above 0.05 on 139 of 320 runs, versus 36 of 320 in Moderate). The median top-three input retains 3 clean trajectories in Healthy and 16 in Moderate. Departure from linearity (curvature or input--input interaction, which this ratio does not separate) is therefore easiest to establish where the engine is most stable, the opposite of where it matters most.

\subsection{Traffic-light baseline: robustness of the male-referenced scoring}\label{sec:s3-traffic}

The rule that colours each cell green, yellow or red is given in Section~\mainref{sec:methods-traffic}{4.5} of the main manuscript and its three criteria in the legend to Figure~2; the $10~\%$ and $30~\%$ deviation bands were locked as item~7 of \texttt{preregistration.md}. Here we check one discretionary choice of that rule: scoring every cell, female cells included, against the Healthy-M baseline. A single baseline keeps male and female cells on one scale, but a female cell's deviation then also contains the expected difference between the two bodies.

This choice does not drive the classification. Re-scoring the 24 female combinations against a sex-specific Healthy-F baseline, with the flag-rate logic unchanged, changes the colour of \emph{0 of 24} (reproducible via \texttt{scripts/baseline\_robustness\_check.py}). Female yellow and red colours are set either by the diverged/implausible flag rate or by ARDS-scale deviations (e.g.\ output heart rate roughly doubling in severe ARDS), which exceed the male--female differences of the two Healthy cells (0.3 to 11.3~\%, largest for PaCO$_2$). Only one female combination, PaCO$_2$ in Healthy-F, crosses a band on the sex difference alone (11.3~\% from Healthy-M, above $10~\%$), and its cell's hard flags already make it yellow.

\subsection{PaO$_2$ ceiling: reclassification under a wider bound}\label{sec:s3-pao2-sensitivity}

Within category (b), the exact numerical choice (e.g.\ a PaO$_2$ lower bound of 20 vs 25~mmHg) is a defensible judgement call within the cited literature rather than a unique biological constant. Section~3.8 of the companion technical monograph applies a sensitivity analysis to the most contestable bound, raising the PaO$_2$ upper bound from 600 to 700~mmHg according to the alveolar-gas-equation ceiling; the 34 exceedances it starts from are reported in Section~\mainref{sec:results-hyperoxia}{2.4} of the main manuscript. Under this wider bound, 32 of the 34 runs flagged \texttt{implausible} by the locked 600~mmHg ceiling no longer exceed the widened PaO$_2$ plausibility bound (31 at FiO$_2 \geq 0.95$; 21 on the female and 11 on the male reference body). The remaining two are Healthy-F runs at maximal minute ventilation (Vt 800~mL $\times$ RR 20 = 16~L/min) whose PaO$_2$ undergoes a genuine numerical runaway within the window, stable at $\approx$480~mmHg for the first 60~s and then climbing to a physically impossible plateau above the 760~mmHg sea-level barometric pressure; they correctly remain implausible under either bound (the same solver-instability mode seen at extended windows, Section~\ref{sec:s9-guidance}).

The wider bound has one limit that bears on how this side-analysis should be read. 700~mmHg is a \emph{flat} ceiling derived at FiO$_2$ = 1.0, whereas the alveolar-gas-equation value scales with each run's own inspired fraction and PaCO$_2$. Of the 32 reclassified runs, 31 are at FiO$_2 \geq 0.95$, where the flat and run-specific ceilings are close; the exception is a Healthy-F run at FiO$_2$ = 0.737, whose final-60-s mean of 679~mmHg far exceeds the $0.737 \times (760 - 47) - 16.2 / 0.8 \approx 505$~mmHg permitted by its own inspired fraction and PaCO$_2$, and which an FiO$_2$-aware ceiling would keep flagged. The flat 700~mmHg bound is therefore a deliberately permissive stress test of the locked classifier, not a better classifier. No locked threshold has been adjusted post hoc; the only amendment to a locked threshold (2026-04-22-b, flatline tolerance 1~\% $\rightarrow$ 0.1~\%) was made before the sweep launched.

\subsection{Sensitivity of the traffic-light map to the preregistered wording}\label{sec:s3-tl-sensitivity}

Section~\mainref{sec:methods-traffic}{4.5} of the main manuscript records a post hoc deviation between the preregistered and implemented traffic-light rules; this subsection quantifies it. The repository establishes that the implemented rule was committed after preregistration, but not whether it was defined before or after the sweep outputs were inspected. All three classifications in Table~\ref{tab:s3-tl-sensitivity} were rebuilt from the eight per-cell run-level artefacts (\texttt{data/runs/full\_sweep/*/aggregated.parquet}) rather than read from the derived classification table, so the comparison does not depend on the chain it audits (\texttt{scripts/audit\_traffic\_light\_rule.py}).

\begin{table}[htbp]
\centering
\caption{\textbf{The traffic-light map under three readings of the preregistered rule.} Under the implemented rule, any hard flag (\texttt{diverged} or \texttt{implausible}) makes a combination at least yellow, and hard flags make it red only when their combined rate exceeds $30~\%$. The literal readings make a combination red as soon as its cell carries any flag, and differ only in whether ``flag'' spans the two hard flags or all four. Deviation thresholds ($10~\%$, $30~\%$) are the preregistered ones throughout.}
\label{tab:s3-tl-sensitivity}
\begin{tabular}{lccc}
\toprule
Reading of the rule & Green & Yellow & Red \\
\midrule
Implemented hard-flag-rate rule (reported) & 7 & 29 & 12 \\
Literal wording, ``flag'' = the two hard flags & 7 & 3 & 38 \\
Literal wording, ``flag'' = all four flags & 0 & 0 & 48 \\
\bottomrule
\end{tabular}
\end{table}

Four properties of Table~\ref{tab:s3-tl-sensitivity} matter more than the counts. First, the rebuilt implemented rule reproduces the published classification for all 48 cell $\times$ output combinations, confirming that the artefact underlying Figure~2 of the main manuscript is correct rather than merely self-consistent. Second, no deviation falls exactly on the $10~\%$ or $30~\%$ threshold, so the readings' different treatment of the bounds (inclusive versus strict) cannot change a single verdict. Third, the green set is invariant: the same seven combinations are green under the implemented rule and the literal hard-flag reading, because green requires zero hard flags either way and only cells Mild-F and Moderate-F carry none; the implemented rule therefore never converted a combination into a favourable verdict. Fourth, all twelve combinations that are red under the implemented rule remain red under the literal reading, so the whole difference lies in the 26 combinations that move from yellow to red at the boundary between ``carries a hard flag'' and ``carries them at above $30~\%$''.

The four-flag reading is reported to document the ambiguity of the preregistered word ``flag'', not as an alternative analysis: every cell contains \texttt{not\_steady\_state} runs (minimum 12, in Moderate-M), so this reading colours all 48 combinations red and carries no information. The list of the 26 transitions, the chronology of the two rules in the repository and the reasons for retaining the implemented rule are recorded in the companion audit register released with the code.

%%%%%%%%%%%%%%%%%%%%%%%%%%%%%%%%%%%%%%%%%%%%%%%%%%%%%%%%%%%%%%%%%%%%%%
%%%%%%%%%%%%%%%%%%%%%%%%%%%%%%%%%%%%%%%%%%%%%%%%%%%%%%%%%%%%%%%%%%%%%%

\subsection{What the composite colour compresses}\label{sec:s3-decomposition}

\begin{figure}[htbp]
\centering
\includegraphics[width=0.95\textwidth]{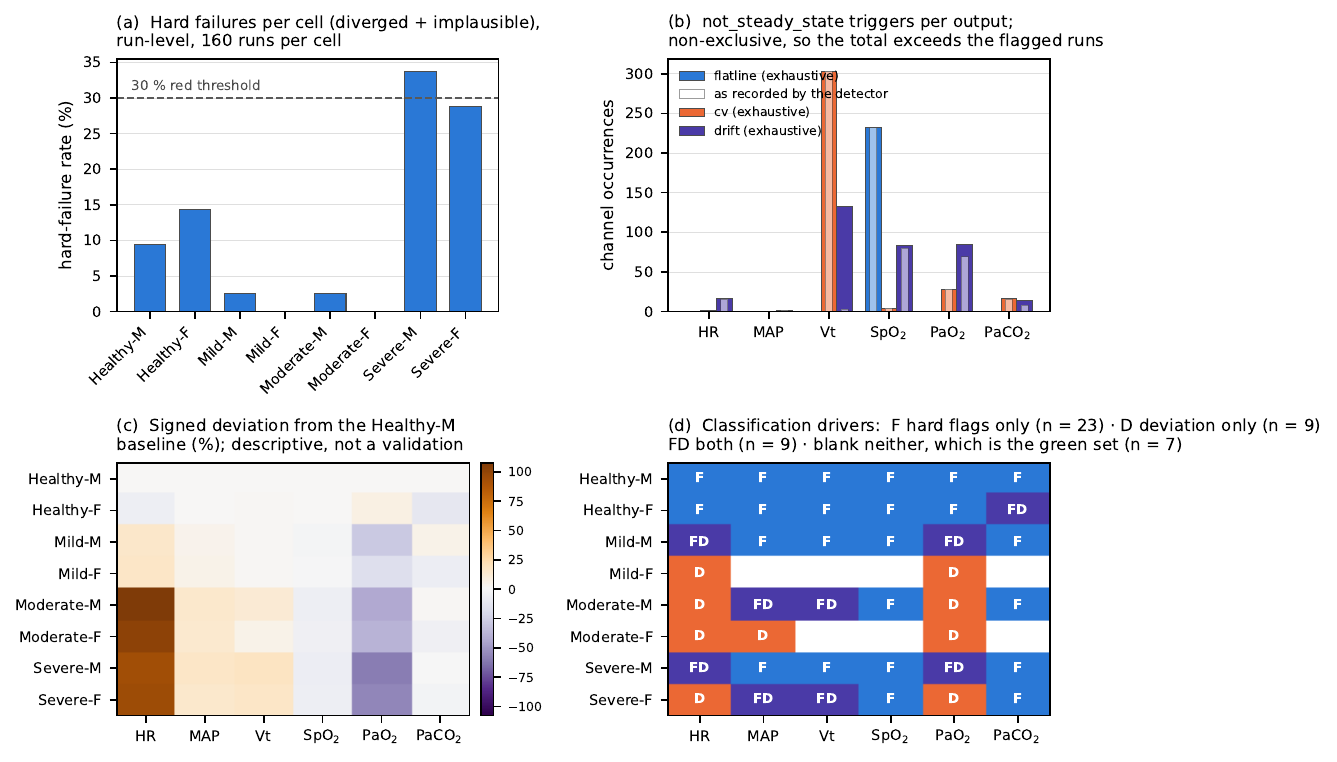}
\caption{\textbf{The three ingredients compressed by the composite map, and the criterion behind each classification.} \textbf{(a)} Hard-failure rate per cell (\texttt{diverged} + \texttt{implausible} over the 160 runs of the cell). These flags are run-level, so there are eight values, deliberately not duplicated across the six outputs. Only Severe-M exceeds the 30~\% rate that makes a cell red on this criterion alone (54 / 160 = 33.75~\%); Severe-F (46 / 160 = 28.75~\%) sits just below it. \textbf{(b)} \texttt{not\_steady\_state} triggers per output. A channel may satisfy several triggers, so these channel-level occurrences (total 918) exceed the number of flagged runs (530) and must not be summed as independent runs. The pale inner bar shows what the classifier actually \emph{recorded}: its checks stop at the first trigger reached per channel (flatline, then CV, then drift), so co-occurring mechanisms are undercounted. For delivered Vt the gap is large: 133 channels exceed the drift threshold but only 3 appear in the recorded reason, because the CV trigger was reached first in 130 of them. \textbf{(c)} Signed deviation of each cell's clean-run mean from the Healthy-M baseline. In every ARDS cell, HR lies above and PaO$_2$ below this baseline, the direction predicted by the pathophysiology. This panel is descriptive: an expected direction does not validate the magnitude, and nothing here constitutes external physiological validation. \textbf{(d)} Criterion behind each of the 48 classifications. \emph{Neither} coincides with the green set by construction and is left blank. Colours are categorical (green / yellow / red are reserved for the composite map), and each tile carries a glyph so that categories survive greyscale printing. Cell-label key: Table~1 of the main manuscript.}
\label{fig:s5-decomposition}
\end{figure}

Figure~\ref{fig:s5-decomposition} separates the ingredients that
Figure~2 of the main manuscript combines into one colour per tile and adds the
per-cell driver matrix that did not fit in the main figure.
The classification thresholds in panel~(d) are class-relative and deliberately
asymmetric: for deviation, yellow above $10~\%$ and red above $30~\%$; for hard
flags, yellow at any positive rate and red above $30~\%$. The
\texttt{not\_steady\_state} attribution in panel~(b) is reconstructed
from the stored per-channel statistics by re-applying the
registered thresholds, which reproduces the stored
flag on 1{,}280 of 1{,}280 runs.

\subsection{Sampling stability at fixed trajectory count}\label{sec:s3-resample}

The canonical sweep draws one Morris sample (seed 42). To test the sensitivity of the rankings to that draw, cell Mild-M was re-run in full with an independent sample (seed 44) of the same design (20 trajectories, 4 levels, 160 points). This varies the sample, not its size: it is a resampling check, not a convergence analysis in the number of trajectories.

The highest-$\mu^{*}$ input was preserved for all six outputs. The complete top-three set was preserved for four of six (MAP, PaCO$_2$, PaO$_2$, SpO$_2$) and its order for three; the observed disagreements were confined to ranks 2 and 3 for heart rate, delivered tidal volume and PaO$_2$.

Two sample-dependent quantities are reported for completeness. First, the Mild-M divergence count rose from 4/160 to 7/160, so the number of divergent configurations a sweep encounters is itself sample-dependent. Second, the $\sigma/\mu^{*}$ point estimate for FiO$_2$ on PaO$_2$ moved from 0.667 to 0.547, crossing the preregistered 0.6 point-estimate promotion trigger, which illustrates the sample dependence of that trigger. This does not contradict the 23-input bootstrap-stable set of \S\ref{sec:s3-sobol}, from which this combination was already excluded because its ratio-aware interval straddled 0.6 under every bootstrap seed (lower bounds 0.287--0.309).
\subsection{Morris terminology: trajectory, elementary effect, $\mu^{*}$ and clean pair}\label{sec:s3-terms}

These four terms are used throughout the main manuscript and are defined here in full; Table~\mainref{tab:terms}{1} of the main manuscript defines the three terms needed to read its figures.

\textbf{Trajectory.} One Morris path through the seven-dimensional input space. From a random starting point, each step changes exactly one input by a fixed Morris step (two-thirds of that input's swept range) until all seven inputs have each been changed once. A trajectory therefore visits $7+1 = 8$ points, and \emph{consecutive} points differ in a single input, apart from the pulse-pressure clip quantified in \S\,\ref{sec:s3-clip-cost}. With 20 trajectories, each cell contains $20 \times 8 = 160$ simulations.

\textbf{Elementary effect (EE).} A finite-difference estimate of one input's influence on one output: the change in steady-state output divided by the size of the step that produced it, expressed as a fraction of the input's swept range. Every step here is the same fixed jump of two-thirds of that range (Morris grid, 4 levels). Each elementary effect comes from one consecutive pair of runs in a trajectory (the run just before and the run just after the step); for each (cell, input, output) combination, elementary effects are then reduced across trajectories into $\mu^{*}$.

\textbf{$\mu^{*}$.} Mean of the absolute elementary effects of one input on one output across all trajectories in a cell. The Morris input-importance metric reported in Figure~\mainref{fig:morris}{3} of the main manuscript; large $\mu^{*}$ indicates strong influence of the input on the output in that cell.

\textbf{Clean pair.} The two runs on either side of one Morris step when \emph{both} are flag-clean, i.e.\ neither fired any of the four flags (Section~\mainref{sec:methods-detector}{4.4} of the main manuscript). Elementary effects are computed only from clean pairs: if either run is flagged, the pair contributes nothing.

\textbf{Two statistical terms of Section~\mainref{sec:methods-stats}{4.6} of the main manuscript.} The \emph{false-discovery rate}, controlled there by the Benjamini--Hochberg procedure, is the expected fraction of false positives among the rejected tests. The \emph{Spearman rank correlation} $\rho$ between the male and female $\mu^{*}$ vectors is $+1$ when the two orderings agree and $-1$ when they are reversed.

%% S4 -- Extended Table 3

\section{Extended Table 3: not\_steady\_state attribution}\label{sec:s4}

Table~3 of the main manuscript reports per-cell counts under the locked four-flag classifier. Each cell's \texttt{not\_steady\_state} count combines three trigger sub-modes: flatline at a hard bound, coefficient of variation above 0.05, and drift above the per-channel threshold (Supplement~\S\,\ref{sec:s3-detector-order}). A post hoc trigger-attribution audit decomposes the 530 \texttt{not\_steady\_state} runs by sub-mode (Table~\ref{tab:s4-nss}). The locked classifier is unchanged; this table is a presentation transform showing in finer detail what the flag captures in each cell.

\begin{table}[htbp]
\centering
\caption{\textbf{Per-cell attribution of the locked \texttt{not\_steady\_state} count by trigger sub-mode.} Two sub-modes are audited explicitly: SpO$_2$ ``flatline at ceiling'' with mean $\geq$ 0.999 (a detector edge case in which a healthy or well-oxygenated patient's physiological hemoglobin saturation at 1.0 numerically satisfies the locked flatline rule), and Vt CV $>$ 0.05 (the tidal-volume oscillation, consistent with patient--ventilator dyssynchrony, characterised in Section~\mainref{sec:results-vcac}{2.3} of the main manuscript). The two ``excl.'' columns count the \texttt{not\_steady\_state} runs that keep at least one other recorded trigger once that sub-mode is set aside; the ``excl.\ both'' column sets both aside simultaneously. The counts use the triggers the locked classifier recorded, at most one per channel (the first to fire, in the order flatline, CV, drift); the classifier is not re-run without the sub-mode. The last column counts the runs then left with no flag at all. Source: \texttt{results/s022/nss\_decomposition.csv}; ledger anchors \texttt{T2.nssC.*}. Cell-label key: Table~1 of the main manuscript.}
\label{tab:s4-nss}
\begin{tabular}{lrrrrrr}
\toprule
Cell & n\_total & n\_nss & excl.\ SpO$_2$ sat. & excl.\ Vt CV & excl.\ both & \shortstack[r]{n\_clean\\(excl.\ both)} \\
\midrule
Healthy-M & 160 & 107 & 91 & 74 & 29 & 119 \\
Healthy-F & 160 & 118 & 88 & 87 & 24 & 122 \\
\midrule
Mild-M & 160 & 69  & 60 & 45 & 23 & 133 \\
Mild-F & 160 & 68  & 52 & 44 & 7  & 153 \\
\midrule
Moderate-M & 160 & 12  & 11 & 9  & 8  & 148 \\
Moderate-F & 160 & 38  & 34 & 10 & 3  & 157 \\
\midrule
Severe-M & 160 & 44  & 44 & 39 & 39 & 67 \\
Severe-F & 160 & 74  & 74 & 52 & 52 & 67 \\
\midrule
\textbf{Total} & \textbf{1{,}280} & \textbf{530} & \textbf{454} & \textbf{360} & \textbf{185} & \textbf{966} \\
\bottomrule
\end{tabular}
\end{table}

Two reading guides apply. First, the SpO$_2$-saturation contribution (locked \texttt{n\_nss} minus the ``excl.\ SpO$_2$ sat.'' column, i.e.\ 76 runs across the eight cells) is mathematically valid but biologically benign: a healthy patient on adequate FiO$_2$ saturates physiologically at SpO$_2$ = 1.0, which the locked flatline rule (\texttt{std/|mean| $<$ 0.01} and mean within 0.001 of a hard bound) catches as ``pinned at the ceiling'', although it was designed to detect solver pinning. Second, the Vt CV $>$ 0.05 contribution (530 minus 360 = 170 runs) is the tidal-volume oscillation identified in Section~\mainref{sec:results-vcac}{2.3} of the main manuscript, where setting the inspiration trigger pressure explicitly brings 78.1~\% of reruns below the threshold. Excluding both sub-modes raises the implied clean count from 663 / 1{,}280 (51.8~\%) to 966 / 1{,}280 (75.5~\%); this is a decomposition for interpretive transparency, not a reclassification, and the locked classifier remains the headline reliability call throughout the main manuscript.

%%%%%%%%%%%%%%%%%%%%%%%%%%%%%%%%%%%%%%%%%%%%%%%%%%%%%%%%%%%%%%%%%%%%%%
%% S5 -- ARDS instability zone full characterisation
%%%%%%%%%%%%%%%%%%%%%%%%%%%%%%%%%%%%%%%%%%%%%%%%%%%%%%%%%%%%%%%%%%%%%%

\section{ARDS impairment-fraction instability zone}\label{sec:s5}

Section~\mainref{sec:results-secondary}{2.6} of the main manuscript refers to a documented ARDS impairment-fraction band (0.91--0.945) in which Pulse's respiratory solver becomes unstable. This section first reports the Berlin-band calibration of the scenario fractions, then describes how Pulse translates the impairment-fraction scalar into respiratory-mechanics and gas-exchange changes, how the band was identified, how the atlas's continuous-severity micro-sweep refines its edges, and how divergence within the band stratifies by sex at midrange Morris inputs.

\subsection{Berlin-band calibration of the scenario impairment fractions}\label{sec:s5-berlin}

Pulse encodes ARDS as an abstract per-lung \texttt{Severity} scalar rather than by the clinical PaO$_2$/FiO$_2$ (P/F) definition, so the four scenario impairment fractions were assigned by \emph{forward calibration}: Pulse was run across severities at a fixed reference ventilation, and the fractions were chosen so that arterial P/F falls near the centres of the Berlin Definition~\cite{ards2012} mild and moderate bands and just inside the upper boundary of the severe band, which is open-ended below and therefore has no centre (mild 200--300, moderate 100--200, severe $\leq 100$~mmHg, at PEEP $\geq 5$~cmH$_2$O). This was the original ARDS severity calibration of the companion translation project, a 28-run impairment sweep of 2026-03 (full log in the companion technical monograph, Appendix~B.3), on the adult male reference body under VC-AC with tidal volume 6~mL/kg actual body weight (403.8~mL), PEEP 10~cmH$_2$O and FiO$_2$ 0.4. Impairment fractions 0.26, 0.60 and 0.95 gave 256, 150 and 96~mmHg (Table~\ref{tab:s5-berlin-calib}). The companion's later E6 experiment (2026-04), with the same ventilator mode, tidal volume, PEEP and FiO$_2$, obtained lower values (252, 148 and 75~mmHg) in the same three Berlin bands.

This calibration has since been reproduced within the present repository (2026-07-26; 40 runs on the atlas's own engine image), so a headline result no longer depends on an external project. Evaluated over a [108, 168]~s window derived from the timing recorded with E6 (a 60-s settle, then a 108-s recording; E6 itself averaged its last five 10-s samples), the replication reproduces the March anchors within 0.6~\% at the three scenario fractions (0.26 $\to$ 256.5, 0.60 $\to$ 150.5, 0.95 $\to$ 96.6~mmHg) and within 2~\% across all five fractions compared. This agreement is empirical rather than a protocol match: the March log records 5-minute runs after a 120-s settle without stating its averaging window, and E6, under its recorded protocol, reported the lower values above. The same runs independently re-derive the instability band: fractions 0.910, 0.920, 0.930, 0.940 and 0.945 all failed to complete, whereas 0.900 and 0.950 completed.

\begin{table}[htbp]
\centering
\caption{\textbf{Reference-ventilation Berlin-band calibration of the scenario impairment fractions (companion technical monograph, Appendix~B.3, based on the \texttt{Pulse\_translation\_new} original ARDS severity calibration, the 28-run impairment sweep of 2026-03; its later \texttt{E6} experiment is an April refinement that re-confirmed the same fractions).} Arterial P/F of the adult male reference body under VC-AC, Vt 6~mL/kg actual body weight, PEEP 10~cmH$_2$O, FiO$_2$ 0.4. Pulse's own documented \{0.3, 0.6, 0.9\} defaults yield 199 / 150 / 119~mmHg at the same settings.}
\label{tab:s5-berlin-calib}
\begin{tabular}{lrl}
\toprule
Impairment fraction & P/F (mmHg) & Berlin band \\
\midrule
0.26 (Mild)     & 256 & mild (band centre) \\
0.60 (Moderate) & 150 & moderate (band centre) \\
0.95 (Severe)   & 96  & severe \\
\bottomrule
\end{tabular}
\end{table}

Because P/F itself depends on ventilation (a recognised limitation of the Berlin definition, since raising FiO$_2$ in the presence of intrapulmonary shunt lowers P/F), the Berlin band a scenario occupies varies across the sweep, whose operational inputs (FiO$_2$ 0.21--1.0, PEEP 0--20~cmH$_2$O, Vt 300--800~mL) change ventilation from run to run. Table~\ref{tab:s5-berlin-sweep} characterises this on the atlas's clean runs meeting the Berlin PEEP $\geq 5$ inclusion criterion. In the mid-FiO$_2$ stratum, nearest the FiO$_2$~=~0.4 calibration point, the scenarios recover their reference bands (Mild, Moderate, Severe). Sweep-wide medians drift because low-FiO$_2$ runs raise P/F, pulling Mild to the mild/not-ARDS boundary, whereas high-FiO$_2$ runs depress it, pulling Severe below its 96~mmHg reference to a median of 75. The two tables are therefore consistent: they measure the same engine at controlled versus varied ventilation, not two conflicting calibrations.

Two distinct effects make the severe assignment at impairment 0.95 protocol-dependent, and the in-repository replication separates them. First, the run has not equilibrated by 300~s: P/F is 87.3~mmHg over the atlas window [60, 120]~s, 96.6~mmHg over the calibration window and 112.6~mmHg over [240, 300]~s. This rise is not a ventilation effect, because alveolar PO$_2$ is essentially flat across these windows (231.3 to 230.7~mmHg) while the alveolar--arterial oxygen gradient narrows from 196.4 to 185.7~mmHg; it reflects the same incomplete equilibration characterised at FiO$_2$ = 1.0 in Supplement~\S\,\ref{sec:s9}.

Second, at any fixed window, the assignment depends on patient triggering. Under Pulse's default behaviour, the simulated patient triggers assisted breaths well above the set rate (30.6 versus a set 16~min$^{-1}$ over [240, 300]~s; each triggered breath is still ventilator-delivered, with a measured tidal volume of 418~mL versus a set 403.8~mL), which holds PaCO$_2$ at 43.6~mmHg. Locking the inspiration trigger at $-2$~cmH$_2$O lowers the rate to 20.0~min$^{-1}$ (still above the set rate), raises PaCO$_2$ to 50.8~mmHg and lowers P/F to 91.5~mmHg, keeping the run severe in all three windows. Because the alveolar--arterial gradient is unchanged between the two arms (185.7 versus 185.1~mmHg), this difference is entirely the alveolar gas equation acting through alveolar ventilation. The atlas measures over [60, 120]~s throughout, where both arms are unambiguously severe (87.3 and 69.4~mmHg); the caveat applies to replications that extend the observation window.

\begin{table}[htbp]
\centering
\caption{\textbf{Sweep-wide arterial P/F distribution per scenario on the atlas's clean runs with PEEP $\geq 5$~cmH$_2$O, stratified by FiO$_2$ (low 0.21--0.40, mid 0.40--0.70, high 0.70--1.00).} This is \emph{not} the reference calibration of Table~\ref{tab:s5-berlin-calib} (which fixes FiO$_2$ = 0.4); it shows how the Berlin-band assignment shifts with FiO$_2$ across the dynamic sweep. Empty cells: Healthy has no clean high-FiO$_2$ runs (transient hyperoxia flags PaO$_2 > 600$ as \texttt{implausible}); Severe has no clean low-FiO$_2$ runs. Reproducible via \texttt{scripts/pf\_berlin\_analysis.py} $\rightarrow$ \texttt{results/pf\_berlin\_by\_scenario.csv}.}
\label{tab:s5-berlin-sweep}
\begin{tabular}{lrrlp{2.6cm}}
\toprule
Scenario & $n$ & Median P/F [IQR] & Band (median) & P/F by FiO$_2$ (low / mid / high) \\
\midrule
Healthy (0.00) & 61  & 431 [388--481] & not-ARDS & 390 / 481 / --- \\
Mild (0.26) & 114 & 304 [264--337] & mild/not-ARDS boundary & 309 / 282 / 287 \\
Moderate (0.60) & 192 & 184 [139--231] & moderate & 228 / 163 / 160 \\
Severe (0.95) & 91  & 75 [69--83]    & severe   & --- / 99 / 74 \\
\bottomrule
\end{tabular}
\end{table}

\subsection{Severity-to-physiology mapping in the Pulse respiratory model}\label{sec:s5-mapping}

Pulse represents ARDS as a single \texttt{Severity} scalar in $[0, 1]$ per lung compartment (the \texttt{AcuteRespiratoryDistressSyndrome} condition), which the atlas sets bilaterally and symmetrically to the scenario impairment fraction. This scalar feeds four hard-coded piecewise-linear interpolation curves in the respiratory-model source (\texttt{RespiratoryModel.cpp}: \texttt{UpdateResistances}, \texttt{UpdateAlveolarCompliances}, \texttt{UpdateDiffusion} and \texttt{UpdatePulmonaryShunt}), each calibrated on the adult standard patient; resistance, compliance and diffusion cannot be set independently. Table~\ref{tab:s5-mapping} gives the resulting multipliers at the three ARDS scenario severities, linearly interpolated from the source calibration nodes.

\begin{table}[htbp]
\centering
\caption{\textbf{How Pulse v4.3.1 maps the single ARDS \texttt{Severity} scalar to respiratory-mechanics and gas-exchange multipliers at the three ARDS scenario severities (Mild = 0.26, Moderate = 0.60, Severe = 0.95).} Each multiplier scales the corresponding adult baseline value; one \texttt{Severity} value drives all four. Columns are linear interpolations of the hard-coded source calibration nodes; the Moderate (0.60) column coincides with exact nodes. Resistance increases with severity, while compliance, diffusion surface area, and shunt-path resistance decrease (a lower shunt-path resistance raises venous admixture). The shunt damage term is non-monotonic across severity and is combined in the engine with an alveolar-recruitment term, so its effective value is not the tabulated damage multiplier alone (see text). Source: \texttt{RespiratoryModel.cpp} (\texttt{UpdateResistances}, \texttt{UpdateAlveolarCompliances}, \texttt{UpdateDiffusion}, \texttt{UpdatePulmonaryShunt}).}
\label{tab:s5-mapping}
\begin{tabular}{llrrr}
\toprule
Respiratory property & Effect of severity & \shortstack[r]{Mild\\(0.26)} & \shortstack[r]{Moderate\\(0.60)} & \shortstack[r]{Severe\\(0.95)} \\
\midrule
Alveolar (distal) resistance      & increases & $8.8\times$  & $15\times$   & $22.5\times$  \\
Alveolar compliance               & decreases & $0.70\times$ & $0.55\times$ & $0.45\times$  \\
Alveolar diffusion surface area   & decreases & $0.57\times$ & $0.20\times$ & $0.075\times$ \\
Pulmonary-shunt path resistance   & decreases & $\approx 0.22\times$ & $0.20\times$ & $\approx 0.125\times$ \\
\bottomrule
\end{tabular}
\end{table}

The four curves act together: in the Severe scenario (impairment 0.95), Pulse multiplies distal alveolar resistance by $\approx 22\times$, reduces alveolar compliance to $\approx 45\,\%$ and the alveolar diffusion surface area to $\approx 7.5\,\%$ of baseline, and opens the intrapulmonary shunt. The collapse of the diffusion surface and the wider shunt provide a plausible model-internal explanation for the severe arterial hypoxemia that accompanies the compensatory tachycardia of Section~\mainref{sec:results-secondary}{2.6} of the main manuscript; shunt fraction was not recorded, so this reading is not directly tested. The shunt row is the least direct: its damage term is non-monotonic in severity and is combined, through a \texttt{MIN} operation, with an alveolar-recruitment term (a function of actual alveolar volume, which PEEP raises) and a left/right calibration factor, so the effective shunt resistance is not simply the tabulated multiplier. Severity also feeds the respiratory driver pressure, lung volumes and the inspiratory:expiratory ratio through smaller restrictive terms. The four curves are smooth; the 0.91--0.945 solver-instability band characterised below is an emergent numerical property of the combined stiff respiratory state, not a discontinuity in any single curve.

\textbf{Adult-only tuning.} Both the baseline resistance, compliance and diffusion-surface values \emph{and} the severity-to-effect curves are tuned to the adult standard patient, and the single-scalar design prevents retuning the three properties independently. Applying them to pediatric mechanics would mis-scale the absolute consequences, which is one reason why Pulse hard-gates age $\geq 18$ at patient setup.

The absence of a surfactant model was also verified directly. A case-insensitive search of the complete Pulse engine tree at \texttt{REL\_4\_3\_2} (commit \texttt{e8a3649}; physiologically identical to the 4.3.1 pin, differing only in seven build-system files) returns no occurrence of \texttt{surfactant} or of the related terms under which such a model would surface (surface tension, Laplace, derecruitment, alveolar stability, opening/closing pressure), and neither the Pulse engineering forum nor the public respiratory methodology mentions it. Pulse reproduces the macroscopic behaviours that surfactant governs without the underlying mechanism: the expected pressure--volume hysteresis of the healthy lung (per Pulse's respiratory-methodology validation) comes from a phenomenological compliance curve, and alveolar recruitment is an empirical acinar-volume-to-shunt-fraction mapping calibrated to PEEP data. Lung compliance is thus a lumped circuit element scaled by disease multipliers, with no surface-tension term, so neonatal respiratory distress syndrome, fundamentally a surfactant-deficiency disease (including its response to surfactant instillation), cannot be represented in the current model.

\subsection{Provenance of the documented band}\label{sec:s5-prov}

Three convergent boundary probes within the companion technical work (same Pulse v4.3.1 engine and male reference body, different impairment-fraction grids, independent runs) characterise the same numerical-bifurcation range in the respiratory solver. A 28-run impairment sweep at the reference ventilation of \S\ref{sec:s5-berlin} found that fractions 0.92 and 0.94 failed to stabilise. A 10-run high-resolution probe (0.90, then 0.91 to 0.95 in 0.005 steps) narrowed the failure range to 0.91--0.945 inclusive (eight consecutive failures, with 0.90 and 0.95 passing on either side). A complementary 9-run refinement at 0.93--0.97 in 0.005 steps (four failures at 0.93--0.945, five passes at 0.95--0.97) confirmed the upper edge. Because all three probes belong to the same project, their convergence shows reproducibility under one engine version and reference body, not generalisation across laboratories. Pulse's documentation does not flag this zone: published Pulse validation work and HowTo examples do not examine impairment fractions within 0.91--0.945.

\subsection{Continuous-severity micro-sweep at midrange inputs}\label{sec:s5-cont}

A 200-run continuous-severity sweep (100 male, 100 female) varied only the ARDS impairment fraction, holding all seven Morris inputs at their midrange values (HR = 80, SBP = 105, DBP = 70, RR = 14, Vt = 550, FiO$_2$ = 0.6, PEEP = 10). The 100 impairment fractions were drawn uniformly at random (seed 2026) in three strata, 50 between 0 and 0.85, 30 between 0.85 and 0.95 and 20 between 0.95 and 1, and the same 100 values were run on both bodies. It refines the empirical band edges and reveals a sex-stratification pattern that differs from the input-extreme regime of Section~\mainref{sec:results-sex}{2.5} of the main manuscript.

\begin{table}[htbp]
\centering
\caption{\textbf{Continuous-severity divergence counts per impairment-fraction bin, sex-stratified.} Runs are distributed non-uniformly across severities, as reflected in the per-bin sample sizes (\texttt{n}); bins are assigned on each run's configured impairment fraction. Divergence concentrates in [0.879, 0.946], the smallest 3-decimal interval containing every diverged run, which is wider than the documented 0.91--0.945 band by 0.031 on the low edge and 0.001 on the high edge. The sex pattern crosses over: female divergence dominates the (0.85, 0.91] bin and male divergence the (0.91, 0.945] bin. Bold cells mark the sex with more divergences in each non-zero bin. Scenario key: Table~1 of the main manuscript.}
\label{tab:s5-cont}
\begin{tabular}{lrrr}
\toprule
Impairment-fraction bin & \texttt{n} per sex & Female diverged & Male diverged \\
\midrule
$[0, 0.5]$         & 31 & 0 / 31 & 0 / 31 \\
$(0.5, 0.85]$      & 19 & 0 / 19 & 0 / 19 \\
$(0.85, 0.91]$     & 19 & \textbf{14 / 19} & 5 / 19 \\
$(0.91, 0.945]$    & 9  & 5 / 9   & \textbf{8 / 9}  \\
$(0.945, 0.95]$    & 2  & 0 / 2   & \textbf{1 / 2}  \\
$(0.95, 1.0]$      & 20 & 0 / 20  & 0 / 20 \\
\midrule
\textbf{Total}     & \textbf{100} & \textbf{19} & \textbf{14} \\
\bottomrule
\end{tabular}
\end{table}

\begin{figure}[htbp]
\centering
\includegraphics[width=0.95\textwidth]{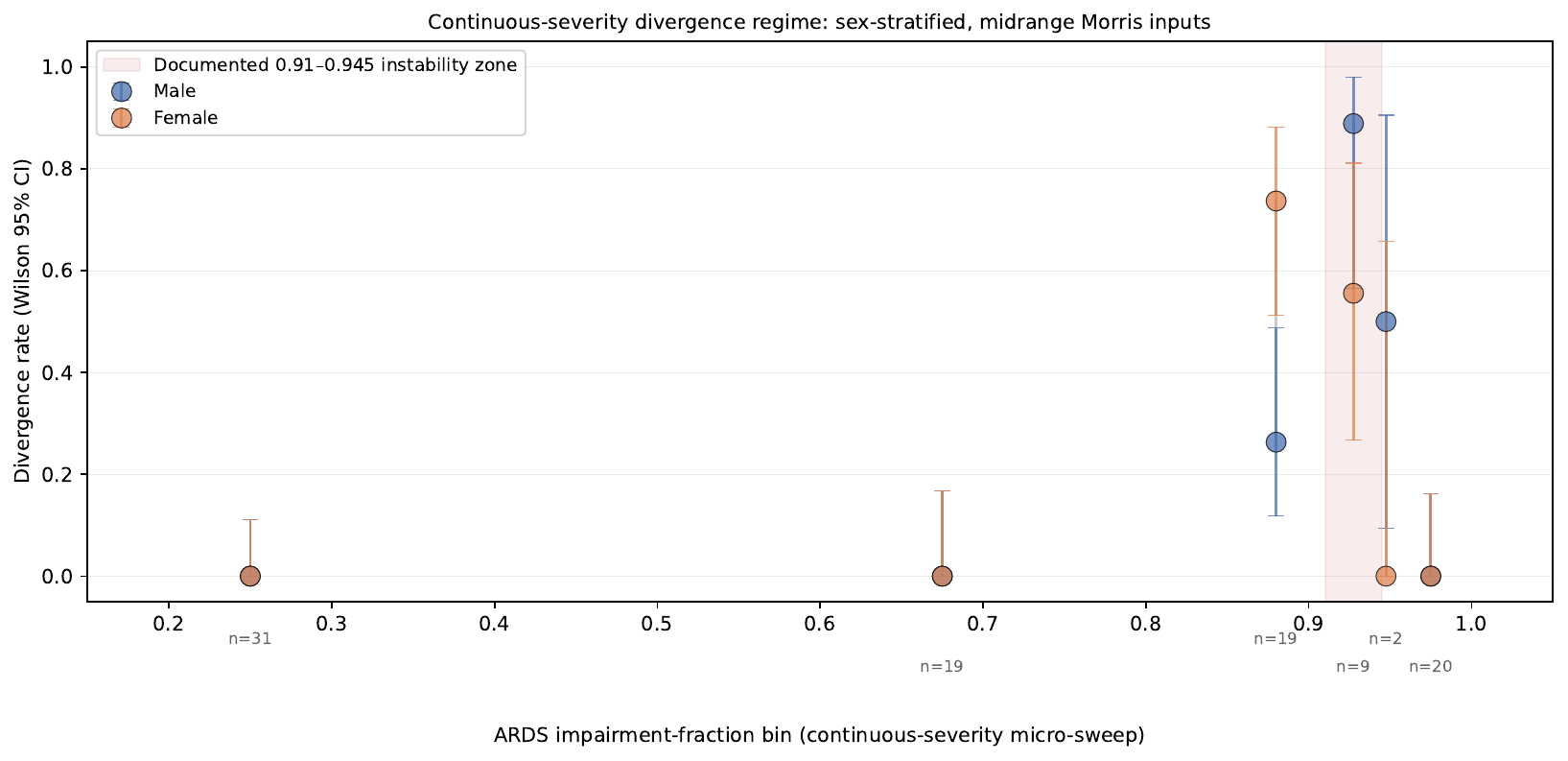}
\caption{\textbf{Sex-stratified divergence rate across the continuous-severity micro-sweep ($n = 200$ runs at midrange Morris inputs).} For each impairment-fraction bin, male (blue) and female (orange) dots are joined by a grey vertical line; error bars are Wilson 95~\% confidence intervals on the per-bin divergence proportion. The (0.945, 0.95] bin has only $n = 2$ per sex, and its very wide intervals ([0.09, 0.91] for the male $1 / 2$, [0, 0.66] for the female $0 / 2$) are shown rather than suppressed. The pink region marks the previously documented 0.91--0.945 instability zone; the continuous sweep extends the empirical band to [0.879, 0.946], almost entirely on the low edge. Female divergence dominates the lower part of the band and male divergence the upper part, so the all-male pattern of the input-extreme regime (Section~\mainref{sec:results-sex}{2.5} of the main manuscript) does not generalise to midrange inputs.}
\label{fig:s5-dumbbell}
\end{figure}

Figure~\ref{fig:s5-dumbbell} shows Table~\ref{tab:s5-cont} as a sex-stratified dumbbell plot with Wilson 95~\% confidence intervals on the per-bin divergence proportion. Two reading guides apply. First, the (0.945, 0.95] bin holds only two runs per sex (one male divergence, at impairment 0.9458; no female divergence), so its Wilson 95~\% CIs are approximately $[0.09, 0.91]$ for $1 / 2$ and $[0, 0.66]$ for $0 / 2$; they are shown because hiding an underpowered stratum at the edge of the band would misrepresent the data. Second, the band identified by this sweep (from impairment 0.879, inside the (0.85, 0.91] bin, to 0.946) is wider than the documented 0.91--0.945 band, almost entirely on the low edge. This is consistent with input-dependence of the instability: the midrange ventilator settings of this sweep reach instability earlier than the reference ventilation of \S\ref{sec:s5-berlin} used in the original boundary probes.

\textbf{Robustness of the discrete scenario calibration.} The continuous sweep also checks the four discrete scenario points of the atlas (Healthy = 0.00, Mild = 0.26, Moderate = 0.60, Severe = 0.95). Mild and Moderate lie in the zero-divergence regime $(0, 0.85]$ of Table~\ref{tab:s5-cont} (0 / 50 divergences for each sex across the two lowest non-trivial bins combined), so calibration shifts of $\pm 0.05$--$0.10$ around these values stay in that regime at midrange inputs; whether the Morris $\mu^{*}$ rankings in Mild and Moderate are robust to such shifts was not tested. Severe (0.95) was preregistered to sit just above the upper edge of the documented band, as a deliberate near-edge stress test of severe ARDS rather than a buffered choice (Section~\mainref{sec:methods-sweep}{4.3} of the main manuscript). The atlas's contribution here, to our knowledge the first systematic characterisation of the instability band itself (\S\ref{sec:s5-prov}; Section~\mainref{sec:results-secondary}{2.6} of the main manuscript), is a property of the engine, not of the calibration values that bracket it; the Mild / Moderate / Severe values are methodological choices anchored to the Berlin PaO$_2$/FiO$_2$ bands, not headline findings.

\subsection{Two divergence regimes}\label{sec:s5-regimes}

Together with Section~\mainref{sec:results-sex}{2.5} of the main manuscript, these data distinguish two operationally separate divergence regimes. The input-driven regime (the main-manuscript finding) concentrates at low HR, low SBP, high RR and a requested FiO$_2$ of 1.0, which is never applied because all 20 divergences occur during patient stabilisation, before the ventilator action is submitted (Section~\mainref{sec:results-overview}{2.1} of the main manuscript); it affects only the male body (20 events in the four male cells, none in the 640 female runs). The severity-driven regime (this sweep) affects both sexes at midrange inputs once the impairment fraction enters [0.879, 0.946], with female divergence concentrated in the lower part of the band and male divergence in the upper part. Both are genuine solver-stability findings. The sex effect of H4 (Section~\mainref{sec:results-sex}{2.5} of the main manuscript) is therefore best described as regime-dependent rather than all-male, and users requiring stable steady-state outputs should avoid impairment fractions of 0.879--0.946.

%%%%%%%%%%%%%%%%%%%%%%%%%%%%%%%%%%%%%%%%%%%%%%%%%%%%%%%%%%%%%%%%%%%%%%
%% S6 -- Sex-stratified reliability: Spearman rho on mu*
%%%%%%%%%%%%%%%%%%%%%%%%%%%%%%%%%%%%%%%%%%%%%%%%%%%%%%%%%%%%%%%%%%%%%%

\section{Sex-stratified comparisons}\label{sec:s6}

\subsection{\texorpdfstring{Cross-sex Spearman $\rho$ on $\mu^{*}$ rankings}{Cross-sex Spearman rho on mu* rankings}}\label{sec:s6-rho}

Section~\mainref{sec:results-sex}{2.5} of the main manuscript reports cross-sex Spearman rank correlations $\rho$ of 0.75--1.00 between Morris $\mu^{*}$ vectors across the 12 (scenario, output) cells where they are computable. The comparison is confined to Mild and Moderate because the two Healthy cells and the Severe-F cell have no clean FiO$_2$ elementary-effect pairs: every Morris step that moves FiO$_2$ between 0.21 and 1.0 has at least one endpoint flagged by SpO$_2$-saturation pinning, transient-hyperoxia drift or implausibility, or, in Severe-F, by an SpO$_2$ still rising at FiO$_2$ = 0.473 (8 of its 20 pairs). Table~\ref{tab:s6-rho} gives the full $\rho$ table with paired-input bootstrap 95~\% confidence intervals from 2{,}000 input resamples.

\begin{table}[htbp]
\centering
\caption{\textbf{Spearman rank correlation $\rho$ between the male and female Morris $\mu^{*}$ input-importance vectors (seven entries per (scenario, output) cell), with paired-input bootstrap 95~\% confidence intervals from 2{,}000 input resamples.} Computed only for Mild and Moderate, because FiO$_2$ has no clean elementary-effect pairs in Healthy and Severe-F (see the Discussion of the main manuscript). Ledger anchors \texttt{T2.sp.S2.*} and \texttt{T2.sp.S3.*}.}
\label{tab:s6-rho}
\begin{tabular}{llcc}
\toprule
Scenario & Output & $\rho$ & 95\% CI (paired-input bootstrap, $n = 2{,}000$) \\
\midrule
Mild & HR    & 0.96 & $[0.70, 1.00]$ \\
Mild & MAP   & 1.00 & $[1.00, 1.00]$ \\
Mild & Vt    & 0.75 & $[-0.06, 1.00]$ \\
Mild & SpO$_2$  & 0.96 & $[0.70, 1.00]$ \\
Mild & PaO$_2$  & 0.93 & $[0.41, 1.00]$ \\
Mild & PaCO$_2$ & 0.89 & $[0.35, 1.00]$ \\
\midrule
Moderate & HR    & 0.96 & $[0.70, 1.00]$ \\
Moderate & MAP   & 0.96 & $[0.70, 1.00]$ \\
Moderate & Vt    & 0.86 & $[0.34, 1.00]$ \\
Moderate & SpO$_2$  & 0.96 & $[0.70, 1.00]$ \\
Moderate & PaO$_2$  & 0.93 & $[0.41, 1.00]$ \\
Moderate & PaCO$_2$ & 0.89 & $[0.41, 1.00]$ \\
\bottomrule
\end{tabular}
\end{table}

\begin{figure}[htbp]
\centering
\includegraphics[width=0.85\textwidth]{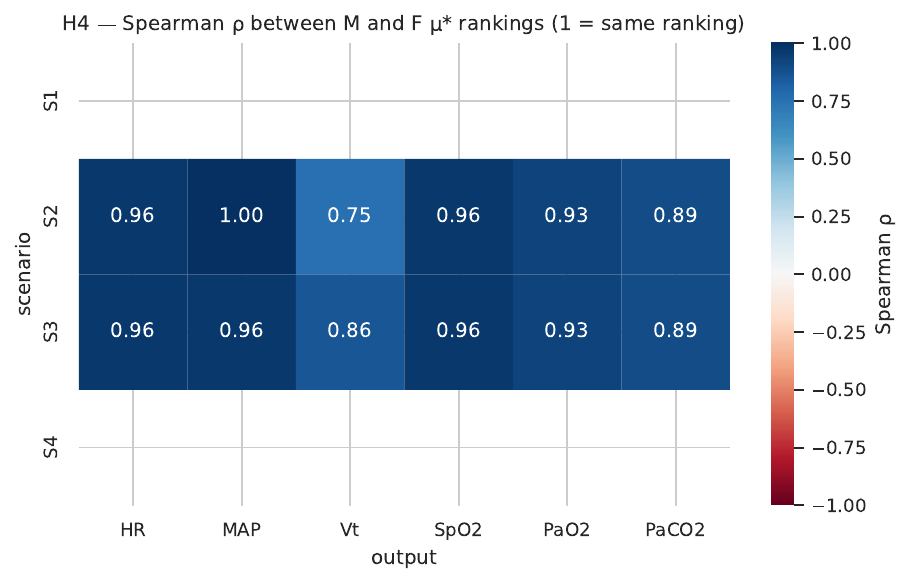}
\caption{\textbf{Spearman rank correlation $\rho$ between male and female Morris $\mu^{*}$ vectors per (scenario, output), with paired-input bootstrap 95~\% confidence intervals ($n = 2{,}000$ input resamples).} Reproduced from the companion technical monograph (its Figure~5). Rows carry the data codes S1--S4 of the Healthy, Mild, Moderate and Severe scenarios; Table~1 of the main manuscript gives the correspondence. The S1 and S4 rows are blank because FiO$_2$ has no clean elementary-effect pairs in Healthy-M, Healthy-F and Severe-F.}
\label{fig:s6-rho}
\end{figure}

Point estimates fall between 0.75 and 1.00 across the 12 (scenario, output) cells, qualitatively supporting the preservation of input-importance rankings across sex in the Mild and Moderate scenarios. Delivered tidal volume is the exception: its Mild lower confidence bound of $-0.06$ means that cross-sex agreement on what drives it cannot be distinguished from chance under this bootstrap design. Figure~\ref{fig:s6-rho} shows the $\rho$ point estimates and confidence intervals.

\subsection{Relative ventilation-dose mismatch between the two reference bodies}\label{sec:s6-dose}

The sweep varies tidal volume in absolute millilitres over the same 300--800~mL range on both reference bodies. A matched input vector therefore delivers a matched \emph{absolute} tidal volume, but not a matched dose relative to body size, as two normalisations show.

Per kilogram of \emph{actual} body weight (67.3 and 56.2~kg), the swept range corresponds to 4.5--11.9~mL\,kg$^{-1}$ on the male body and 5.3--14.2~mL\,kg$^{-1}$ on the female body, so at every matched setting the female body receives 19.8~\% more tidal volume per kilogram.

Per kilogram of \emph{ideal} body weight, the gap is larger. Pulse computes ideal body weight at patient setup with the sex-specific Devine relation ($50.0 + 2.3\,(\text{height in inches} - 60)$~kg for a male body and $45.5 + 2.3\,(\text{height in inches} - 60)$~kg for a female body), giving 74.99~kg at 180~cm and 55.10~kg at 163~cm. The swept range then corresponds to 4.0--10.7~mL\,kg$^{-1}$ of ideal body weight on the male body and 5.4--14.5~mL\,kg$^{-1}$ on the female body, a 36.1~\% relative excess for the female body.

This is an interpretive constraint, not a defect in the sweep: sweeping tidal volume in absolute units keeps the Morris design identical across all eight cells, the property on which the balanced factorial rests. However, any male--female contrast on a gas-exchange output then combines three contributions that the design does not separate: the sex flag, body morphology and ventilatory dose. This matters most for PaCO$_2$, whose dominant determinant is alveolar ventilation and which accounts for three of the five surviving Wasserstein distances reported in Section~\mainref{sec:results-sex}{2.5} of the main manuscript. Separating the three would require both bodies to retain the same grid points \emph{and} receive matched relative ventilation doses; matching grid points alone does not suffice.

%%%%%%%%%%%%%%%%%%%%%%%%%%%%%%%%%%%%%%%%%%%%%%%%%%%%%%%%%%%%%%%%%%%%%%
%% S7 -- Cornish Table 2 expanded
%%%%%%%%%%%%%%%%%%%%%%%%%%%%%%%%%%%%%%%%%%%%%%%%%%%%%%%%%%%%%%%%%%%%%%

\section{Cornish et al.\ 2026 per-channel cross-mapping}\label{sec:s7}

The Discussion of the main manuscript summarises the complementarity between the atlas and the causal-falsification study of Cornish et al.~\cite{cornish2023}: 80 of their 85 Hoeffding rejections fall on channels the atlas does not assess. Table~\ref{tab:s7-cornish} gives the full 15-channel per-quantity cross-mapping, grouped by physiological system. ``Hoeffding'' counts come from Cornish's Appendix Table~4, which lists all 15 assessed quantities with family-wise-error-rate-controlled rejections under Holm--Bonferroni adjustment; ``Bootstrap'' counts come from their Appendix Table~5 (reverse-percentile bootstrap rejections at the same significance level). The rightmost column maps each channel to an atlas output, and bold rows mark the four directly overlapping channels (HR, MeanBP $\approx$ atlas MAP, PaO$_2$, PaCO$_2$).

\begin{table}[htbp]
\centering
\caption{\textbf{Cornish et al.\ 2026 per-quantity rejection counts (Cornish Appendix Tables~4 and~5) mapped onto the atlas's six-output panel and grouped by physiological system.} Bold rows are the four directly overlapping channels. Of Cornish's 85 Hoeffding rejections, 80 fall on the eight channels the atlas does not assess (electrolyte, metabolic, acid--base, dermal), 2 on the partially overlapping pressure channel SysBP, and only 3 on the four overlapping channels (all on PaCO$_2$). The two studies' failure surfaces are therefore complementary rather than redundant.}
\label{tab:s7-cornish}
\begin{tabular}{lrrl}
\toprule
Cornish 2026 channel & Hoeffding & Bootstrap & Maps to atlas output? \\
\midrule
\multicolumn{4}{l}{\textit{Electrolytes}} \\
Chloride            & 24 / 94 & 47 / 94 & No (electrolyte) \\
Sodium              & 21 / 94 & 46 / 94 & No (electrolyte) \\
Potassium           & 13 / 94 & 33 / 94 & No (electrolyte) \\
Calcium             & 5 / 88  & 44 / 88 & No (electrolyte) \\
\midrule
\multicolumn{4}{l}{\textit{Metabolic and acid--base}} \\
Glucose             & 5 / 96  & 19 / 96 & No (metabolic) \\
HCO$_3$ (Bicarbonate) & 2 / 90 & 8 / 90  & No (acid--base) \\
Arterial pH         & 0 / 80  & 0 / 80  & No (acid--base) \\
\midrule
\multicolumn{4}{l}{\textit{Dermal}} \\
Skin Temperature    & 10 / 86 & 43 / 86 & No (dermal) \\
\midrule
\multicolumn{4}{l}{\textit{Cardiovascular}} \\
\textbf{HR}         & \textbf{0 / 162} & \textbf{1 / 162}  & \textbf{Yes (direct)} \\
\textbf{MeanBP}     & \textbf{0 / 92}  & \textbf{3 / 92}   & \textbf{Yes ($\approx$ atlas MAP)} \\
SysBP               & 2 / 154 & 8 / 154 & Partial (MAP input) \\
DiaBP               & 0 / 72  & 0 / 72  & Partial (MAP input) \\
\midrule
\multicolumn{4}{l}{\textit{Respiratory and gas exchange}} \\
\textbf{PaCO$_2$}   & \textbf{3 / 70}  & \textbf{13 / 70}  & \textbf{Yes (direct)} \\
\textbf{PaO$_2$}    & \textbf{0 / 78}  & \textbf{4 / 78}   & \textbf{Yes (direct)} \\
RR                  & 0 / 172 & 12 / 172 & Atlas input, not output \\
\midrule
\textbf{Total} & \textbf{85} & \textbf{281} & \\
\bottomrule
\end{tabular}
\end{table}

Cornish's rejections cannot be cross-tabulated against the atlas's per-cell traffic light, because they are indexed by (time step, action sequence, sex--age--quantile bin) over sepsis trajectories rather than by (scenario, sex) over the operational envelope, and neither indexing maps onto the other. No ordinal comparison of the two studies' channel rankings is therefore drawn; the supported conclusion is the complementarity shown in Table~\ref{tab:s7-cornish}.

%%%%%%%%%%%%%%%%%%%%%%%%%%%%%%%%%%%%%%%%%%%%%%%%%%%%%%%%%%%%%%%%%%%%%%
%% S8 -- VC-AC trigger lockout
%%%%%%%%%%%%%%%%%%%%%%%%%%%%%%%%%%%%%%%%%%%%%%%%%%%%%%%%%%%%%%%%%%%%%%

\section{VC-AC trigger lockout}\label{sec:s8}

Section~\mainref{sec:results-vcac}{2.3} of the main manuscript reports that setting Pulse's \texttt{InspirationPatientTriggerPressure} to $-2$~cmH$_2$O on the \texttt{SEMechanicalVentilatorVolumeControl} action reduces the median delivered-tidal-volume coefficient of variation from 0.351 to 0.016 across 64 paired reruns, with 78.1~\% of locked-trigger reruns falling below the 0.05 steady-state threshold. This section describes the protocol and shows per-breath waveform evidence that the tidal-volume oscillation is real (not a sampling artefact) and that the intervention strongly reduces it.

\subsection{Protocol design}\label{sec:s8-protocol}

The 64 paired reruns sampled seven of the eight (scenario, sex) cells (none from Moderate-M), mostly the Healthy (26 reruns) and Mild (19) cells. Each rerun reproduced the original configuration exactly, with a single modification: the \texttt{InspirationPatientTriggerPressure} field of the VC-AC action was set to $-2$~cmH$_2$O (the constant declared in \texttt{scripts/run\_s014\_trigger\_lockout.py} and regex-extracted into ledger anchor \texttt{T1.06d.s14\_trigger\_pressure}). All other settings (Morris input vector, simulation duration, scenario, reference body) were unchanged, so the intervention is a clean one-parameter manipulation and any change in the steady-state Vt distribution is attributable to the trigger pressure. Two configurations share the highest pre-lock delivered-Vt coefficient of variation, 1.826 (S1\_F\_0117 and S1\_F\_0118, run indices 117 and 118 in the Healthy-F cell); the lock reduces them to 0.018 and 0.024.

\begin{figure}[htbp]
\centering
\includegraphics[width=0.95\textwidth]{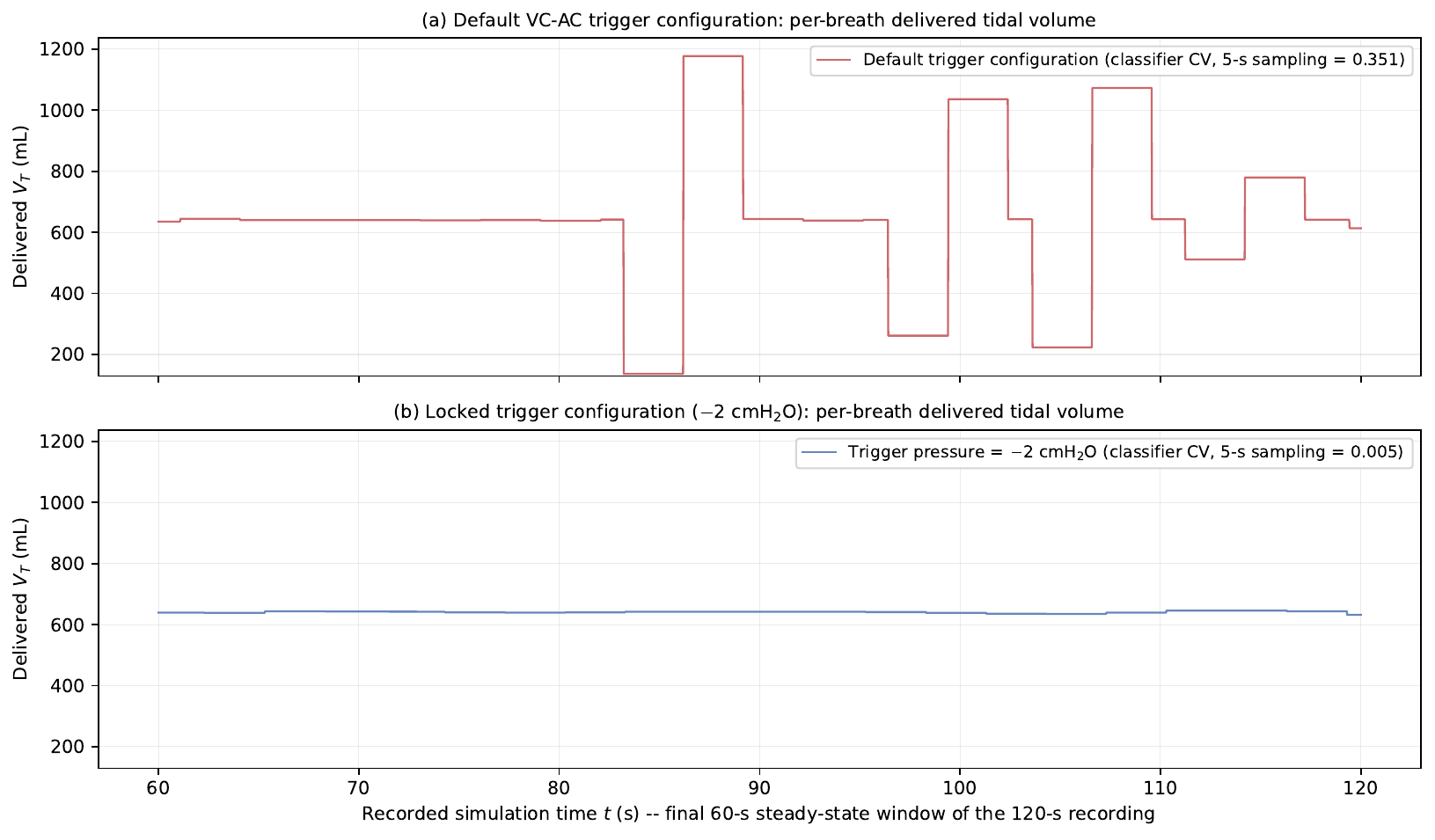}
\caption{\textbf{Per-breath delivered tidal volume at Pulse's native 50~Hz sampling rate, for a deterministically selected S014 paired configuration (S4\_F\_0003).} This run's pre-lock classifier Vt CV is closest to the median of the 64 paired reruns (ties broken on case identifier), so it could not be chosen after inspecting the waveforms; such flags are more frequent in Healthy than in Severe, and no claim about that distribution is made here. Both panels plot the 50~Hz waveform over the classifier's [60, 120]~s window in recording-relative time; each quoted CV is the classifier statistic from the 5-second samples of that window, not of the trace. \textbf{(a)} Default VC-AC trigger: delivered volume holds near 640~mL but is interrupted by cycles of collapsed and then oversized volume (classifier CV = 0.351), an oscillation consistent with patient--ventilator dyssynchrony and flagged \texttt{not\_steady\_state}. \textbf{(b)} Locked trigger ($-2$~cmH$_2$O): Vt is metronomic across the same window (classifier CV = 0.005), so in this configuration the lock removes the oscillation without disrupting volume delivery. The two worst-affected configurations (\S\ref{sec:s8-protocol}; the two rightmost points of Figure~4a of the main manuscript) are not shown. Pulse exposes airway pressure ($P_{\text{aw}}$) and inspiratory and expiratory flow ($\dot{Q}$) as queryable properties, but the atlas data request did not include them, so per-breath delivered volume is the directly observable proxy of the trigger-coupling dynamics; 50~Hz sampling is Nyquist-safe for the $\sim$0.25~Hz breath-rate signal.}
\label{fig:s8-vt}
\end{figure}

\subsection{Native 50 Hz waveform evidence (Figure~\ref{fig:s8-vt})}\label{sec:s8-waveform}

In the atlas-aggregated parquet, delivered tidal volume is a breath-integrated time series sampled every 5 seconds. For waveform-level inspection, the native Pulse output of each run (\texttt{output\_pulse\_native.csv}) provides the same signal every 0.02~s (50~Hz). At adult respiratory rates of about 0.25~Hz (one breath every four seconds), this is a 200-fold oversampling of the breath rate, well above the Nyquist criterion of about $2 \times 0.25 = 0.5$~Hz, so the per-breath delivered volume in Figure~\ref{fig:s8-vt} is the directly observable breath-by-breath dynamic of the ventilator coupling, not an aliasing artefact.

\subsection{Cycle-resolved check of the 5-second classifier statistic}\label{sec:s8-cycle}

The delivered-Vt coefficient of variation used throughout this work is the locked classifier statistic, computed from 5-second samples over the final 60-second window; it is not a direct measure of the physiological amplitude of patient--ventilator dyssynchrony. At a breath rate near 0.25~Hz, sampling at 0.2~Hz cannot reliably resolve breath-to-breath dynamics, so a post hoc cycle-resolved analysis was performed on the native 50~Hz traces already recorded for every run; no additional simulation was required.

Across the full sweep, the two resolutions reach the same steady-state verdict on 1{,}239 of 1{,}260 runs (98.3~\%): 297 runs are flagged by both, 942 by neither, 15 cross the 0.05 threshold only at cycle resolution, and 6 only under the 5-second statistic. Of the 15 cycle-only crossings, ten fall below a factor of 3; five are substantial (factors of 4.0 to 113.9), and all five fall on the female reference body. Nine runs deliver a numerically constant volume across the window and carry zero cycle-level variability at the resolution of the recorded trace.

The effect on the locked flag count is small. Six of the 15 cycle-only crossings are already flagged by another channel or criterion, and three of the six 5-second-only crossings would lose their flag entirely. A counterfactual reclassification using cycle-resolved Vt variability would therefore change the \texttt{not\_steady\_state} total from 530 to 536, an increase of 1.1~\%. The locked classification and every count reported elsewhere in this work are unchanged.

On the 64 paired trigger-lockout configurations of \S\,\ref{sec:s8-protocol}, delivered-Vt variability decreases after lockout in every pair, at both resolutions. Four descriptive summaries were computed on the 63 configurations with a non-zero post-lockout cycle-level CV (Table~\ref{tab:s8-estimands}). The qualitative effect is consistent, but quantitative agreement between resolutions depends on the estimand: the relative gap ranges from 0.6~\% to 8.6~\%, and its direction varies (two summaries make the 5-second statistic overstate the effect and two make it understate it). Ratios are particularly sensitive to the small post-lockout CV values. Configuration S1\_F\_0016 is excluded from all four summaries: after lockout its delivered volume is identical at every breath to the precision of the recorded trace (819.027985~mL), so its post-lockout CV is zero and no finite reduction ratio is defined, and the two difference summaries use the same 63 configurations so that the four rows are comparable. This zero is observed, not imputed, and the run remains in the 64-pair medians and counts given at the opening of this section.

\begin{table}[htbp]
\centering
\caption{\textbf{Four descriptive summaries of the trigger-lockout effect, at both sampling resolutions.} Computed on the same 63 paired configurations (of 64) that carry a non-zero post-lockout cycle-level CV. Ratios are dimensionless; differences are in CV units. Within each pair of summaries (the two ratio-based and the two difference-based rows), the summaries disagree on which resolution overstates the effect, which is why all four are reported.}
\label{tab:s8-estimands}
\begin{tabular}{lrr}
\toprule
Summary statistic & 5-second & Cycle-resolved \\
\midrule
Ratio of group medians & 21.42 & 22.66 \\
Median of paired ratios & 27.84 & 25.63 \\
Difference of group medians & 0.3345 & 0.3224 \\
Median of paired differences & 0.3256 & 0.3275 \\
\bottomrule
\end{tabular}
\end{table}

One limitation of the cycle-resolved statistic was measured rather than assumed. Recovering cycles as change points of the recorded volume merges two consecutive breaths that deliver numerically identical volumes, so a repeated volume is counted once rather than twice. Across the 1{,}251 runs whose delivered volume is not numerically constant, the recovered cycle count and the engine's reported respiration rate agree to a median of 5.4~\%, but the 95th percentile of the disagreement is 38~\% and the maximum 95~\%. Of the 126 runs that disagree by more than 20~\%, four are cycle-only crossings and none is a 5-second-only crossing.

A sensitivity analysis bounds the consequence, and its direction was measured rather than reasoned about. Scoring the same window on every native 50~Hz sample weights each recorded volume by its duration, which at fixed breath length is per-breath weighting and is therefore unaffected by merging. The two estimators are nearly identical in magnitude, run for run: on the 1{,}250 runs whose two values differ by more than a relative $10^{-6}$ (the remaining ten, nine of them constant-volume runs, are identical to the precision of the trace), the change-point value is the larger on 640 and the smaller on 610, with a median ratio of 1.0005. Merging therefore does not systematically inflate or deflate the statistic; it only affects which runs fall on either side of the threshold.

Against the 5-second statistic, the all-sample estimator agrees on 295 flagged and 945 unflagged runs, crosses the threshold alone on 12 and is crossed alone by the 5-second statistic on 8 (98.4~\% agreement, versus 98.3~\% above), and it disagrees with the change-point estimator on 9 of the 1{,}260 runs. Its counterfactual reclassification would change the \texttt{not\_steady\_state} total from 530 to 533, versus 530 to 536 above. The 15 / 6 split is therefore not an artefact of the segmentation rule, and the reported counterfactual is the larger of the two, so it overstates rather than understates the cost of changing resolution.

These results support the intervention effect on simulated breath-to-breath volume variability, while showing that the classifier CV should not be read as a validated physiological measure of dyssynchrony, nor as a universally reliable measure of variability magnitude at the individual-run level.

\subsection{Mechanism and downstream consequence}\label{sec:s8-mech}

The default \texttt{SEMechanicalVentilatorVolumeControl} action in \texttt{AssistedControl} mode enables Pulse's respiratory-model-driven patient trigger when no explicit \texttt{InspirationPatientTriggerPressure} or \texttt{InspirationPatientTriggerFlow} is supplied (\texttt{SEMechanicalVentilatorVolumeControl.cpp}, lines 120--128 of the upstream Pulse source at \texttt{REL\_4\_3\_1}). This spontaneous-drive trigger couples breath delivery to the modelled respiratory drive and produces effort-dependent breath-to-breath tidal-volume variance consistent with patient--ventilator dyssynchrony~\cite{figueroacasas2016}, but it breaks the steady-state-mean assumption on which the locked classifier relies. Setting the trigger pressure explicitly to $-2$~cmH$_2$O replaces it with a fixed pressure threshold, which sufficiently strong efforts can still cross, and reduces the Vt CV more than twentyfold in the median paired comparison.

This setting is a data-generation and exploration lever, not a clinical recommendation. In a real patient, the threshold is a clinical choice: with any fixed threshold, inspiratory efforts too weak to reach it fail to trigger a breath (ineffective triggering), a form of patient--ventilator dyssynchrony~\cite{thille2006} that is clinically undesirable~\cite{blanch2015}; such dyssynchrony is detectable on ventilator flow and pressure waveforms but most directly confirmed with specialised monitoring such as esophageal pressure~\cite{akoumianaki2014} or diaphragm electrical activity~\cite{sinderby1999}. Users running atlas-style VC-AC sensitivity sweeps that need lower tidal-volume variability can set the trigger pressure explicitly on every action, with this caveat in mind. Dyssynchrony itself is well documented in ICU patients receiving lower-tidal-volume volume-control ventilation~\cite{figueroacasas2016}; the atlas contributes a Pulse tidal-volume oscillation consistent with it and the $-2$~cmH$_2$O trigger lever for synthetic-data generation, not the clinical discovery of dyssynchrony.

%%%%%%%%%%%%%%%%%%%%%%%%%%%%%%%%%%%%%%%%%%%%%%%%%%%%%%%%%%%%%%%%%%%%%%
%% S9 -- Hyperoxia 300-second convergence
%%%%%%%%%%%%%%%%%%%%%%%%%%%%%%%%%%%%%%%%%%%%%%%%%%%%%%%%%%%%%%%%%%%%%%

\section{Hyperoxia 300-second convergence}\label{sec:s9}

Section~\mainref{sec:results-hyperoxia}{2.4} of the main manuscript reports that the locked 120-second window is too short for PaO$_2$ convergence at FiO$_2$ = 1.0 in the Healthy cohort: in 300-second reruns, 40 of 44 (90.9~\%) converged to a stable plateau, with a median plateau PaO$_2$ of 643~mmHg over the 40 drift-converged runs, about 20~mmHg below the simple alveolar-gas-equation prediction at sea level ($\approx$~663~mmHg)~\cite{west2020respiratory}. This section provides the kinetic anchor for that claim and shows the convergence trajectories.

\begin{figure}[htbp]
\centering
\includegraphics[width=0.95\textwidth]{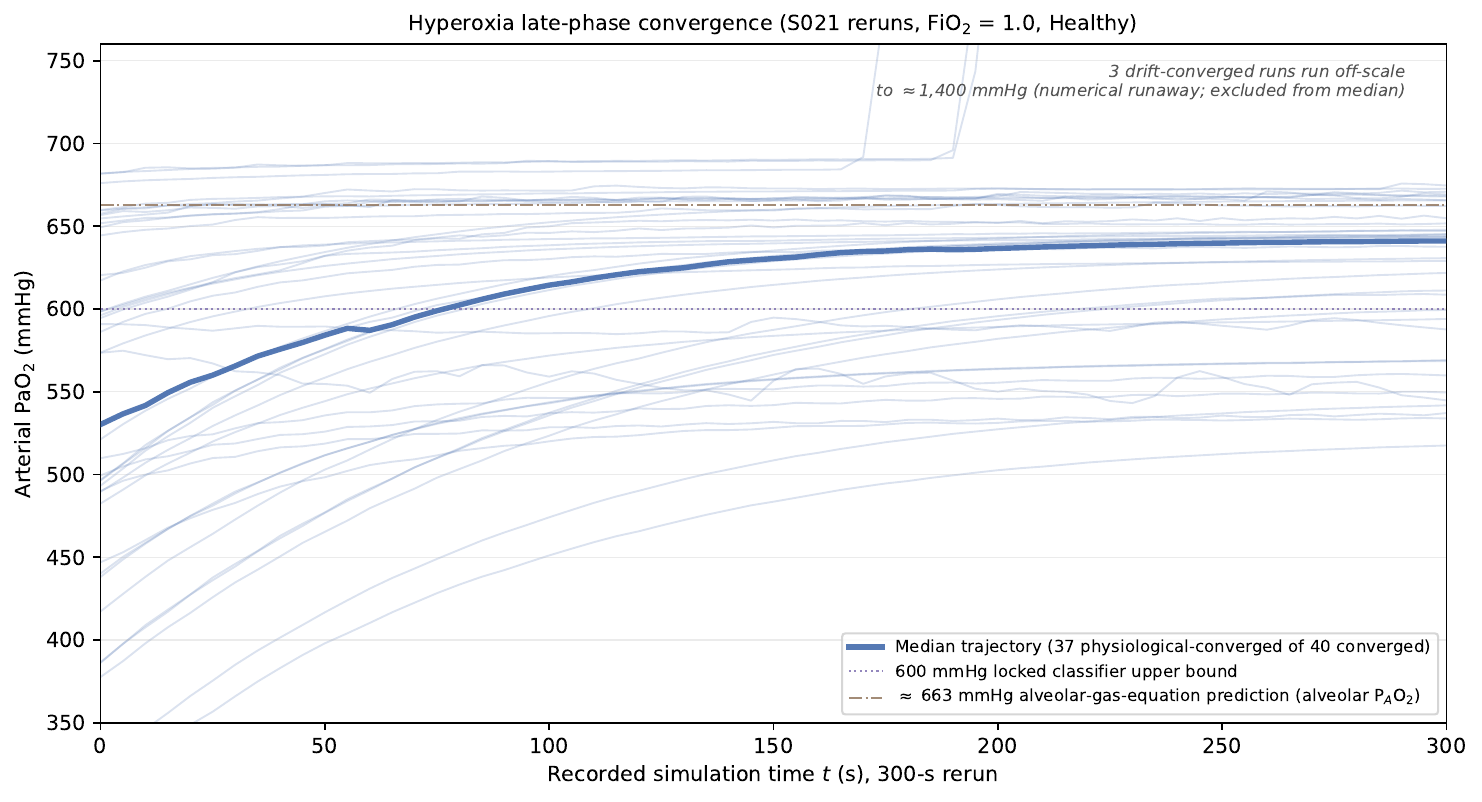}
\caption{\textbf{Late-phase PaO$_2$ convergence in the S021 300-second reruns at FiO$_2$ = 1.0 in the Healthy cohort (40 of 44 converged under the preregistered drift criterion).} The x-axis is the rerun's recorded simulation time $t$, from 0 to 300~s; the rerun is a re-execution from time zero, not a continuation of the original run. PaO$_2$ is already hyperoxic at $t = 0$ because FiO$_2$ = 1.0 applies during the preceding 120-second settling phase, so $t = 0$ is not a room-air baseline. Thin blue traces are the 40 drift-converged trajectories; three of them relax to unphysical $>700$~mmHg plateaus and leave the panel (vertical axis limited to $[350, 760]$~mmHg; these traces continue to $\approx$1{,}400~mmHg, see Section~\ref{sec:s9-guidance}), so the thick blue line is the median of the remaining 37 \emph{physiological}-converged runs. The dotted purple line at 600~mmHg is the locked classifier bound; the dash-dotted brown line at $\approx$~663~mmHg is the alveolar-gas-equation sea-level prediction (alveolar PAO$_2$). The median trajectory rises from about 530~mmHg at $t = 0$ to about 641~mmHg at $t = 300$~s, with 95~\% of this late gain reached at $t = 180$~s (audit anchor \texttt{T1.06u.s021\_time\_to\_95pct\_plateau}), which explains why the 120-second window underestimates plateau PaO$_2$ at FiO$_2$ = 1.0.}
\label{fig:s9-pao2}
\end{figure}

\subsection{Late-phase convergence trajectories (Figure~\ref{fig:s9-pao2})}\label{sec:s9-trajectories}

The 60 reruns comprise the 22 Morris points requested at FiO$_2$ = 1.0 that diverged in none of the eight cells, run in both Healthy cells (44 reruns), and the first four of these points in sample order (run indices 6, 7, 12 and 13), run in the two Moderate and two Severe cells (16 reruns). Each S021 rerun re-executes the same configuration from time zero with the recorded duration raised from 120 to 300 seconds, sampled every 5 seconds into the atlas-aggregated CSV. The 120-second ventilator-settling phase that precedes recording is unchanged, so one rerun spans 420~s of Pulse-internal time; all reported times use recorded simulation time ($t = 0$--$300$~s). Because Pulse is deterministic, the aggregated CSVs of all 60 reruns are byte-identical to those of the corresponding 120-second runs over the shared interval $t = 0$--$120$~s; the reruns are deterministic re-executions, not saved-state continuations. PaO$_2$ is already hyperoxic at $t = 0$ in this cohort, because FiO$_2$ = 1.0 applies throughout the preceding settling phase, so the figure shows the \emph{late-phase convergence} of PaO$_2$ toward its plateau, not a wash-in from room air. Each thin trace is one of the 40 drift-converged Healthy reruns; the thick line is the median of the 37 with a physiological plateau (the three runaways are excluded; \S\ref{sec:s9-guidance}).

\subsection{Practical guidance}\label{sec:s9-guidance}

The kinetic anchor (95~\% of the late-plateau gain reached by $t = 180$~s on the 37 physiological-converged runs) gives an objective threshold for hyperoxia simulations: recording at least 300 seconds and averaging over the final 60 seconds ($t = 240$--$300$~s) places the PaO$_2$ estimate within about 0.1~\% of the value at $t = 300$~s on the median trajectory, and within 1.6~\% for any individual run. The 120-second window of the locked classifier suits HR, MAP and PaCO$_2$, whose drift-threshold exceedances are uncommon at FiO$_2$ $\geq$ 0.95 (\S\ref{sec:s16-window}), but is too short for PaO$_2$ at the inspired-oxygen boundary, where the classifier flags runs as \texttt{implausible} (PaO$_2$ above 600~mmHg) under exactly the conditions where the slow late-phase relaxation has not yet completed.

A separate solver-instability mode appears in extended runs on healthy lungs at FiO$_2$ = 1.0; two runs at FiO$_2$ = 0.737 show the same runaway within the 120-s window (\S\ref{sec:s3-pao2-sensitivity}). In the S021 cohort, 6 of 44 Healthy reruns (14~\%) developed unphysical PaO$_2$ plateaus above 700~mmHg (four near 1{,}400~mmHg, above the 760~mmHg sea-level barometric pressure), all at the single tidal-volume level Vt~=~633.3~mL (6 of the 20 Healthy reruns at that level; none of the 24 at the other three levels, including 0 of 8 at Vt~=~800~mL). Minute ventilation does not separate them: five of the six have RR~=~20 ($\dot{V}_E \approx 12.7$~L/min, about twice the physiological norm of 6--8~L/min) but three other runs at that $\dot{V}_E$ do not run away, and the sixth has RR~=~16 at PEEP~=~0 ($\dot{V}_E \approx 10.1$~L/min), so no input-side ventilation threshold separates affected from unaffected runs. Three of the six stay below the drift threshold and are counted among the 40 drift-converged runs, and three are non-converged; the 44 Healthy reruns thus partition into 37 physiological-converged, 3 runaway-converged, 3 runaway-non-converged and one non-runaway non-converged run, and all six runaways are excluded from the median trajectory of Figure~\ref{fig:s9-pao2}. The locked 120-second window ends before the runaway emerges in these six runs; users running extended-window hyperoxia simulations should therefore screen each PaO$_2$ trajectory individually rather than rely on an input-side ventilation threshold.

What does separate the runaways is the reference body. Across the five input-identical male/female pairs at Vt~=~633.3~mL with RR~$\geq$~16, the female body ran away in 5 of 5 and the male in 1 of 5, and the single male case plateaus at 770~mmHg against 1{,}126--1{,}408~mmHg for the five female cases. This is the opposite direction to the all-male input-extreme divergence cluster of Section~\mainref{sec:results-sex}{2.5} of the main manuscript and a further instance of the regime-dependence established in \S\ref{sec:s5-cont}. It is consistent with the systematically higher PaO$_2$ the female body reaches at identical inputs (mean 686 versus 663~mmHg in the locked 120-second window across the four RR~=~20 pairs) and with the sex effect on PaO$_2$ in the Healthy scenario measured by the controlled 2$\times$2 sex$\times$mass design of S011, which held body mass and height fixed across sexes.

The two arms of that design should be weighted differently: $+13.7$~mmHg in the 56.2~kg comparison, where both phenotypes lie within Pulse's stated height and BMI ranges, versus $+21.3$~mmHg in the 67.3~kg comparison, whose synthetic 180~cm female exceeds Pulse's nominal female height ceiling of 175.5~cm and triggers the engine's own model-validity warning; the smaller, fully in-envelope figure is the one to weigh. In the Severe scenario, however, the same in-envelope 56.2~kg comparison reverses: female-minus-male PaO$_2$ = $-1.97$~mmHg ($-3.8$~\%), versus $+3.97$~\% in Healthy, and the 67.3~kg female run diverged outright in Severe, leaving that arm unmeasured. S011 therefore demonstrates a scenario-dependent sensitivity to the sex-conditioned setup at the tested operating point, not a scenario-invariant direction or magnitude of effect. Even the in-envelope arm has three limits: the two bodies are not percentile-matched (163~cm is both Pulse's female 50th-percentile standard and its male 3rd-percentile floor), it was run at a single midrange operating point at FiO$_2$ = 0.6, and Pulse's sex flag switches parameters across several subsystems at once (\S\ref{sec:s10}). The S021 hyperoxia cohort, which contrasts the two standard reference bodies, therefore cannot separate sex from the 11.1~kg body-mass difference in this regime, and no mechanism is claimed.

\subsection{Final-60-s oxygenation across ARDS severity, and the clean-regime FiO$_2$ dose response}\label{sec:s9-shunt}

The 300-second rerun experiment also replayed four paired Morris input vectors on both reference bodies in the Moderate and Severe scenarios at FiO$_2$ = 1.0, alongside the 44-run Healthy hyperoxia set. The per-run mean PaO$_2$ over the final 60~s ($t = 240$--$300$~s) falls monotonically with severity (Table~\ref{tab:s9-shunt}), nearly ninefold from the Healthy median to 73~mmHg in Severe (144~mmHg in Moderate). This pattern is consistent with worsening gas exchange as the alveolar diffusion surface collapses and intrapulmonary-shunt behaviour changes (\S\ref{sec:s5-mapping}), increasingly separating arterial PaO$_2$ from the calculated alveolar oxygen tension at unchanged FiO$_2$ = 1.0. It supports interpreting the elevated Healthy final-60-s oxygenation as a property of the healthy-lung configuration in these simulations rather than a generic FiO$_2$ = 1.0 artefact.

\begin{table}[htbp]
\centering
\caption{\textbf{Final-60-s PaO$_2$ by ARDS scenario in the S021 300-second reruns at FiO$_2$ = 1.0.} Healthy is the full 44-run hyperoxia set; Moderate and Severe are the four paired Morris input vectors replayed on both reference bodies (8 runs each). Values are per-run means over the final 60~s of the 300~s rerun ($t = 240$--$300$~s), summarised as median and interquartile range (IQR) per scenario. They include every run, converged or not, and are therefore not called plateaus (four of the 44 Healthy reruns do not meet the preregistered drift criterion; \S\ref{sec:s9-trajectories}).}
\label{tab:s9-shunt}
\begin{tabular}{lrrr}
\toprule
Scenario & $n$ & Final-60-s median (mmHg) & Final-60-s IQR (mmHg) \\
\midrule
Healthy        & 44 & 643.2 & [592, 668] \\
Moderate  &  8 & 143.6 & [93, 218]  \\
Severe    &  8 &  72.6 & [71, 76]   \\
\bottomrule
\end{tabular}
\end{table}

Within the \emph{clean} operating regime, the oxygenation channel is well behaved and strongly FiO$_2$-driven. Across the 91 flag-clean Healthy runs, FiO$_2$ alone explains 90~\% of the variance in steady-state PaO$_2$ (ordinary least squares, $R^2 = 0.90$), with mean PaO$_2$ rising from $81 \pm 11$~mmHg at FiO$_2$ = 0.21 to $214 \pm 30$~mmHg at FiO$_2$ = 0.47, i.e.\ about 5~mmHg per 0.01 FiO$_2$ (fitted line $\mathrm{PaO}_2 \approx 505\,\mathrm{FiO}_2 - 25$~mmHg). Because clean Healthy runs survive only at the two lower Morris FiO$_2$ levels (the two higher levels are flagged \texttt{implausible}, \texttt{not\_steady\_state} or \texttt{diverged}), this establishes the dose slope and the FiO$_2$ dominance seen in the Morris screen, not the curvature of the relation: linear extrapolation would predict $\approx$480~mmHg at FiO$_2$ = 1.0, well below the healthy-lung plateau actually reached (Table~\ref{tab:s9-shunt}); the relation is therefore not linear over the full FiO$_2$ range in these simulations, and the cause of this curvature is not analysed here.

%%%%%%%%%%%%%%%%%%%%%%%%%%%%%%%%%%%%%%%%%%%%%%%%%%%%%%%%%%%%%%%%%%%%%%
%% S10 -- Sex-stratified BV mechanism: absolute + mL/kg
%%%%%%%%%%%%%%%%%%%%%%%%%%%%%%%%%%%%%%%%%%%%%%%%%%%%%%%%%%%%%%%%%%%%%%

\section{Blood-volume dependence and sex-conditioned setup parameters}\label{sec:s10}

Section~\mainref{sec:results-sex}{2.5} of the main manuscript reports that all 20 input-extreme divergences (of 1{,}280 runs) fall on the male reference body, that Pulse's allometric blood-volume formula gives 4{,}802.63~mL for the 67.3~kg male body versus 3{,}996.09~mL for the 56.2~kg female body (a 16.8~\% absolute gap), and that lowering the male \texttt{BloodVolumeBaseline} toward its 4{,}082.24~mL setup floor eliminates the divergence cluster (20 / 20 at the default $\rightarrow$ 0 / 20 at both 4{,}400 and 4{,}082.24~mL). This section separates the absolute-volume framing from the weight-normalised framing, because the two answer different methodological questions and a relevant confounder lies between them.

\subsection{Absolute volumes (the framing used in the main manuscript)}\label{sec:s10-abs}

The Pulse allometric formula~\cite{pulsepatientmethod} is $\text{BV} = 65.6 \cdot \text{weight}_{\text{kg}}^{1.02}$~mL. On the two reference patient bodies this yields:

\begin{itemize}
\item Male (67.3~kg): $\text{BV} = 65.6 \cdot 67.3^{1.02} = 4{,}802.63$~mL.
\item Female (56.2~kg): $\text{BV} = 65.6 \cdot 56.2^{1.02} = 3{,}996.09$~mL.
\item Absolute-mL difference: $4{,}802.63 - 3{,}996.09 = 806.54$~mL, or 16.8~\% of the male default.
\end{itemize}

Pulse's setup floor for the male body is 4{,}082.24~mL (the allometric blood volume $\times$ 0.85 in \texttt{SetupPatient.cpp}, a 15~\% reduction below which Pulse requires hypovolemia to be modelled by an explicit condition), 720.39~mL below the male default and 86.15~mL above the female default. A one-way \texttt{BloodVolumeBaseline} transplant therefore cannot make the male body blood-volume-equivalent to the female body.

The same gate has an upper arm: \texttt{SetupPatient.cpp} rejects any baseline above the allometric value $\times$ 1.15, with an \texttt{[ERROR]} and exit, exactly as it rejects one below the floor. For the 56.2~kg female reference body, this ceiling is 4{,}595.50~mL, 207.13~mL below the 4{,}802.63~mL male default at which the divergence cluster appears, so the reciprocal transplant is refused too and neither body can be moved to the other's default. This is a property of that body's mass-dependent range, not of female-configured bodies in general: a female-configured body of 67.3~kg computes the same 4{,}802.63~mL default and admissible range, because Pulse's blood-volume relation is not sex-branched.

\subsection{Weight-normalised volumes (the confounder-controlled framing)}\label{sec:s10-norm}

Comparing absolute blood volumes across bodies of different mass conflates sex differences with morphology. Dividing each absolute volume by body weight gives a weight-normalised blood-volume index (mL/kg) that controls for morphology:

\begin{itemize}
\item Male: $4{,}802.63 / 67.3 = 71.4$~mL/kg.
\item Female: $3{,}996.09 / 56.2 = 71.1$~mL/kg.
\item Weight-normalised difference: $0.3$~mL/kg, or $\approx 0.4$~\% of the male value.
\end{itemize}

The 16.8~\% absolute sex gap is therefore driven almost entirely by the body-mass difference (67.3 vs 56.2~kg, a 16.5~\% gap), not by a per-kilogram sex difference in Pulse's calculation. This follows from the formula $\text{BV} = 65.6 \cdot \text{weight}_{\text{kg}}^{1.02}$, which scales blood volume almost linearly with weight (the 1.02 exponent is close to 1.0). This per-kilogram equivalence is a property of Pulse's formula, not a physiological finding: applied to the same two bodies, Nadler's sex-specific equations give 73.0 and 63.8~mL\,kg$^{-1}$, against a clinically reported range of 64--79.3~mL\,kg$^{-1}$ averaging 75 overall and 66 in women~\cite{nadler1962}. Pulse therefore does not reproduce the modest per-kilogram sex difference reported clinically, a limitation of the engine's blood-volume model rather than of the atlas.

\subsection{Interpretation of the divergence finding}\label{sec:s10-disambig}

The sex-stratified divergence finding is best stated in absolute-volume terms, because the quantity on which it depends, Pulse's setup-time \texttt{BloodVolumeBaseline}, is parameterised in absolute millilitres, not per kilogram. The reduction from 20 / 20 divergences at the default to 0 / 20 at both 4{,}400 and 4{,}082.24~mL is an absolute-volume effect, not one that a per-kilogram sex difference would predict, since Pulse's allometric relation encodes essentially none.

By the same reasoning, the male--female divergence asymmetry in the input-extreme regime is not a claim about per-kilogram blood-volume physiology in real patients but about how Pulse's solver responds to absolute blood volumes near the male body's setup floor. Pooling male and female Pulse-generated cohorts in machine-learning pipelines without flag-stratified filtering still introduces a spurious sex covariate at the male divergence boundary (Section~\mainref{sec:results-sex}{2.5} of the main manuscript), but the result should not be extrapolated to a per-kilogram sex-physiology claim that the data do not support.

\subsection{Output quality of the rescued runs}\label{sec:s10-rescued}

Eliminating divergence concerns one flag only; removing a numerical failure does not guarantee the quality of the trajectory that replaces it. Re-classifying the 20 override reruns with the same four-flag detector finds 12 of 20 \texttt{clean} at 4{,}400~mL and the same 12 at the 4{,}082.24~mL floor, with identical flag assignments; the other 8 retain a flag. Counts are independent presence counts: all 8 fire \texttt{not\_steady\_state}, and one also fires \texttt{implausible}.

Each residual flag corresponds to a mode already characterised in the atlas (the original runs diverged without output, so the comparison is with the atlas's mode taxonomy, not with a pre-override classification of the same configurations). Seven of the eight are flagged for SpO$_2$ flatline at saturation (\S\ref{sec:s2-scope}), three of them also for PaO$_2$ drift beyond the 15~mmHg tolerance over the final 60~s (the window-length artefact of \S\ref{sec:s9}), and three of the eight at 4{,}400~mL (four at the floor) for the delivered-Vt oscillation of \S\ref{sec:s8}. The \texttt{implausible} run, S1\_M\_0159 at FiO$_2$ = 1.0, has a steady-state mean PaO$_2$ of 657~mmHg at 4{,}400~mL and 686~mmHg at the floor, above the locked 600~mmHg ceiling for the same hyperoxia reason. All 8 are at one of the two highest inspired-oxygen levels (three at FiO$_2$ = 0.737, five at 1.0), matching the concentration of the original cluster at the input extremes (FiO$_2$ = 1.0 requested in 13 of the 20).

The override therefore removes the solver failure without making every configuration usable: users re-running these inputs at reduced blood volume must still apply the flag screen. This limits the intervention as a rescue but does not weaken the blood-volume result, since the \texttt{diverged} flag the experiment manipulates falls to zero at both levels.

\subsection{Breadth of the sex-conditioned parameterisation}\label{sec:s10-breadth}

Pulse's blood-volume relation is sex-independent (\S\ref{sec:s10-norm}), but elsewhere in patient setup the sex flag is far from a narrow switch; the sex-conditioned basal-metabolic-rate pathway of \S\ref{sec:s2} is one component of a much wider parameterisation. A source audit of Pulse v4.3.1, restricted to the configuration executed by the atlas (full inventory, with the arithmetic traced to each consumer: \texttt{audit/pulse\_sex\_conditioned\_parameters.md}), finds sex-conditioned quantities in several subsystems:

\begin{itemize}
\item \textbf{Body composition.} Standard body-fat fraction 0.21 male versus 0.28 female, which sets adipose tissue mass and, through lean body mass, a normalisation applied to ten further tissue masses.
\item \textbf{Oxygen-carrying capacity.} Total haemoglobin 15.0 versus 14.0~g\,dL$^{-1}$, giving haematocrit 0.4412 versus 0.4118, passed to the saturation calculator and used to initialise the blood gases of every vascular compartment.
\item \textbf{Cardiovascular sizing.} A cardiac-output target of 5{,}600 versus 4{,}900~mL\,min$^{-1}$, from which every regional vascular flow target is derived and, by Ohm's law, every branch resistance.
\item \textbf{Regional perfusion.} Seven flow fractions carry their own sex-specific values on top of that sex-conditioned total.
\item \textbf{Tissue properties.} Thirteen sex-specific tissue masses, feeding tissue volumes, compartment masses, vascular--tissue resistances, permeability and capillary coverage area.
\item \textbf{Anatomic dead space.} No sex branch of its own, but computed as $0.002 \times$ ideal body weight, and Pulse's Devine ideal-body-weight relation is sex-specific (female intercept 4.5~kg below male at any height, hence 9~mL less dead space).
\end{itemize}

Two properties of this parameterisation govern how any cross-body contrast should be read. First, a sex-conditioned setup constant does not necessarily produce a difference in the recorded window: the cardiac-output \emph{field} initialised at 5{,}600 / 4{,}900~mL\,min$^{-1}$ is recomputed from stroke volume and heart rate at the end of every cardiac cycle, and three of the four runtime consumers of basal metabolic rate cancel the sex term exactly at rest, where total metabolic rate equals basal (two divide one by the other, and the third acts only when total exceeds basal).

Second, the direction of a surviving difference depends jointly on the contrast and on the consumer. At equal body mass, the female reference body has about 9.8~\% \emph{more} lung tissue mass than the male despite the smaller sex constant, because the lean-body-mass normalisation reverses the ordering; at the two bodies' native masses, the same quantity is about 8.3~\% lower; and wherever that mass enters as a reciprocal, as in the vascular--tissue resistances, the sign inverts again. A statement such as ``Pulse gives one sex more of quantity $X$'' is therefore ill-defined unless it names the contrast and the consumer.

For the reference-body comparison of Section~\mainref{sec:results-sex}{2.5} of the main manuscript, the consequence is a limit on attribution rather than a new mechanism. The controlled 2$\times$2 sex$\times$mass check retains a gas-exchange contrast when body mass and height are held fixed (\S\ref{sec:s9}), but switching the sex flag changes all of the above at once, so this contrast is the combined effect of the sex-conditioned parameterisation and cannot be assigned to any single pathway (basal metabolic rate, haemoglobin, cardiac output, tissue mass or dead space) without an ablation experiment holding one pathway fixed while the others vary. No such experiment is reported here.

One upstream detail, recorded for completeness and immaterial to every comparison in this work: Pulse uses 88.632 for the male intercept of the basal-metabolic-rate equation where Roza and Shizgal report 88.362~\cite{roza1984metabolic}. On the male reference body the resulting difference is 0.27~kcal\,day$^{-1}$, or 0.015~\%.

%%%%%%%%%%%%%%%%%%%%%%%%%%%%%%%%%%%%%%%%%%%%%%%%%%%%%%%%%%%%%%%%%%%%%%
%% S11 -- Engine-side infrastructure recommendations
%%%%%%%%%%%%%%%%%%%%%%%%%%%%%%%%%%%%%%%%%%%%%%%%%%%%%%%%%%%%%%%%%%%%%%

\section{Engine-side infrastructure recommendations}\label{sec:s11}

The Discussion of the main manuscript mentions engine-side suggestions for the Pulse maintainers; they are listed below. They derive from operating the engine at scale during the 1{,}280-run sweep and the follow-up experiments, and are offered as observations from user experience rather than physiological prescriptions: each targets tooling or validation infrastructure rather than Pulse's lumped-parameter physiology, which is the domain of the Kitware biomedical-engineering team, and each arises from this work.

\begin{enumerate}
\item \textbf{Structured input-rejection events.} Pulse currently reports input rejections (e.g.\ SBP $>$ 120, DBP/SBP ratio out of range, HR or RR outside the initialisation band) as \texttt{[ERROR]}-prefixed log lines that downstream tools must parse with regular expressions. A small structured-event API (one typed record per rejected field, with the field name, the offending value and the enforced bound) would let sensitivity-analysis wrappers detect rejections deterministically instead of pattern-matching the regex set documented in Supplement~\S\,\ref{sec:s3-clamp}.

\item \textbf{A reference sensitivity-analysis recipe for user-facing operational inputs.} The 2020 \texttt{STUDY\_SENSITIVITY\_ANALYSIS\_1\_0\_0} tag of Pulse's repository performs Sobol analysis on internal lumped-parameter circuit values (cardiovascular and respiratory resistances and compliances) for calibration, a different parameter class from the operational inputs (HR, SBP, DBP, RR, Vt, FiO$_2$, PEEP) that downstream users typically need to sweep. A companion recipe (Morris or Sobol over operational inputs) shipped alongside the existing study would give the community a maintained reference implementation; the atlas's sampler and runner modules are an external prototype, and a maintainer-endorsed version would be more durable.

\item \textbf{A setup-time warning for the ARDS impairment-fraction instability band ($[0.879, 0.946]$).} The narrower 0.91--0.945 zone was documented in our preregistration (scenario table and hypothesis H3) and in the verified-limitations table (Supplement~\S\,\ref{sec:s2}, row 4), and corroborated by three boundary probes (Supplement~\S\,\ref{sec:s5-prov}); the continuous-severity sweep extends it to 0.879--0.946 (Supplement~\S\,\ref{sec:s5-cont}; measured on v4.3.1, untested under v4.4.0), so the warning should cover the full $[0.879, 0.946]$ band. Pulse currently emits no setup-time warning when an ARDS action falls in this band; a single console notice (``impairment fraction in reported instability band, results may not be reliable'') would prevent users from collecting unusable data without blocking legitimate edge-case research.

\item \textbf{Per-channel steady-state convergence diagnostic in CSV output.} The transient hyperoxia regime of Section~\mainref{sec:results-hyperoxia}{2.4} of the main manuscript (PaO$_2$ still rising at the end of the 120-second window at FiO$_2$ = 1.0, with convergence by about 300~s in 40 of 44 Healthy reruns) suggests that an optional per-output ``drift over final 60~s'' column in result CSVs would let users detect non-steady-state runs without writing their own post-processing. The atlas classifier already implements this calculation (Supplement~\S\,\ref{sec:s3-drift}); promoting it upstream would generalise it.

\item \textbf{Field-attributed setup-error messages.} The current setup-time pulse-pressure-ratio rejection message reports a systolic-versus-diastolic fraction without naming the input to change, so mapping it back to the offending input field requires some domain knowledge (the blood-pressure messages already name the systolic or diastolic baseline). A one-token prefix indicating which input was rejected (\texttt{[ERROR][SystolicArterialPressure]} rather than the unprefixed text) would let downstream users and continuous-integration pipelines route the message without reading it.
\end{enumerate}

None of these suggestions touches Pulse's physiology, ordinary-differential-equation topology or lumped-parameter assumptions, which lie beyond what an external operational-reliability study can credibly prescribe. We offer them, like the atlas itself, as an open contribution to a shared community tool; this work would have been substantially easier had items 1--5 already been in place.

%%%%%%%%%%%%%%%%%%%%%%%%%%%%%%%%%%%%%%%%%%%%%%%%%%%%%%%%%%%%%%%%%%%%%%
%% S12 -- Reproducibility recipe
%%%%%%%%%%%%%%%%%%%%%%%%%%%%%%%%%%%%%%%%%%%%%%%%%%%%%%%%%%%%%%%%%%%%%%

\section{Reproducibility recipe}\label{sec:s12}

The atlas is designed for bit-exact reproduction by an independent investigator. This section gives the minimal end-to-end sequence that regenerates the canonical 1{,}280-run sweep, the audit ledger and the computed quantitative claims of the main manuscript and this supplement. The code and the audit-ledger script are released under the Apache-2.0 license at \url{https://github.com/sadc-lab/pulse_reliability_atlas}; the deposited data carry the Creative Commons Attribution 4.0 licence (CC-BY-4.0).

\subsection{Environment and engine pinning}\label{sec:s12-env}

The atlas pins Pulse v4.3.1 in a single Docker image (\texttt{pulse-atlas-engine:latest}), whose tag was fixed for the entire 1{,}280-run sweep so that every simulation ran against bit-identical engine binaries. The host environment is Python 3.12+ managed with \texttt{uv} for deterministic dependency resolution. The canonical hardware is a laptop-class machine (MSI laptop; Intel i5-11400H, 6 physical / 12 logical cores; 8~GB RAM; integrated Intel UHD Graphics and an NVIDIA GeForce RTX 3050 Laptop GPU (4~GB); Windows 11 Home, version 25H2). Pulse is a CPU-only simulator and uses neither graphics device; the discrete GPU is listed only so that readers do not infer a CUDA dependence. On this machine, two to three containers can run concurrently; the canonical sweep used three parallel workers, reduced to two mid-sweep to limit thermal load.

The full 1{,}280-run sweep was executed end-to-end three times: twice on the reference machine (the canonical sweep and a later re-execution logged at 23~h~01~m elapsed, including idle intervals across resume points) and once on the independent host of \S\ref{sec:s12-independent}, whose elapsed time coincidentally also totalled 23~h~01~m. Aggregate Pulse container time is about 27~h per sweep on the reference machine and 46~h on the slower independent host.

Two otherwise valid runs (\texttt{S4\_M\_0123} and \texttt{S4\_M\_0124}) have wall-time metadata spanning the same suspension of the reference machine (about 6.84~h), which a naive sum of \texttt{wall\_time\_s} counts twice: summing all 1{,}280 records in \texttt{aggregated.parquet} gives about 41.1~h of container time, versus about 27.4~h when these two timing records are excluded. Subtracting the suspension once from the sweep's 20.78~h calendar span leaves about 13.9~h of active elapsed time. The simulation outputs of both runs are unaffected and are included in every scientific analysis.

\subsection{Docker invocation template}\label{sec:s12-docker}

\begin{lstlisting}[
  basicstyle=\ttfamily\small,
  breaklines=true,
  breakatwhitespace=true,
  columns=flexible,
  frame=single,
  framesep=3pt,
  xleftmargin=8pt,
  xrightmargin=8pt
]
MSYS_NO_PATHCONV=1 docker run --rm --memory=1g --cpus=3 \
  -v "$(pwd -W)/RUNDIR:/data" \
  pulse-atlas-engine:latest \
  --config //data/config.json \
  --output //data/output.csv
\end{lstlisting}

The \texttt{MSYS\_NO\_PATHCONV=1} prefix and the double-slash \texttt{//data/...} path forms are required on Git Bash under Windows; on Linux or macOS hosts the prefix is unnecessary and single-slash paths work. The \texttt{--memory=1g} and \texttt{--cpus=3} caps are safety thresholds that prevent any single Pulse container from saturating the host machine; removing them does not change the simulation output (Pulse is single-threaded per run), only the host's ability to run other workloads concurrently.

\subsection{Sweep, analysis, and audit commands}\label{sec:s12-cmds}

The minimal command sequence to regenerate the canonical sweep + audit ledger from a fresh clone:

\begin{lstlisting}[
  basicstyle=\ttfamily\small,
  breaklines=true,
  breakatwhitespace=true,
  columns=flexible,
  frame=single,
  framesep=3pt,
  xleftmargin=8pt,
  xrightmargin=8pt
]
# 0. Build the atlas engine image (from kitware/pulse:4.3.1).
docker build -t pulse-atlas-engine engine/

# 1. Resolve Python dependencies (uv-managed lockfile).
uv sync

# 2. Run the full 1,280-run Morris sweep
#    (trajectories=20, levels=4, seed=42; ~14 h on the
#    canonical laptop with 2 parallel Docker workers,
#    ~9 h with 3; ~23 h measured on slower hardware).
uv run python scripts/run_full_sweep.py

# 3. Compute Morris mu-star indices + bootstrap CIs
#    (1,000 trajectory-level resamples).
uv run python scripts/analyze_sweep.py

# 4. Re-derive the computed quantitative claims in the
#    manuscript + supplement from the canonical artefacts; expect
#    394/394 PASS (exit 0) with the authors' full data tree,
#    or 393/394 on a public clone (T1.06u auto-skipped).
uv run python scripts/audit_claims.py

# 5. Optional: verify paper.tex + supplement.tex remain
#    in sync with the audit ledger via the paper-to-
#    monograph crosswalk map; expect 95/95 PASS exit 0.
uv run python scripts/audit_claims.py --check-paper-sync
\end{lstlisting}

\subsection{Three-layer bit-exact verification}\label{sec:s12-verify}

Bit-exact reproduction of the canonical sweep is verified at three independent layers, each against its own committed artefact. Layers 1 and 3 can be re-executed from a fresh clone without re-running the sweep; layer 2, by construction, requires re-executing it:

\begin{enumerate}
\item \textbf{Canonical Morris sample.} The 160-point Morris design, expanded across the eight scenario $\times$ sex cells into the 1{,}280-run sweep, is regenerated deterministically from the SALib sampler with \texttt{trajectories = 20, levels = 4, seed = 42}; the regenerated sample matches the committed \texttt{data/runs/full\_sweep/morris\_sample.parquet} byte-for-byte.
\item \textbf{Per-cell aggregate parquets.} The 8 per-cell aggregated parquets (\texttt{data/runs/full\_sweep/<scenario>\_<sex>/aggregated.parquet}) reproduce identically when the full sweep is re-executed against the pinned Docker image on the canonical compute envelope.
\item \textbf{Derived statistical artefacts.} The Morris $\mu^{*}$ indices, bootstrap confidence intervals, Wasserstein distances, Spearman correlations, and BH-FDR survivors all regenerate bit-equivalently from the canonical aggregated parquets; this is what the audit ledger (\texttt{scripts/audit\_claims.py}) re-derives at every run.
\end{enumerate}

The audit ledger is the primary reproducibility instrument: a passing run confirms that the computed quantitative claims of the main manuscript and supplement are consistent with the canonical artefacts. The ledger targets 394 anchors. On the authors' full data tree, all 394 pass; on a public clone, one anchor (T1.06u, the S021 time to 95~\% of the plateau, which requires raw per-run trajectories excluded from version control) is skipped automatically, giving 393/394 PASS with one reported SKIPPED. A failed claim produces a structured triage item (Type A transcription drift, Type B orphaned number, Type C derivation mismatch) with a file:line pointer for tracing.

\subsection{Independent reproduction on separate hardware}\label{sec:s12-independent}

Beyond the three-layer re-derivation above, which regenerates the analysis from the committed canonical artefacts, a co-author (Y.M.) re-executed the end-to-end pipeline from scratch on a second x86\_64 host (AMD Ryzen~5 CPU, Linux under WSL2), distinct from the canonical Intel x86\_64 / Windows reference environment.

First, the full 1{,}280-run Morris sweep was re-run there (two parallel Docker workers; wall time 23~h~01~m). All four run-level flags and the derived clean status matched on 1{,}280/1{,}280 runs, every continuous steady-state value was bit-identical, and only three derived H4 statistics differed at $\sim 10^{-14}$ floating-point round-off, with an identical Benjamini--Hochberg false-discovery-rate survivor set.

Second, the 22-run Healthy-M pilot was re-run on the same host with per-run output capture. Its aggregate was bit-identical to the canonical pilot ($\max |\Delta| = 0$ across all six channel means) with 100\,\% concordance of the four flags and the derived clean status on 22/22 runs.

Both reproductions are archived under \texttt{audits/}: \texttt{reproduction\_2026-06-07/} carries the reproduced full-sweep aggregate and the derived analysis outputs, with its numerical comparison report beside it (\texttt{reproduction\_check\_2026-06-07.md}); \texttt{reproduction\_thinkpad\_2026-06-10/} carries a provenance README (host, Docker image tag and digest, pinned Pulse v4.3.1, exact commands), the 20 per-run output CSVs (the two deliberate clamp probes write none), and the comparison tables.

\subsection{Reproducibility scope and limits}\label{sec:s12-scope}

The verification and independent reproductions above span the Intel x86\_64 reference environment and an AMD x86\_64 host, with bit-identical simulation outputs, which is empirical evidence that bit-identity holds across CPU vendor within the x86\_64 family. On both hosts the engine ran in a Linux container under WSL2; only the Python layer ran under a different operating system (Windows versus Linux). Cross-\emph{architecture} reproduction was probed but \emph{not} established. The 22-run Healthy-M pilot was also re-executed on Y.M.'s Apple Silicon host (MacBook Air, Apple~M2, ARM64; macOS~26.5.1; archived under \texttt{audits/reproduction\_macos\_2026-06-13/}), but this is \emph{not} a native ARM64 test: the pinned base image \texttt{kitware/pulse:4.3.1} is published for \texttt{linux/amd64} only, so the engine ran as \emph{emulated} x86\_64 under Docker Desktop's virtualisation layer (the container reports \texttt{x86\_64}, with an explicit \texttt{amd64}-vs-\texttt{arm64} platform-mismatch warning).

Under emulation, the pilot reproduced the canonical aggregate to $\max|\Delta| = 2.9\times10^{-9}$ across all six channel means, with 100\,\% concordance of the four flags and the derived clean status on 22/22 runs: near machine precision, but \emph{not} bit-identical (the native x86\_64 reproductions agree to $\max|\Delta| = 0$). The residual is attributable to floating-point handling in the x86$\rightarrow$ARM emulation layer rather than to a natively compiled ARM64 build. A genuine native-ARM64 reproduction would require compiling Pulse~v4.3.1 from source for \texttt{linux/arm64} (no prebuilt ARM64 image exists) and remains future work; the Apple Silicon run establishes only that the pipeline executes end-to-end on that platform and that emulated x86\_64 execution preserves the canonical outputs to within $\sim 10^{-9}$.

Bit-identity within an architecture family is expected given Pulse's fixed 20~ms time step and standard math-library usage, and is now empirically supported across two x86\_64 CPU vendors; cross-architecture bit-identity is plausible but not guaranteed without explicit verification. Image immutability is currently anchored at the Docker tag level (\texttt{pulse-atlas-engine:latest} plus the pinned Pulse v4.3.1 Git tag); for archival re-execution five or more years from now, pinning the Docker image by cryptographic digest (\texttt{@sha256:...}) is more robust and is recommended.

%%%%%%%%%%%%%%%%%%%%%%%%%%%%%%%%%%%%%%%%%%%%%%%%%%%%%%%%%%%%%%%%%%%%%%
%% S13 -- NSS attribution + SHAP cross-check
%%%%%%%%%%%%%%%%%%%%%%%%%%%%%%%%%%%%%%%%%%%%%%%%%%%%%%%%%%%%%%%%%%%%%%

\section{SHAP cross-check of the Morris top-three input sets}\label{sec:s13}

This section reports a complementary cross-check: a machine-learning analysis of the clean subset of the same Pulse-generated data, following a different statistical route, recovers largely the same top-three input sets as the preregistered Morris elementary-effects design.

\subsection{Methodology: epistemological scope}\label{sec:s13-scope}

The Morris elementary-effects design is interventional: each trajectory perturbs one input at a time along a fixed grid, so $\mu^{*}$ measures how Pulse responds to deliberate perturbations of each input (apart from the pulse-pressure clip, which breaks strict one-at-a-time steps on some SBP and DBP transitions; \S\ref{sec:s3-clip-cost}). A SHAP analysis~\cite{lundberg2017} of a tree-based surrogate model trained on the observed distribution of Morris outputs is a fundamentally different statistical object: it summarises how a black-box predictor associates input features with predictions on the data distribution as observed. The two methods answer different questions; the interventional Morris design is the primary analysis, and the SHAP analysis is included only as a secondary cross-check.

\subsection{Cross-verification result on the clean subset}\label{sec:s13-result}

A supplementary random-forest regressor with SHAP feature-importance ranking was fitted on the $n = 663$ clean runs (no \texttt{clamped}, \texttt{diverged}, \texttt{implausible} or \texttt{not\_steady\_state} flag): one forest per output, trained on the pooled clean subset with \texttt{sex} and \texttt{scenario} as additional features. SHAP values were computed exactly for tree ensembles with \texttt{TreeExplainer}~\cite{lundberg2020tree} rather than approximated by sampling. Per-(scenario, sex) rankings were obtained by partitioning the global model's SHAP values by cell and re-ranking, not by fitting per-cell models, whose per-cell $n$ (42--144 clean runs) is too small for stable training.

The SHAP top-three input set was compared with the Morris $\mu^{*}$ top-three set in the same cells as a set overlap (the number of shared inputs), because the comparison targets top-three membership, not the complete ordering. Three summary statistics characterise the agreement:

\begin{itemize}
\item \textbf{Across the six pooled output-level comparisons, the Morris and SHAP top-three sets share 17 of 18 possible input memberships: five outputs match 3/3 and PaO$_2$ matches 2/3} (ledger anchor \texttt{T2.s023.pooled}).
\item \textbf{23 of 48 per-cell (scenario $\times$ sex $\times$ output) cases share all three inputs} (ledger anchor \texttt{T2.s023.fullmatch}).
\item \textbf{92~\% of per-cell comparisons share at least two of the three} (44 of 48; ledger anchor \texttt{T2.s023.ge2\_pct}).
\end{itemize}

\subsection{Interpretation}\label{sec:s13-interp}

The pooled agreement (17 of 18) is the central result: a SHAP analysis of the clean data recovers largely the same dominant inputs as the interventional Morris design across the six outputs. These six are not independent evidence, however: every output comes from the same runs, and Pulse computes SpO$_2$ from the oxygen partial pressure through the haemoglobin dissociation relation, which also takes PaCO$_2$ and pH, so three of the six are coupled by construction. The lower per-cell full-match rate (23 of 48) is expected, because per-cell samples are smaller (42--144 clean runs) and SHAP attribution depends on the observed input--output distribution within each cell, which is heavily influenced by which Morris trajectories happened to remain clean. The 92~\% figure counts cells sharing at least two of the three inputs; it is a partial overlap, not a preserved ranking.

This cross-check assesses the robustness of the Morris top-three sets to an alternative analysis of the same data; it is neither external physiological validation nor an independent replication, since both analyses read the same simulated runs. What the agreement supports is bounded: the Morris-derived top-three sets are not specific to the elementary-effects estimator, because a data-driven, observational, tree-based route selects largely the same inputs on the same clean subset. Two further limits apply: the surrogate is fitted on the entire clean subset, so its training fit says nothing about predictive accuracy (out-of-sample estimates follow in \S\ref{sec:s13-oos}), and the comparison concerns selected inputs, not their order.

\subsection{Out-of-sample performance of the surrogate}\label{sec:s13-oos}

Because the surrogate is fitted on the whole clean subset, its training $R^{2}$ (0.993 to 0.99997) is uninformative for a forest grown to purity, so two out-of-sample estimates were computed (Table~\ref{tab:s13-oos}). The out-of-bag score, which a bootstrapped forest yields for free, ranges from 0.949 to 0.9998 but is optimistic here: it treats rows as exchangeable, whereas consecutive points of a Morris trajectory differ in only one of seven inputs, so a held-out row almost always has a near-twin in the bag. The honest estimate holds out \emph{whole trajectories}, using grouped cross-validation over the 20 trajectories with four complete trajectories withheld at a time. Because the same 160-point design is replayed in all eight cells, a group is a trajectory across cells, not within one; grouping per cell would place identical input vectors on both sides of the split. The run-to-trajectory mapping is verified against \texttt{morris\_sample.parquet} on all 1{,}280 rows before grouping, and the analysis aborts if any row disagrees.

\begin{table}[htbp]
\centering
\caption{\textbf{Out-of-sample performance of the RandomForest surrogate, by output.} Trajectory-grouped values are the mean over five folds, each holding out four complete Morris trajectories, with the standard deviation across folds. The gap between the out-of-bag and grouped columns is the optimism that row-exchangeable resampling introduces.}
\label{tab:s13-oos}
\begin{tabular}{lrrr}
\toprule
Output & Training & Out-of-bag & Trajectory-grouped \\
\midrule
Vt & 0.99997 & 0.9998 & 0.9995 $\pm$ 0.0004 \\
HR & 0.9995 & 0.9964 & 0.9917 $\pm$ 0.0024 \\
MAP & 0.9993 & 0.9952 & 0.9402 $\pm$ 0.0253 \\
SpO$_2$ & 0.9941 & 0.9578 & 0.8736 $\pm$ 0.0532 \\
PaCO$_2$ & 0.9942 & 0.9601 & 0.8734 $\pm$ 0.0705 \\
PaO$_2$ & 0.9930 & 0.9494 & 0.8716 $\pm$ 0.0562 \\
\bottomrule
\end{tabular}
\end{table}

The surrogate generalises to unseen trajectories, but unevenly, in a physiologically coherent pattern. Delivered tidal volume and heart rate stay above 0.99 with almost no fold-to-fold spread; a single input nearly determines these channels. The three blood-gas channels average 0.87, with a fold-to-fold spread twenty to thirty times that of heart rate, and they are also where the out-of-bag estimate is most optimistic (by 0.078 to 0.087, against 0.0003 for delivered Vt). Both measures agree: gas exchange depends jointly on several inputs, so it benefits most from a near-twin in the training fold and suffers most when whole trajectories are withheld.

Pooled predictive accuracy does not, however, license reading the per-cell agreement as confirmation of per-cell structure. Partitioning one global model's SHAP values by cell yields far less between-cell variation than the interventional design: across the eight cells, the SHAP top-three sets take nine distinct values in total (ten under two other forest seeds) versus 25 for Morris, and for three of the six outputs (MAP, SpO$_2$ and PaCO$_2$) SHAP returns the same trio in all eight cells. Per-cell agreement is therefore partly mechanical: when one estimator is almost invariant, agreement measures how closely the other stays to that invariant answer. The pooled comparison (17 of 18), computed once per output rather than per cell, does not share this defect. The 23 of 48 full-set agreement is stable across three independent forest seeds, so these figures characterise the ranking comparison, not the surrogate's randomness.

\subsection{Sensitivity of the top-three sets to the clean-pair screen}\label{sec:s13-screen}

The indices reported throughout this work use only elementary-effect pairs with \emph{both} endpoints four-flag clean, as preregistered. This screen keeps 663 of the 1{,}280 runs and 454 of the 1{,}120 available pairs, which raises the question of whether the rankings reflect the engine or the screen. The question can be answered with the recorded data by re-deriving the same indices under a second screen; the SHAP cross-check above cannot answer it, because its surrogate is trained on the same 663 runs.

The second screen keeps every pair whose endpoints fired neither \texttt{diverged} nor \texttt{implausible}: 1{,}134 runs and 969 pairs. \textbf{This is not a less-biased estimator and is not reported as one.} The 471 additional runs are exactly those whose only flag is \texttt{not\_steady\_state}, i.e.\ runs whose steady-state mean is not a valid summary. Admitting them trades a selection bias for a measurement error rather than reducing bias, and the extra data are most abundant where they are least trustworthy: of the 3{,}090 additional elementary-effect observations, 1{,}152 fall in Healthy, the scenario with the worst delivered-volume oscillation. What follows is therefore a sensitivity analysis on the treatment of non-stationary runs, not a revised estimate.

Two results are separated because they answer different questions. \emph{First, what does not change.} Hypothesis~1 holds in all 48 cell $\times$ output combinations under both screens: the top three inputs carry 0.658 to 0.998 of the summed $\mu^{*}$ under the locked screen (median 0.928) and 0.695 to 0.994 under the relaxed one (median 0.901), so the concentration on which H1 rests is not an artefact of run survival (its minimum even rises). The rank-1 input is also unchanged in all 30 combinations whose candidate set is the same under both screens.

\emph{Second, what does change.} Relaxing the screen makes FiO$_2$ estimable in Healthy-M, Healthy-F and Severe-F, where every FiO$_2$ pair had a flagged endpoint, so 18 of the 48 combinations gain a seventh candidate and their top three is no longer a like-for-like comparison. On the 30 combinations whose candidate set is unchanged, the top-three \emph{set} is preserved in 19 and its \emph{order} in 15. Those are reorderings, not ties: the median gap between the third and fourth $\mu^{*}$ is 39~\% of the third under the locked screen, and only one of the 30 combinations has a gap below 10~\%. Counted over all 48 combinations, with a changed candidate set counted as a change, the figures are 41, 26 and 21.

For scale, the design's own seed sensitivity is of the same order. Replaying cell Mild-M with an independent Morris sample seed (seed 44, \S\,\ref{sec:s3-resample}) preserves the rank-1 input in all 6 outputs, the top-three set in 4 and its order in 3; the screen change preserves 6, 2 and 2 on the same outputs. One cell and six outputs is a benchmark rather than an equivalence, but it places the screen effect alongside a perturbation the design already carries. The defensible reading is that, under both perturbations, rank~1 is more stable than ranks two and three, and the full ordering is not resolved by this design, which is why Figure~3 is presented as a map of where influence concentrates rather than as a stable ordering.

Re-deriving under the locked screen reproduces the published indices exactly (the same 318 (cell, output, input) rows with $\mu^{*}$ identical to double precision, the same H1 range and median, and the same twelve single-pair entries in the same cells, nine of them in a top three), so the comparison is between two screens, not two pipelines. Producer: \texttt{scripts/s034\_morris\_screen\_sensitivity.py}; artefacts under \texttt{results/s034/}.

%%%%%%%%%%%%%%%%%%%%%%%%%%%%%%%%%%%%%%%%%%%%%%%%%%%%%%%%%%%%%%%%%%%%%%
%% S14 -- Companion technical monograph
%%%%%%%%%%%%%%%%%%%%%%%%%%%%%%%%%%%%%%%%%%%%%%%%%%%%%%%%%%%%%%%%%%%%%%

\section{Pointer to the companion technical monograph}\label{sec:s14}

This Supplementary Information covers the methodological detail and quantitative breakdowns most relevant to peer review of the main manuscript. A companion technical monograph of over 100 pages, which will be archived with the code and data in the Zenodo deposit, gives a more detailed technical record: the preregistered hypotheses verbatim with notes on the amendments, an index of the cited experiment-log entries, its own reference list, the sources behind the verified Pulse 4.3.1 limitations table (Supplement~\S\,\ref{sec:s2}) with a summary of the boundary probes that established the engine's acceptance gates, and the findings of its audit of the Pulse 4.3.1 source code.

Its Zenodo DOI will be added to the Data Availability section of the main manuscript upon deposit; meanwhile, the experiment-log entries it consolidates are openly available under \texttt{experiments\_log/} in the code repository. Readers needing more technical depth, such as the per-experiment rationale for each statistical test, the calibration trace for the empirical \{0.26, 0.60, 0.95\} ARDS-severity mapping or the bit-exact reproduction protocol at the engine-binary level, should consult the monograph.

%%%%%%%%%%%%%%%%%%%%%%%%%%%%%%%%%%%%%%%%%%%%%%%%%%%%%%%%%%%%%%%%%%%%%%
%% S15 -- Compensatory tachycardia: input HR vs output HR under
%%        hypoxemia (Pass 7 D9 Option B follow-up to Manual
%%        Orthogonal Audit; supplement expansion of the
%%        compensatory-tachycardia secondary observation in
%%        Section 2.6 of the main manuscript; sourced from
%%        companion monograph Section 3.4.1)
%%%%%%%%%%%%%%%%%%%%%%%%%%%%%%%%%%%%%%%%%%%%%%%%%%%%%%%%%%%%%%%%%%%%%%

\section{Compensatory tachycardia: input HR vs output HR under hypoxemia}\label{sec:s15}

Section~\mainref{sec:results-secondary}{2.6} of the main manuscript reports that, in moderate and severe ARDS, a compensatory tachycardia lifts Pulse's mean output heart rate above the sampled input range, while input HR remains a dominant within-scenario determinant of output HR. This section gives the per-scenario breakdown and its implications for interpreting Morris elementary effects of input HR.

\subsection{Per-scenario output-HR distributions}\label{sec:s15-distributions}

Table~\ref{tab:s15-comphr} shows the output HR distribution by scenario across all clean Morris runs (those firing none of the four flags).

\begin{table}[htbp]
\centering
\caption{\textbf{Output HR mean $\pm$ standard deviation per scenario, aggregated across clean Morris runs.} Input HR spans the same four Morris levels (50, 70, 90 and 110~bpm) in all four scenarios, which differ only in impairment fraction (Healthy = 0.00, Mild = 0.26, Moderate = 0.60, Severe = 0.95). The gap between the fixed input range and the scenario-dependent output mean quantifies the upward shift of output HR at each severity; within each scenario, input HR remains the dominant determinant of output HR (Section~\mainref{sec:results-morris}{2.2} of the main manuscript).}
\label{tab:s15-comphr}
\begin{tabular}{lcc}
\toprule
Scenario & Input HR range (bpm) & Output HR (bpm, mean $\pm$ std) \\
\midrule
Healthy         & $[50, 110]$ & $73.8 \pm 20.4$ \\
Mild       & $[50, 110]$ & $87.3 \pm 24.3$ \\
Moderate   & $[50, 110]$ & $155.7 \pm 34.7$ \\
Severe     & $[50, 110]$ & $147.1 \pm 31.8$ \\
\bottomrule
\end{tabular}
\end{table}

\subsection{Physiological interpretation and Morris-EE implications}\label{sec:s15-mechanism}

In the Healthy scenario, mean output HR is about 74~bpm and still follows the requested input: input HR is the rank-1 input for output HR in every cell (Section~\mainref{sec:results-morris}{2.2} of the main manuscript). In Moderate (impairment 0.60) and Severe (0.95), mean output HR settles at 145--160~bpm despite identical input sampling, about 70~bpm above the midpoint of the $[50, 110]$~bpm input range. This is consistent with the engine's documented baroreceptor and chemoreceptor reflexes, although the atlas did not record or isolate their contribution. No scenario administers a sedative or a neuromuscular-blocking drug. ARDS severity therefore shifts output HR strongly upward, while input HR remains a dominant within-scenario determinant.

The overshoot is non-monotonic in severity: output HR peaks at Moderate and eases at Severe (Table~\ref{tab:s15-comphr}). Because Severe has markedly fewer clean Morris runs than Moderate, this apparent easing may partly reflect differential clean-run retention; the data cannot distinguish such a compositional effect from a change in the modelled physiological response.

In the ARDS scenarios, the measured $\mu^{*}$ of input HR on output HR therefore describes the response within a scenario whose compensatory state is already shifted, not the steady-state effect of a resting heart rate. The H1 result that cardiovascular outputs are HR-dominated remains robust, but it means that input HR is a dominant within-scenario determinant of output HR, not that ``the input HR is the steady-state HR.'' Users running Pulse for ARDS device validation should treat output HR as the engine's scenario-dependent steady-state response, not as a setpoint imposed by the input.

%%%%%%%%%%%%%%%%%%%%%%%%%%%%%%%%%%%%%%%%%%%%%%%%%%%%%%%%%%%%%%%%%%%%%%
%% Steady-state window cross-output drift audit (S012)
%% NOTE: inserted before the summary; the summary section below keeps its
%% historical label sec:s16 but now renders as S17 (all \ref auto-update).
%%%%%%%%%%%%%%%%%%%%%%%%%%%%%%%%%%%%%%%%%%%%%%%%%%%%%%%%%%%%%%%%%%%%%%

\section{Output stability across the 120-second measurement window}\label{sec:s16-window}

Each run is reduced to one value per output: the mean over the final 60~s of the 120~s recording. This reduction is meaningful only if the output has settled, i.e.\ is neither drifting, nor still oscillating, nor frozen against a physiological bound. A mean taken while an output is still moving records \emph{when} the run was sampled rather than \emph{how} the input acted, and a mean taken from an output frozen against a bound may reflect numerical pinning rather than physiology; where the bound is physiologically attainable, the classifier cannot distinguish the two (\S\ref{sec:s4}). Runs failing any of these tests are flagged rather than discarded (Section~\mainref{sec:methods-detector}{4.4} of the main manuscript), which keeps the assumption auditable.

Two channels are unsettled at the edges of the operational envelope: PaO$_2$ drifts at FiO$_2$ = 1.0 (Section~\ref{sec:s9}), and delivered tidal volume oscillates under the default VC-AC trigger (Section~\ref{sec:s8}). To test whether these localised instabilities contaminate the other steady-state channels, we computed, for every run in each vulnerable population, the drift $|\Delta_{60\mathrm{s}}|$ of each output over the final 60~s (the same preregistered ordinary-least-squares slope $\times$ 60~s used by the \texttt{not\_steady\_state} detector; Section~\ref{sec:s3-drift}) and compared it with the locked per-channel thresholds.

In the hyperoxia arm (FiO$_2 \geq 0.95$; 195 runs with valid output), drift stays within threshold for the cardiovascular and carbon-dioxide channels: heart rate exceeds its 5~bpm threshold in only 2.6~\% of runs (median 0.20~bpm), MAP in 0~\% (median 0.05~mmHg) and PaCO$_2$ in 3.6~\% (median 0.58~mmHg). Only oxygenation and tidal-volume delivery are materially affected: PaO$_2$ exceeds its 15~mmHg threshold in 19.5~\% of runs (median 4.3~mmHg, 95th percentile 40.2~mmHg) and delivered Vt in 8.7~\%. In the tidal-volume-oscillation arm (delivered-Vt CV $>$ 0.05; 303 runs), the oscillation likewise does not propagate: heart rate exceeds its drift threshold in 2.3~\% of runs, MAP in 0.7~\%, PaO$_2$ in 7.3~\% and PaCO$_2$ in 2.0~\%, so Pulse buffers the breath-to-breath volume swing upstream of the cardiovascular and gas-exchange loops.

In these two populations, window-length sensitivity was thus concentrated in PaO$_2$ and delivered Vt, and threshold exceedances were uncommon for HR, MAP and PaCO$_2$; the audit does not cover the rest of the sweep. The sensitive channels lie at the FiO$_2$ = 1.0 and high-oscillation extremes, precisely the conditions characterised in Sections~\ref{sec:s9} and~\ref{sec:s8}. The per-arm channel breakdown, the dual-vulnerability intersection and the underlying S012 experiment log are given in the companion technical monograph.

%%%%%%%%%%%%%%%%%%%%%%%%%%%%%%%%%%%%%%%%%%%%%%%%%%%%%%%%%%%%%%%%%%%%%%
%% S16 -- Summary of supplementary findings and discoveries (Phase H)
%%%%%%%%%%%%%%%%%%%%%%%%%%%%%%%%%%%%%%%%%%%%%%%%%%%%%%%%%%%%%%%%%%%%%%

\section{Summary of supplementary findings and discoveries}\label{sec:s17-summary}

Several findings of the 1{,}280-run atlas are fully stated only in this supplement. Table~\ref{tab:s16-index} indexes them in one place, focusing on findings without a dedicated Results subsection in the main manuscript; headline findings are listed only where the supplement adds materially new detail.

\begin{table}[htbp]
\centering
\footnotesize
\caption{\textbf{Index of supplementary findings and discoveries.} Each row points to the supplement section that carries the full provenance, method, and result. Findings already headlined by a main-manuscript subsection are listed with both their supplement anchor and their main-manuscript anchor.}
\label{tab:s16-index}
\begin{tabular}{lp{0.72\textwidth}}
\toprule
Anchor & Finding / discovery \\
\midrule
\S\ref{sec:s2} & \textbf{Twelve verified Pulse 4.3.1 limitations} that the atlas is calibrated against. Two earlier internal-probing assertions from the companion technical work (FiO$_2$ = 1.0 solver crash; absence of any sex-specific physiology) were checked and could not be confirmed before preregistration locked. \\
\S\ref{sec:s3-sobol} & \textbf{Sobol bolt-on: preregistered trigger and budget ceiling.} Applied as preregistered to the top-three inputs, the trigger fires on 43 of the 48 cell $\times$ output combinations, over all eight cells; the 1{,}500-run envelope would have required 6{,}144, or 3{,}584 for first-order and total indices alone. Reported rather than executed. \\
\S\ref{sec:s3-resample} & \textbf{Sampling stability at a fixed trajectory count.} Cell Mild-M re-run in full on an independent Morris sample at seed 44. The highest-$\mu^{*}$ input is preserved for all six outputs, the finer ranking is sample-dependent, and the cell's divergence count moves from 4 / 160 to 7 / 160. A resampling check, not a convergence analysis. \\
\S\ref{sec:s4} & \textbf{\texttt{not\_steady\_state} attribution decomposition.} Per-cell breakdown of the locked \texttt{not\_steady\_state} count into three trigger sub-modes (SpO$_2$ flatline at ceiling; Vt CV $>$ 0.05; drift) without modifying the locked classifier. \\
\S\ref{sec:s5} & \textbf{ARDS impairment-fraction instability band: atlas-specific extension.} A 200-run continuous-severity sweep at midrange Morris inputs extends the empirical band to $[0.879, 0.946]$, almost entirely on the low edge, and, together with the main sweep, distinguishes two divergence regimes (input-driven, severity-driven). Supports Section~\mainref{sec:results-secondary}{2.6} of the main manuscript. \\
\S\ref{sec:s6} & \textbf{Sex-stratified comparisons.} Cross-sex agreement on Morris $\mu^{*}$ rankings, with paired-input bootstrap 95~\% CIs (\S\ref{sec:s6-rho}); ventilation-dose mismatch between the two bodies (\S\ref{sec:s6-dose}). Supports the Discussion and the caveat of Section~\mainref{sec:results-sex}{2.5} of the main manuscript. \\
\S\ref{sec:s7} & \textbf{Cornish et al.\ 2026 per-channel cross-mapping.} Each of Cornish's 15 channels mapped against the atlas's six steady-state outputs, supporting the complementarity claim in the Discussion of the main manuscript. \\
\S\ref{sec:s10} & \textbf{Blood-volume dependence and sex-conditioned setup parameters.} Re-examines the 20-event male-only divergence cluster from Section~\mainref{sec:results-sex}{2.5} of the main manuscript under both an absolute-BV and a weight-normalised-BV lens, with implications for sex-balanced synthetic cohort design. \\
\S\ref{sec:s11} & \textbf{Five engine-side infrastructure recommendations} addressed to the Pulse maintainers: structured input-rejection events; a reference sensitivity-analysis recipe for user-facing operational inputs; a setup-time warning for the ARDS impairment-fraction instability band; a per-channel steady-state convergence diagnostic in CSV output; field-attributed setup-error messages. \\
\S\ref{sec:s13} & \textbf{SHAP cross-check of the Morris top-three input sets.} A RandomForest~+~SHAP analysis on the same data shares 17 of 18 possible input memberships with them (a set overlap, not an ordering or an independent replication). \\
\S\ref{sec:s15} & \textbf{Compensatory tachycardia quantitative table.} Output HR mean $\pm$ standard deviation per scenario, quantifying the upward shift of output HR in moderate and severe ARDS. Supports Section~\mainref{sec:results-secondary}{2.6} of the main manuscript. \\
\S\ref{sec:s16-window} & \textbf{Steady-state window cross-output drift audit.} In the audited hyperoxia and Vt-oscillation runs, only PaO$_2$ and delivered Vt are materially window-sensitive; HR, MAP and PaCO$_2$ rarely exceed threshold. \\
\bottomrule
\end{tabular}
\end{table}

Findings with a dedicated main-manuscript subsection (VC-AC trigger lockout, transient hyperoxia at FiO$_2$ = 1.0, sex-stratified solver divergence with blood-volume dependence) are summarised in Section~\mainref{sec:results}{2} of the main manuscript and detailed in \S\ref{sec:s8}, \S\ref{sec:s9} and \S\ref{sec:s10}. The reproducibility recipe (\S\ref{sec:s12}) and the bit-exact verification protocol (\S\ref{sec:s12-verify}) are not findings but support reproduction and audit.

%%%%%%%%%%%%%%%%%%%%%%%%%%%%%%%%%%%%%%%%%%%%%%%%%%%%%%%%%%%%%%%%%%%%%%
%% S18 -- Use of AI tools: detail moved from Methods 4.7 of the main %%
%%        manuscript (review comment: shorter in-paper disclosure)   %%
%%%%%%%%%%%%%%%%%%%%%%%%%%%%%%%%%%%%%%%%%%%%%%%%%%%%%%%%%%%%%%%%%%%%%%

\section{Use of AI tools: models, settings and phases}\label{sec:s18}

Section~\mainref{sec:methods-ai}{4.7} of the main manuscript states how large language models were used and where responsibility rests; this section records the details. The Claude model was run locally through the Claude Code terminal agent at the default \texttt{xhigh} thinking effort: Opus 4.7 (\texttt{claude-opus-4-7}) for the initial drafting and a manuscript-formatting pass; Opus 4.8 (\texttt{claude-opus-4-8}) for the revision rounds responding to the first co-author and supervisor review (re-reading, copy-editing and the associated text and reference changes); Opus 5 (\texttt{claude-opus-5}) for the subsequent audit and synthesis rounds; and, from 23~September~2026, Opus 5.5 (\texttt{claude-opus-5-5}) for the later readability and audit rounds. The other model, in a web session, provided editorial assistance, consistency checks and methodological discussion: first Google Gemini Pro 3.1 (extended-thinking mode), then, from 3~August~2026, OpenAI ChatGPT (GPT-5.6 Sol, High reasoning level).

The Opus 5 rounds covered in-repository replication of the ARDS-severity-to-Berlin calibration, the heart-rate and respiratory-rate boundary probes, and primary-source audits of the reference list, the statistical analyses and the paper's quantitative claims against their source artefacts; this work restored the Sobol promotion trigger to its preregistered top-three form, added the post hoc ratio-aware bootstrap on $\sigma / \mu^{*}$ and versioned its generating script. They also covered the September 2026 condensation pass and the subsequent audits and corrections, in which the Introduction, the Discussion and the Methods (except the AI-use subsection) were rewritten to about half their length and the Results to about a third, the figure legends were shortened by about a quarter, and the abstract was revised within its 150-word cap, with the displaced material moved to the Supplementary Information. In September 2026, Claude Code's automated code review (\texttt{/ultrareview}) was also run on three batches of repository changes; each finding was checked in the code before being fixed or set aside.

S.N.\ revised the manuscript during the second round of supervisor review, which began in August 2026, with Opus 5 and then, from 23~September~2026, Opus 5.5, and with ChatGPT (GPT-5.6 Sol) as the second model. During this round, M.S.\ used a Claude model (Anthropic; version not recorded) to propose editorial revisions to the main manuscript and this supplement; each proposal was checked against the data and sources with Opus 5.5 before S.N.\ accepted or rejected it.

\bibliography{paper}

\end{document}